\documentclass[prc,superscriptaddress,showpacs,twocolumn]{revtex4}
\usepackage{graphicx,dcolumn,array,bm,amsmath}
\usepackage{epsfig}

\newcommand{\beq}{\begin{equation}}
\newcommand{\eeq}{\end{equation}}
\newcommand{\ba}{\begin{array}}
\newcommand{\ea}{\end{array}}
\newcommand{\bn}{\begin{eqnarray}}
\newcommand{\en}{\end{eqnarray}}
\newcommand{\bnl}{\begin{mathletters}\begin{eqnarray}}
\newcommand{\enl}{\end{eqnarray}\end{mathletters}}
\newcommand{\bml}{\begin{mathletters}}
\newcommand{\eml}{\end{mathletters}}
\newcommand{\bc}{\begin{center}}
\newcommand{\ec}{\end{center}}
\newcommand{\bi}{\begin{itemize}}
\newcommand{\ei}{\end{itemize}}
\newcommand{\bt}{\begin{tabular}}
\newcommand{\et}{\end{tabular}}
\newcommand{\bnll}[1]{\begin{subequations}\label{#1}\begin{eqnarray}}
\newcommand{\enll}{\end{eqnarray}\end{subequations}}

\newcommand{\half}{\frac{1}{2}}

\newcommand{\thalf}{\tfrac{1}{2}}

\newcommand{\tquart}{\tfrac{1}{4}}

\newcommand{\rmd}{{\rm d}}

\newcommand{\intsum}{\int\hspace{-1.4em}\sum}
\newcommand{\tintsum}{{\textstyle\int}\hspace{-0.7em}{\mbox{\raisebox{0.1em}{$\scriptstyle\sum$}}}}

\renewcommand{\bbox}[1]{\bm{#1}}
\newcommand{\ofbboxofr}{(\bbox{r})}
\renewcommand{\ofbboxofr}{}

\newcommand{\tra}{^{(t)}}

\newcommand{\particle}{p-h}
\newcommand{\pairing} {p-p}

\begin{document}

\title{Self-consistent symmetries in
the proton-neutron Hartree-Fock-Bogoliubov approach}
\author{S.G. Rohozi\'nski}
\affiliation{Institute of Theoretical Physics,
University of Warsaw,
ul. Ho\.za 69, PL-00681, Warsaw, Poland}
\author{J. Dobaczewski}
\affiliation{Institute of Theoretical Physics,
University of Warsaw,
ul. Ho\.za 69, PL-00681, Warsaw, Poland}
\affiliation{Department of Physics, University of Jyv\"askyl\"a,
P.O. Box 35 (YFL), FI-40014 University of Jyv\"askyl\"a, Finland}
\author{W. Nazarewicz}
\affiliation{Institute of Theoretical Physics,
University of Warsaw,
ul. Ho\.za 69, PL-00681, Warsaw, Poland}
\affiliation{Department of Physics and Astronomy,
The University of Tennessee,
Knoxville, Tennessee 37996}
\affiliation{Physics Division,
Oak Ridge National Laboratory,
P.O. Box 2008, Oak Ridge, Tennessee 37831}

\date{\today}

\begin{abstract}
Symmetry properties of densities and mean fields
appearing in  the nuclear Density Functional Theory with pairing  are studied.
We consider  energy functionals that depend only on
local densities and their derivatives. The most important self-consistent symmetries are discussed:
spherical, axial, space-inversion, and mirror symmetries. In each case,
the consequences of breaking or conserving the time-reversal and/or
proton-neutron symmetries are discussed and summarized in a tabulated form, useful in practical applications.
\end{abstract}

\pacs{21.60.Jz, 
21.30.Fe, 
21.10.Hw,  
11.30.Er, 
71.15.Mb 
}
\maketitle

\section{Introduction}\label{sec1}

The nuclear Density Functional Theory (DFT)
\cite{[Ben03],[Lal04],[Sto06]} is a theoretical tool of choice for
describing complex, open-shell  nuclei, for which the dimension of
the configuration space becomes intractable for other methods of
theoretical nuclear structure, such as {\it ab initio} and
configuration interaction (shell model) techniques \cite{[Ber07a]}.
The main building blocks of nuclear DFT are the effective fields,
often represented by local proton and neutron densities and their
derivatives. When compared to the electronic DFT, the unique features
of the nuclear variant are: (i) the presence of two kinds of
fermions, protons and neutrons; (ii) the essential role of nucleonic
pairing; and (iii) the absence of external potential and the need for
symmetry restoration in a self-bound system.

At the heart of the DFT lies  the energy density functional $\overline{H}$
that is built from the nucleonic intrinsic density matrices. The
requirement that the total energy be minimal under a variation of the
densities leads to the Hartree-Fock-Bogoliubov (HFB; or Bogoliubov-de
Gennes) equations. They form a set of non-linear integro-differential
equations which has to be solved iteratively for the self-consistent
densities.

The quasi-particle vacuum associated with the DFT solution is a
highly correlated state. This is partly because the DFT description
is performed in a frame of reference attached to the nucleus, the
intrinsic frame, in which the nuclear mean field may  spontaneously
break the symmetry of the original Hamiltonian, or energy density.
While the resulting {\it deformed} solutions do not obey symmetries
present in the laboratory system, they acquire lower binding  through
long-range  polarization effects. Additional correlations may be
gained by means of symmetry restoration. Such a  strategy, rooted in
the nuclear Jahn-Teller effect \cite{[Rei84],[Naz94a]}, has proven to
be very effective in nuclear mean-field calculations
\cite{[RS80],[Ben08]}.

Since the symmetry breaking  is  essential for determining the
optimal mean field of the nucleus, the   self-consistent symmetries
present in the model
\cite{[Rip67],[Rip68],[Sau68],[Fae68],[Pra72],[Goo74],[Gra79a]} may
often determine physics. A self-consistent symmetry (SCS)  is a
unitary or antiunitary transformation ${\cal S}$, which commutes with
the HFB  Hamiltonian. Due to self-consistency, ${\cal S}$ also
commutes with  DFT densities. The associated DFT energy density
${\cal H}$ is referred to as {\it symmetry invariant} \cite{[Car08]}:
\begin{equation}\label{SCS}
{\cal H}^{\cal S}(\bbox{r})={\cal H}(\bbox{r}).
\end{equation}
The  above definition of SCS that can often be found in the literature
\cite{[RS80]} is too limiting when it comes to the DFT. Indeed,
invariance of the energy density itself is not a prerequisite for the
invariance of the energy density functional (EDF). Actually, the EDF
is invariant with respect to ${\cal S}$  also when the energy density
is covariant with ${\cal S}$, i.e.,
\begin{equation}\label{covariant}
{\cal H}^{\cal S}(\bbox{r})={\cal H}({\cal S}^+ \bbox{r}{\cal S}).
\end{equation}
The energy density that meets  (\ref{covariant}) is referred to as
{\it symmetry covariant}, see discussion in Appendix~A of
Ref.~\cite{[Car08]}. The existence of SCS has a profound impact on
self-consistent solutions. If the initial density matrix employed at
the first iteration of HFB equations  contains a SCS, then that
symmetry will propagate through to the  final DFT solution.
Therefore, the introduction of SCS restricts  the generality of the
self-consistent density matrix and  may lead to an erroneous estimate of
the DFT energy and deformation of the system.

A considerable literature exists on nuclear collective modes
associated with spontaneously broken symmetries. Ground-state
deformations of nuclei, including those in the pairing channel,  have
been reviewed in Refs.~\cite{[Abe90a],[Naz96]}. In the presence of
angular momentum, new deformations of magnetic character may appear
in the rotating nuclear mean field. High-spin particle-hole (p-h) and
particle-particle (p-p or pairing) deformations, both isoscalar and
isovector,  have been discussed in detail in
Refs.~\cite{[Fra00],[Fra05]}, which also  contain a general
discussion  of spontaneous symmetry-breaking phenomena in rotating
nuclei  (see also \cite{[Sat05]} for a recent update).

Symmetry
properties of Hartree-Fock (HF) densities were studied in
Ref.~\cite{[Dob00a]} in the context of a  double point group
D$_{2\mathrm{h}}^\mathrm{TD}$ that contains three mutually
perpendicular symmetry axes of the second order, space inversion, and
time reversal. The associated symmetry-breaking schemes have been
outlined in the following Ref.~\cite{[Dob00b]}.

In this paper, we extend the discussion of
Refs.~\cite{[Dob00a],[Dob00b]} to the pairing channel and transition densities using the
coordinate-space HFB  theory which incorporates an arbitrary mixing
between protons and neutrons in the p-h and p-p channels
\cite{[Per04]}. The constraints imposed on  DFT densities by the
time-reversal and proton-neutron (p-n)
symmetries   are studied for various spatial geometries (spherical,
axial, space-inversion, mirror, and D$_{2\mathrm{h}}$). The details
pertaining to the p-n HFB theory in the Local Density
Approximation (LDA) can be found in Ref.~\cite{[Per04]}. Throughout
our work, we refer to this previous paper  as I and to, e.g., Eq.~(1)
therein as (I-1). However, to make the present paper self-contained,
some of the definitions given in I are also repeated here.

Our interest in a general HFB formalism that incorporates an
arbitrary mixing between proton and neutron quasiparticles is
motivated by numerous phenomena that are present in medium-mass and
heavy nuclei with $N$$\approx$$Z$. These include: p-n
pairing correlations; alpha decay and alpha clustering; local
increase in binding (Wigner energy); interplay between isospin $T$=0
and 1 states in $N$=$Z$ nuclei at low and high angular momenta;
isospin mixing and mirror symmetry breaking; and beta decays and
superallowed beta decays in particular, just to mention a few. All
these cases involve isospin in one way or another, and since we are
often interested in nuclear states with  nonzero spin, proper
treatment of time-reversal symmetry is crucial. A considerable amount
of literature exists on a mean-field treatment of $N$$\approx$$Z$
nuclei.  For instance, for  a concise overview of p-n pairing, we
refer the reader to Ref.~\cite{[Per04]}.

The paper is organized as follows. Section~\ref{sec2} introduces  the
local DFT densities and defines transformation rules for density
matrices. The symmetries of interest, both in the position-spin space
and in the  isotopic space are discussed in Sec~\ref{sec7}. Symmetry
properties of densities are studied in Sec.~\ref{sec12} for (i)
spherical symmetry (with and without space inversion); (ii) axial
symmetry (with and without space inversion); and  (iii)
D$_{2\mathrm{h}}$ symmetry. When going beyond the mean-field
approximation, e.g., by using the Generator Coordinate Method or
projection techniques, multi-reference transition densities appear.
The  associated  symmetry properties are summarized in
Sec.~\ref{sec14}. Previous self-consistent calculations of p-n pairing
are commented upon in Sec.~\ref{sec15} in the context of our
findings. Section ~\ref{sec11} summarizes the main results of our
study. Finally, Appendix~A is devoted to  the generalized
Cayley-Hamilton theorem for  irreducible spherical tensors and tensor
fields.

\section{Densities and fields}\label{sec2}

\subsection{Density matrices and mean fields in the {\particle} and
{\pairing} channels}\label{sec2a}

To fix the notation, we begin with a brief recapitulation of
definitions and basic properties of the one-body HFB
density matrices in {\particle} and {\pairing} channels, see I for details.
The
{\particle} and {\pairing} density matrices are defined, respectively, as
\begin{eqnarray}
\label{rho}
\hat{\rho}(\bbox{r}st,\bbox{r}'s't')
&=&\langle \Psi |a_{\bbox{r}'s't'}^{+}a_{\bbox{r}st}|\Psi \rangle , \\
\label{rhobreve}
\hat{\breve{\rho}}(\bbox{r}st,\bbox{r}'s't')
&=& 4s't'\langle \Psi |a_{\bbox{r}'-s'-t'}a_{\bbox{r}st}|\Psi \rangle ,
\end{eqnarray}
where $a_{\bbox{r}st}^{+}$ and $a_{\bbox{r}st}$ create and
annihilate, respectively, nucleons at point $\bbox{r}$, spin
$s$=$\pm\thalf$, and isospin $t$=$\pm\thalf$, while $|\Psi \rangle $
is the HFB independent-quasiparticle state.

The {\particle} and {\pairing} density matrices
form together the projective generalized ``breve"
density matrix,
\begin{equation}
\hat{\breve{\mathcal R}}(x,x')
=\left(\begin{array}{cc}\hat{\rho}(x,x')&\hat{\breve{\rho}}(x,x')\\
\hat{\breve{\rho}}^{+}(x,x')    & \hat{1}-\hat{\rho}^{TC}(x,x')\end{array}\right),
\label{genden}
\end{equation}
where we abbreviated the position-spin-isospin variables by
$x$$\equiv$$\{\bbox{r}st\}$ and $\hat{1} := \delta(x-x') := \delta
(\bbox{r}-\bbox{r}')\delta_{ss'}\delta_{tt'}$.

In I,
we found that instead of using the usual
antisymmetric pairing tensor
\cite{[RS80]}, it is more convenient to introduce the above
{\pairing} (or anomalous) density matrix, $\hat{\breve{\rho}}$
(\ref{rhobreve}). The relation between the standard density matrix
$\hat{\mathcal R}(x,x')$ \cite{[RS80]} and
$\hat{\breve{\mathcal R}}(x,x')$ is given by a unitary transformation
\cite{[Per04]}
\begin{equation}
\hat{\mathcal W}
=\left(\begin{array}{cc}\hat{1} & 0 \\
0 & \hat{\sigma}^y\hat{\tau}^{(2)}\end{array}\right).
\label{Wtran}
\end{equation}
The quantities expressed in the representation (\ref{Wtran})
are indicated by a ``breve" symbol in the following.

Throughout our paper we apply the following naming convention.
The matrices are denoted by ``hat"; the quantities expressed
in the representation (\ref{Wtran}) are marked by ``breve'';
the matrices in a double HFB space are denoted by
the calligraphic capital letters.
As in I, we label space vectors by boldface symbols
and their scalar products by the central dot, e.g.,
$\bbox{r}$$\cdot$$\bbox{\nabla}$;
the components of vectors
and tensors are labeled with indices $a, b, c$, and the names
of axes are $x, y$, and $z$, e.g., $\bbox{r}=(\bbox{r}_x, \bbox{r}_y, \bbox{r}_z)$.
Here we note that similarly  as in I, the individual  vector components are
also shown in boldface.

Vectors in isospace (isovectors)
are labeled  by arrows with their
scalar products in the isospace denoted by the circle, e.g.,
$\vec{v}\circ\vec{w}$.
The components of isovectors are labeled with
indices $i$, $k$, and the names of iso-axes are 1, 2, and 3.
Isoscalars are marked with subscript
``0", and we often combine formulae for isoscalars
and isovectors by letting the indices run through all the
four values, e.g., $k$=0,1,2,3.

The  symbol
 $\tintsum\rmd{x}$ represents  integration over spatial
coordinates and summation over spin and isospin indices,
and  $\bullet$ denotes the matrix multiplication and integration/summation
$\tintsum\rmd{x}$.
The asterisk stands for the complex conjugation.
The spin and isospin  Pauli matrices are, respectively,
$\hat{\bbox{\sigma}}_{ss'}
=(\hat{{\sigma}}^x_{ss'},\hat{{\sigma}}^y_{ss'},\hat{{\sigma}}^z_{ss'})$ and
$\hat{\vec{\tau}}_{tt'}=(\hat{\tau}^{(1)}_{tt'},\hat{\tau}^{(2)}_{tt'},
\hat{\tau}^{(3)}_{tt'})$, and the corresponding unity matrices are
$\hat{\tau}^{(0)}_{tt'}=\delta _{tt'}$ and $\hat{\sigma}^u_{ss'}=\delta
_{ss'}$.

Under  $TC$, the product of  time reversal $T$ and
charge reversal $C$ (a rotation by $\pi$
 in isospace around the second axis), the
 density matrices (\ref{rho}) and (\ref{rhobreve}) become:
\bnll{timechargerev}
\hat{\rho}^{TC}(\bbox{r}st,\bbox{r}'s't')&=&
16ss'tt'\hat{\rho}^*(\bbox{r}\,\mbox{$-s$}\,\mbox{$-t$},\bbox{r}'\,\mbox{$-s$}'\,\mbox{$-t$}'),\\
\hat{\breve{\rho}}^{TC}(\bbox{r}st,\bbox{r}'s't')&=&
16ss'tt'\hat{\breve{\rho}}^*(\bbox{r}\,\mbox{$-s$}\,\mbox{$-t$},\bbox{r}'\,\mbox{$-s$}'\,\mbox{$-t$}').
\enll
The symmetries of  $\hat\rho$ and $\hat{\breve{\rho}}$ can be conveniently
expressed in terms of just the Hermitian conjugation and $TC$:
\bnll{hc}
\label{hc1}
\hat{\rho}^{+} &=& \hat{\rho},
\\ \label{hc2}
\hat{\breve{\rho}}^{+} &=& -\hat{\breve{\rho}}^{TC}.
\enll

Expressed in terms of spin-isospin components, the density matrices can
be written as:
\begin{widetext}
\bnll{izo}
\hat{\rho}(\bbox{r}st,\bbox{r}'s't')
&=&\tquart\rho_0(\bbox{r},\bbox{r}')\delta_{ss'}\delta_{tt'}
+ \tquart\delta_{ss'}\vec{\rho}(\bbox{r},
\bbox{r}')\circ \hat{\vec{\tau}}_{tt'}
+ \tquart\bbox{s}_0(\bbox{r},\bbox{r}')\cdot \hat{\bbox{\sigma}}_{ss'}\delta_{tt'}
+ \tquart\vec{\bbox{s}}(\bbox{r},\bbox{r}')\cdot \hat{\bbox{\sigma}}_{ss'}\circ \hat{\vec{\tau}}_{tt'},
\label{izo01} \\
\hat{\breve{\rho}}(\bbox{r}st,\bbox{r}'s't')
&=&\tquart\breve{\rho}_0(\bbox{r},\bbox{r}')\delta_{ss'}\delta_{tt'}
+ \tquart\delta_{ss'}\vec{\breve{\rho}}(\bbox{r},\bbox{r}')\circ \hat{\vec{\tau}}_{tt'}
+ \tquart\breve{\bbox{s}}_0(\bbox{r},\bbox{r}')\cdot \hat{\bbox{\sigma}}_{ss'}\delta_{tt'}
+ \tquart\vec{\breve{\bbox{s}}}(\bbox{r},\bbox{r}')\cdot\hat{\bbox{\sigma}}_{ss'}\circ \hat{\vec{\tau}}_{tt'}.
\label{izo02}
\enll
\end{widetext}
To avoid confusion, the functions of two position vectors
$\bbox{r}$ and $\bbox{r}'$
appearing on the right-hand sides of Eqs.~(\ref{izo}) will be
called the {\it nonlocal} density functions or, simply, densities, unlike
the density matrices (\ref{rho}) and (\ref{rhobreve})
appearing on  the left-hand sides.

Since the {\particle} density matrix and the Pauli matrices are both
Hermitian, according to (I-16) all  {\particle} densities are Hermitian
as well; hence, their real parts are symmetric, while the
imaginary parts are antisymmetric with respect to exchanging
 $\bbox{r}$ and $\bbox{r}'$. Similarly,
transformation properties of Pauli matrices under time-reversal and
charge conjugation (I-17) make the {\pairing} densities either
symmetric (scalar-isovector and vector-isoscalar) or antisymmetric
(scalar-isoscalar and vector-isovector) with respect to exchanging
 $\bbox{r}$ and $\bbox{r}'$, see Eq.~(I-18).
 These properties are fulfilled
independently of any other symmetries conserved by the system; they
are consequences of definitions of density matrices $\hat{\rho}$ and
$\hat{\breve{\rho}}$.

In the HFB theory with the zero-range
Skyrme interaction \cite{[Sky59],[Vau72]},
or in the local density approximation
(LDA; cf.~Refs.~\cite{[Neg72],[RS80]}), the energy functional depends
only on {\it local}
 densities, and on local densities built from derivatives
up to a given order, see Refs.~\cite{[Dob96b],[Car08]} for systematic
constructions. The local densities are denoted by having only one spatial
argument to distinguish them from the nonlocal densities.
Following the standard definitions \cite{[Flo75],[Eng75]},
in the present study we employ definitions of local
{\particle} and {\pairing}  densities according to
Ref.~I. For the sake of completeness, we repeat them here:

\vspace{1ex}\noindent$\bullet$
scalar densities:
\begin{itemize}
\item[--]  particle and pairing densities:
\bnll{scalar-p}
\label{scalar-p-ph}
{\rho}_k(\bbox{r})
&=&           {\rho}_k(\bbox{r},\bbox{r}')_{\bbox{r}=\bbox{r}'},\\
\label{scalar-p-pp}
\vec{\breve       {\rho}} (\bbox{r})
&=&\vec{\breve{\rho}} (\bbox{r},\bbox{r}')_{\bbox{r}=\bbox{r}'},
\enll
\item[--]  {\particle} and {\pairing} kinetic densities:
\bnll{scalar-k}
\label{scalar-k-ph}
{\tau}_k(\bbox{r})
&=&\big[        (\bbox{\nabla}\cdot\bbox{\nabla}')
{\rho}_k(\bbox{r},\bbox{r}')\big]_{\bbox{r}=\bbox{r}'}, \\
\label{scalar-k-pp}
\vec{\breve {\tau}} (\bbox{r})
&=&\big[        (\bbox{\nabla}\cdot \bbox{\nabla}')
\vec{\breve {\rho}} (\bbox{r},\bbox{r}')\big]_{\bbox{r}=\bbox{r}'},
\enll
\end{itemize}

\noindent$\bullet$
vector densities:
\begin{itemize}
\item[--]   {\particle} and {\pairing} spin (pseudovector) densities:
\bnll{vector-s}
{\bbox{s}}_k(\bbox{r})
&=&      {\bbox{s}}_k(\bbox{r},\bbox{r}')_{\bbox{r}=\bbox{r}'}, \\
\breve{\bbox{s}}_0(\bbox{r})
&=&\breve{\bbox{s}}_0(\bbox{r},\bbox{r}')_{\bbox{r}=\bbox{r}'}\label{isoscdens},
\enll
\item[--]   {\particle} and {\pairing} spin-kinetic (pseudovector) densities:
\bnll{vector-T}
\label{vector-T-ph}
{\bbox{T}}_k(\bbox{r})
&=&\big[             (\bbox{\nabla}\cdot\bbox{\nabla}')
{\bbox{s}}_k(\bbox{r},\bbox{r}')\big]_{\bbox{r}=\bbox{r}'}, \\
\label{vector-T-pp}
\breve{\bbox{T}}_0(\bbox{r})
&=&\big[             (\bbox{\nabla}\cdot\bbox{\nabla}')
\breve{\bbox{s}}_0(\bbox{r},\bbox{r}')\big]_{\bbox{r}=\bbox{r}'},
\enll
\item[--]  {\particle} and {\pairing} current (vector) densities:
\bnll{vector-j}
\label{vector-j-ph}
{\bbox{j}}_k(\bbox{r})
&=&\tfrac{1}{2i}\big[ (\bbox{\nabla} - \bbox{\nabla}')
{\rho}_k(\bbox{r},\bbox{r}')\big]_{\bbox{r}=\bbox{r}'}, \\
\label{vector-j-pp}
\breve{\bbox{j}}_0(\bbox{r})
&=&\tfrac{1}{2i}\big[ (\bbox{\nabla} - \bbox{\nabla}')
\breve    {\rho}_0(\bbox{r},\bbox{r}')\big]_{\bbox{r}=\bbox{r}'},
\enll
\item[--]   {\particle} and {\pairing} tensor-kinetic (pseudovector) densities:
\bnll{vector-F}
\label{vector-F-ph}
{\bbox{F}}_k(\bbox{r})
&\!=\!&\thalf\big[       (\bbox{\nabla} \!\otimes\!\bbox{\nabla}'
\!+\!\bbox{\nabla}'\!\otimes\!\bbox{\nabla})\!\cdot\!
{\bbox{s}}_k(\bbox{r},\bbox{r}')\big]_{\bbox{r}=\bbox{r}'},\! \\
\label{vector-F-pp}
\breve{\bbox{F}}_0(\bbox{r})
&\!=\!&\thalf\big[       (\bbox{\nabla} \!\otimes\!\bbox{\nabla}'
\!+\!\bbox{\nabla}'\!\otimes\!\bbox{\nabla})\!\cdot\!
\breve{\bbox{s}}_0(\bbox{r},\bbox{r}')\big]_{\bbox{r}=\bbox{r}'},\!
\enll
\end{itemize}

\noindent$\bullet$
tensor densities:
\begin{itemize}
\item[--] {\particle} and {\pairing} spin-current (pseudotensor) densities:
\bnll{tensor-J}
\label{tensor-J-ph}
{{\mathsf J}}_k(\bbox{r})
&=&\tfrac{1}{2i}\big[ (\bbox{\nabla} - \bbox{\nabla}')\otimes
{\bbox{s}}_k(\bbox{r},\bbox{r}')\big]_{\bbox{r}=\bbox{r}'}, \\
\label{tensor-J-pp}
\vec{\breve{\mathsf J}}  (\bbox{r})
&=&\tfrac{1}{2i}\big[(\bbox{\nabla} - \bbox{\nabla}')\otimes
\vec {\breve{\bbox{s}}}  (\bbox{r},\bbox{r}')\big]_{\bbox{r}=\bbox{r}'},
\enll
\end{itemize}
where $k$=0,1,2,3, and $\otimes$ stands for the tensor product of vectors in the
physical space, e.g.,
$(\bbox{v}$$\otimes\,$$\bbox{w})_{ab}$$\,\equiv\,$$\bbox{v}_{a}\bbox{w}_{b}$ and
$[(\bbox{v}$$\otimes\,$$\bbox{w})$$\cdot$$\bbox{z}]_{a}$$\,\equiv\,$$\bbox{v}_{a}(\bbox{w}$$\cdot$$\bbox{z})$.\

The kinetic, spin-kinetic, and tensor-kinetic densities are, in fact, equal
to contractions of the following 2nd-order and 3rd-order tensor densities:
\bnll{T1}
\tau_{kbc}(\bbox{r})&=&\big[\bbox{\nabla}_b\bbox{\nabla}'_c\rho_k(\bbox{r},\bbox{r}')\big]_{\bbox{r}=\bbox{r}'}, \label{tau1}\\
\vec{\breve{\tau}}_{bc}(\bbox{r})&=&\big[\bbox{\nabla}_b\bbox{\nabla}'_c
\vec{\breve{\rho}}(\bbox{r},\bbox{r}')\big]_{\bbox{r}=\bbox{r}'}, \label{tau3a} \\
T_{kbcd}(\bbox{r})&=&\big[\bbox{\nabla}_b\bbox{\nabla}'_c\bbox{s}_{kd}(\bbox{r},\bbox{r}')
\big]_{\bbox{r}=\bbox{r}'},\label{T1a}\\
\breve{T}_{0bcd}(\bbox{r})&=&\big[\bbox{\nabla}_b\bbox{\nabla}'_c\breve{\bbox{s}}_{0d}
(\bbox{r},\bbox{r}')\big]_{\bbox{r}=\bbox{r}'},\label{T1b}
\enll
namely,
\bnll{TF}
\tau_{k}(\bbox{r})&=&\sum_{b=x,y,z}\tau_{kbb}(\bbox{r}),\label{tau4a}\\
\vec{\breve{\tau}}(\bbox{r})&=&\sum_{b=x,y,z}\vec{\breve{\tau}}_{bb}(\bbox{r}), \label{tau4b} \\
\bbox{T}_{kd}(\bbox{r})&=&\sum_{b=x,y,z}T_{kbbd}(\bbox{r}),\label{TF1}\\
\breve{\bbox{T}}_{0d}(\bbox{r})&=&\sum_{b=x,y,z}\breve{T}_{0bbd}(\bbox{r}),\label{TF2}\\
\bbox{F}_{kb}(\bbox{r})&=&\half\sum_{c=x,y,z}\big(T_{kbcc}(\bbox{r})+
T_{kcbc}(\bbox{r})\big),\label{TF3}\\
\breve{\bbox{F}}_{0b}(\bbox{r})&=&\half\sum_{c=x,y,z}\big(\breve{T}_{0bcc}(\bbox{r})
+\breve{T}_{0cbc}(\bbox{r})\big).\label{TF4}
\enll

All pseudotensor densities can be decomposed into trace,
antisymmetric, and symmetric components (I-26)-(I-28), that is, into   pseudoscalar
${J}_k(\bbox{r})$ and $\vec{\breve{J}}(\bbox{r})$; vector
${\bbox{J}}_{k}(\bbox{r})$ and $\vec{\breve{\bbox{J}}}(\bbox{r})$;
and pseudotensor $\underline{{\mathsf J}}_{k} (\bbox{r})$ and
$\underline{\vec{\breve{{\mathsf J}}}}(\bbox{r})$ densities.

In the case of the Skyrme  effective interaction, as well as in the
framework of the energy density functional approach, the energy functional is a
three-dimensional spatial integral,
\begin{equation}
\overline{H}=\int \rmd^3\bbox{r}{\mathcal H}(\bbox{r}) ,\label{lenden}
\end{equation}
of the
local energy density ${\mathcal H}(\bbox{r})$ that is supposed to be a real, scalar,
time-even, and isoscalar function of local densities and their
derivatives.

Minimization of the energy functional with
respect to the {\particle} and {\pairing} density matrices under auxiliary conditions
\begin{equation}\label{Acond}
\int {\rmd}^3\bbox{r}\rho_0(\bbox{r})= A
\end{equation}
and
\begin{equation}\label{Tzcond}
\int {\rmd}^3\bbox{r}\rho_3(\bbox{r}) =  N-Z=2T_3,
\end{equation}
leads to the common eigenvalue problem for the generalized density
matrix $\hat{\breve{\mathcal R}}$   and
the generalized mean-field Hamiltonian matrix defined as
\begin{equation}
\hat{\breve{\mathcal H}}
=\left(\begin{array}{cc}\hat{h}-\hat{\lambda}&\hat{\breve{h}}\\
\hat{\breve{h}}^{+} & -\hat{h}^{TC}+\hat{\lambda}\end{array}\right) ,
\label{genha}
\end{equation}
with the Lagrange multiplier matrix given by
\begin{equation}
\hat{\lambda}=\thalf(\lambda_n+\lambda_p) \hat{I}+
\thalf(\lambda_n-\lambda_p)\hat{\tau}^{(3)} =
\lambda_0\hat{I}+
\lambda_3\hat{\tau}^{(3)},
\end{equation}
where $\lambda_n$ and $\lambda_p$ are the neutron and proton Fermi
levels, respectively.

\subsection{Transformation rules for the density matrices}\label{sec6}

A general Hermitian one-body operator in the Fock space can be written as
\begin{equation}
G= \int \rmd^3\bbox{r}'\sum_{s't'}\int\rmd^3\bbox{r}\sum_{st}\hat{g}(\bbox{r}'s't',\bbox{r}st)
a_{\bbox{r}'s't'}^{+}a_{\bbox{r}st}, \label{sp}
\end{equation}
where
\begin{equation}
\hat{g}(\bbox{r}'s't',\bbox{r}st)=\langle \bbox{r}'s't'|g|\bbox{r}st\rangle
\label{me}
\end{equation}
is the matrix element of the single-particle operator $g$ acting in the
space of one-body wave functions. Let us now consider a unitary
 transformation  $U$ in the Fock space generated by $G$:
\begin{equation}
U = e^{i\alpha G},\label{u}
\end{equation}
where $\alpha$ is a real parameter.
By making use of the Baker-Campbell-Hausdorff relations,
the annihilation and creation operators transform under $U$ as:
\bnll{tra}
{}U^+a    _{\bbox{r}st}U &=& \int \rmd^3\bbox{r}'\sum_{s't'}\hat{u}  (\bbox{r}st,\bbox{r}'s't')a_{\bbox{r}'s't'},\\
  U^+a^{+}_{\bbox{r}st}U &=& \int \rmd^3\bbox{r}'\sum_{s't'}\hat{u}^+(\bbox{r}'s't',\bbox{r}st)a^{+}_{\bbox{r}'s't'},
\enll
where
\begin{equation}
\hat{u}(\bbox{r}'s't',\bbox{r}st)=\langle \bbox{r}'s't'|e^{i\alpha g}|\bbox{r}st\rangle .
\label{meu}
\end{equation}

From Eqs.~(\ref{tra}) it follows that
the density matrices calculated for the
transformed state $U|\Psi\rangle$, i.e., the transformed
density matrices, are:
\begin{widetext}
\bnll{trrho}
\hat{\rho}^U(x_1,x'_1)
&=&\intsum\rmd{x_2}\rmd{x'_2}\hat{u}(x_1,x_2)\hat{\rho}(x_2,x'_2)\hat{u}^+(x'_2,x'_1),\\
\hat{\breve{\rho}}^U(x_1,x'_1)
&=&4s'_1t'_1
\intsum\rmd{x_2}\rmd{x'_2}\,4s'_2t'_2\hat{u}(x_1,x_2)\hat{\breve{\rho}}(x_2,x'_2)
\hat{u}(\overline{x'_1},\overline{x'_2}),
\enll
where
$\overline{x}\equiv \{\bbox{r},-s,-t\}$. Using a shorthand notation,
Eqs.~(\ref{trrho}) can be written as
\begin{equation}
\label{matmul}
{}\hat{\rho}^U = \hat{u}\bullet\hat{\rho}\bullet\hat{u}^+,\quad
\hat{\breve{\rho}}^U = \hat{u}\bullet\hat{\breve{\rho}}\bullet\hat{\breve{u}}^+,
\end{equation}
where $\hat{\breve{u}}$ is defined as
\begin{equation}
\hat{\breve{u}}(\bbox{r}st,\bbox{r}'s't')=16ss'tt'\hat{u}^{\ast}(\bbox{r}-s-t,\bbox{r}'-s'-t')
=\sum_{s'',s'''}\sum_{t'',t'''}(-i\hat{\tau}^{(2)}_{tt''})(-i\hat{{\sigma}}^y_{ss''})
\hat{u}^{\ast}(\bbox{r}s''t'',\bbox{r}'s'''t''')(i\hat{{\sigma}}^y_{s'''s'})(i\hat{\tau}^{(2)}_{t'''t'}),
\label{utc}
\end{equation}
\end{widetext}
or
\begin{equation}
\hat{\breve{u}}=(-i\hat{\tau}^{(2)})(-i\hat{{\sigma}}^y)\hat{u}^{\ast}
(i\hat{{\sigma}}^y)(i\hat{\tau}^{(2)}),
\end{equation}
\begin{equation}
\hat{\breve{u}}=\hat{u}^{TC} ,
\end{equation}
cf.~Eqs.~(\ref{timechargerev}).
It immediately follows from
Eqs.~(\ref{matmul})  that the generalized density matrix
(\ref{genden}) transforms under  $U$ as:
\begin{equation}
\hat{\breve{\mathcal R}}^U= \hat{\breve{\mathcal U}}\bullet\hat{\breve{\mathcal R}}\bullet
\hat{\breve{\mathcal U}}^+,
\label{trgen}
\end{equation}
where the transformation matrix in the doubled-dimension space is
defined as
\begin{equation}
\hat{\breve{\mathcal U}}
=\left(\begin{array}{cc}\hat{u}& 0\\
 0    & \hat{\breve{u}}\end{array}\right).\label{ugen}
\end{equation}

Similar definitions can be introduced for any unitary {\em antilinear}
transformation operator, $U_K$, which can always be presented in the form:
\begin{equation}
U_K=UK,
\label{antlin}
\end{equation}
where $U$ is a linear unitary operator and $K$ is the operator of complex
conjugation in the position-spin-isospin representation,
in which the basis states $|\bbox{r}st\rangle$ are assumed to be real: $K|\bbox{r}st\rangle=|\bbox{r}st\rangle$.
Let us recall that
the action of $K$ on a single-particle wave function $\Phi(\bbox{r}st)$ expressed
in a basis $\phi_i(\bbox{r}st)$,
$\Phi(\bbox{r}st)=\sum_i c_i \phi_i(\bbox{r}st)$,
 is defined as
\begin{equation}
K\Phi(\bbox{r}st) \equiv \Phi^*(\bbox{r}st) = \sum_i  c_i^* \phi_i^*(\bbox{r}st).
\label{antlin2}
\end{equation}
The
generalized density matrix (\ref{genden}) transformed under
 $U_K$ reads:
\begin{equation}
\hat{\breve{\mathcal R}}^{U_K}= \hat{\breve{\mathcal U}}_{K}\bullet\hat{\breve{\mathcal R}}\bullet
\hat{\breve{\mathcal U}}_K^+,
\label{trant}
\end{equation}
where the transformation matrix is
\begin{equation}
\hat{\breve{\mathcal U}}_K
=\left(\begin{array}{cc}\hat{u}_K& 0\\
 0    & \hat{\breve{u}}_K\end{array}\right),\label{agen}
\end{equation}
with
\bnll{au}
\hat{u}_K&=&\hat{u}\hat{K}, \\
\hat{\breve{u}}_K&=&\hat{\breve{u}}\hat{K},
\enll
and $\hat{K}$ is the matrix complex conjugation operator associated
with the position-spin-isospin representation. One has to remember that
the decomposition (\ref{antlin}) and definition of $\hat{K}$
(\ref{antlin2}) does not depend on any specific choice of the
single-particle basis $\phi_i(\bbox{r}st)$. The advantage of such a
choice is that properties of antilinear symmetries (like the
time reversal) directly translate into the complex-conjugation
properties of densities. However, other choices of $K$ can be useful
when the complex-conjugation properties of matrix elements of
operators in the given basis $\phi_i(\bbox{r}st)$ are considered.

\section{Symmetries}\label{sec7}

Let us suppose that $U$ (or $U_K$) is a symmetry transformation of
the nuclear many-body Hamiltonian, $H$, i.e.,
\begin{equation}
UHU^+ = H .
\label{sh}
\end{equation}
The generalized density matrix (\ref{genden}) and mean-field
Hamiltonian (\ref{genha}), obtained through the minimization
procedure, may, but need not, obey the symmetry $U$. It can be only
proved \cite{[RS80]} that if $U$ is a symmetry of $H$ then the transformed
generalized mean-field Hamiltonian depends functionally on the
transformed generalized density matrix in the same way as the
original Hamiltonian on the original density:
\begin{equation}
\hat{\breve{\mathcal H}}^U\left\{\hat{\breve{\mathcal R}}\right\}
 = \hat{\breve{\mathcal U}}\bullet\hat{\breve{\mathcal H}}
   \left\{\hat{\breve{\mathcal R}}\right\}
   \bullet\hat{\breve{\mathcal U}}^+
 = \hat{\breve{\mathcal H}}\left\{\hat{\breve{\mathcal R}}^U\right\}.
\label{selfsym}
\end{equation}
This means that in order to understand the symmetries of the mean field,
it suffices to  analyze the  symmetries
of the underlying density matrix. The nuclear Hamiltonian is
 supposed to conserve
numbers of protons and neutrons, and to be invariant under space
rotations
$D({\alpha},{\beta},{\gamma})$,
space inversion $P$, time reversal $T$, and rotations in the
isotopic space (isorotations)
$D_\tau({\alpha_\tau},{\beta_\tau},{\gamma_\tau})$.
(Throughout the present paper, to denote rotations we use
symbols $D$, $d$, and ${\mathcal D}$ instead of the usual letter $R$.
The symbol ${\mathcal R}$ is reserved for the generalized density
matrix.)
In the
space-spin-isospin basis, the single-particle matrix
elements of corresponding operators $d$, $p$, $t$, and $d_\tau$ are:
\bnll{symat}
\hat{d}^{\alpha\beta\gamma}(\bbox{r}'s't',\bbox{r}st)&\!\!=\!\!&\delta (\bbox{r}'-\bbox{r})\delta_{t't}
e^{i\gamma \bbox{j}_z(\bbox{r})}
e^{i\beta \bbox{j}_y(\bbox{r})}
e^{i\alpha \bbox{j}_z(\bbox{r})}\nonumber \\
&\!\!=\!\!&\delta (\bbox{r}'-\bbox{r})\delta_{t't}
e^{i\gamma \bbox{l}_z(\bbox{r})}
e^{i\beta \bbox{l}_y(\bbox{r})}
e^{i\alpha \bbox{l}_z(\bbox{r})}\nonumber \\
&&\times\hat{a}_{s's}(\hat{\bbox{\sigma}},\alpha\beta\gamma),\label{romat}\\
\hat{p}(\bbox{r}'s't',\bbox{r}st)&\!\!=\!\!&\delta (\bbox{r}'+\bbox{r})\delta_{s's}\delta_{t't},\label{invmat}\\
\hat{t}(\bbox{r}'s't',\bbox{r}st)&\!\!=\!\!&\delta (\bbox{r}'-\bbox{r})(-i\hat{{\sigma}}^y_{s's})\delta_{t't}\hat{k},\label{timat} \\
\hat{d}_\tau^{{\alpha_\tau}{\beta_\tau}{\gamma_\tau}}(\bbox{r}'s't',\bbox{r}st)&\!\!=\!\!&\delta (\bbox{r}'-\bbox{r})\delta_{s's}
\hat{a}_{t't}(\hat{\vec{\tau}},\alpha_\tau\beta_\tau\gamma_\tau),\label{chamat}
\enll
respectively, where $\bbox{l}(\bbox{r})$ is the single-particle
orbital-angular-momentum  operator,
$\bbox{j}(\bbox{r})=\bbox{l}(\bbox{r})+\tfrac{\hbar}{2}\bbox{\sigma}$ is the
total single-particle angular-momentum operator, and $\alpha$,
$\beta$, and $\gamma$ ($\alpha_\tau$, $\beta_\tau$, and $\gamma_\tau$) are
the  Euler angles of rotations in space (isospace).
The spin rotation matrix $\hat{a}_{s's}$, being the function of the Pauli
matrices and Euler angles, reads
\bn
&&\!\!\!\!\!\!\!\!\!\!\!\!\!\!\!\!\hat{a}_{s's}(\hat{\bbox{\sigma}},\alpha\beta\gamma)
  =(e^{i\thalf\gamma\hat{\bbox{\sigma}}^z}e^{i\thalf\beta\hat{\bbox{\sigma}}^y}e^{i\thalf\alpha\hat{\bbox{\sigma}}^z})_{s's} \nonumber \\
&&=\cos{\frac{\beta}{2}}\left(\cos{\frac{\gamma +\alpha}{2}}\delta_{s's}+i\sin{\frac{\gamma +\alpha}{2}}\hat{\bbox{\sigma}}^z_{s's}\right) \nonumber \\
&&+i\sin{\frac{\beta}{2}}\left(\cos{\frac{\gamma -\alpha}{2}}\hat{\bbox{\sigma}}^y_{s's}
+\sin{\frac{\gamma -\alpha}{2}}\hat{\bbox{\sigma}}^x_{s's}\right),
\label{spinrot}
\en
and the isospin rotation matrix $\hat{a}_{t't}(\hat{\vec{\tau}},\alpha_\tau\beta_\tau\gamma_\tau)$ is defined analogously.

Below, rotations by angle $\pi$ about
the three axes $x$, $y$, and $z$, which are called signature
operators, will be of particular interest:
\bnll{sigmat}
\hat{r}_x(\bbox{r}'s't',\bbox{r}st)&\!\!=\!\!&\hat{d}^{0\pi\pi}(\bbox{r}'s't',\bbox{r}st), \\
\hat{r}_y(\bbox{r}'s't',\bbox{r}st)&\!\!=\!\!&\hat{d}^{0\pi0}  (\bbox{r}'s't',\bbox{r}st), \\
\hat{r}_z(\bbox{r}'s't',\bbox{r}st)&\!\!=\!\!&\hat{d}^{\pi00}  (\bbox{r}'s't',\bbox{r}st).
\enll
Products of signature operators and the space-inversion operator, which are
called simplex operators, correspond to reflections with respect to
the $y$--$z$, $z$--$x$, and $x$--$y$ plains, respectively:
\bnll{simmat}
\hat{s}_x&\!\!=\!\!&\hat{r}_x\bullet\hat{p}, \\
\hat{s}_y&\!\!=\!\!&\hat{r}_y\bullet\hat{p}, \\
\hat{s}_z&\!\!=\!\!&\hat{r}_z\bullet\hat{p}.\label{signz}
\enll

Symmetry operations (\ref{symat}) form well-known group structures:
\begin{itemize}
\item
Proper rotations (\ref{romat}): they  belong to the orthogonal unimodular
group in three dimensions SO(3).
\item
Improper (or mirror)  rotations, i.e., rotations (\ref{romat})  combined
with space inversion (\ref{invmat}):  they
 belong to the full orthogonal group O(3).
\item
Improper rotations supplemented by time reversal (\ref{timat}): they
form the group called O$^T$(3).
\item
Together with the group of isorotations
(\ref{chamat}), the symmetry operations (\ref{symat}) constitute the
group O$^T$(3)$\times$SO(3).
\item
Space-inversion (\ref{invmat}) together with three signatures (\ref{sigmat})
and three simplexes (\ref{simmat}) constitute the point group of
symmetries of a parallelepiped, called D$_{2\mathrm{h}}$, which is of interest for
triaxial nuclei.
\end{itemize}
Due to the fact that the rotation of
a spin-$\thalf$ system by $2\pi$ changes sign of the wave function,
one has to, in fact, double these groups
(see Refs.~\cite{[Dob00a],[Dob00b]} for
details).  The doubling of groups has no
bearing on the results of the present study; hence, in what
follows we do not refer to it.

Although strong nuclear forces are charge-independent, i.e.,
invariant under rotations in isospace, the nuclear Hamiltonian is
not, at least because of electromagnetic forces. Within the HFB
theory, the charge independence is additionally broken by the
auxiliary condition (\ref{Tzcond}) which is manifestly isovector.
This constraint gives rise to HFB product states which violate
isospin even for charge-independent Hamiltonians (see
Ref.~\cite{[Sat09a]} for a recent discussion). If the density matrix
in the {\pairing} channel does not vanish, particle number is also
violated in the HFB theory. The local energy density ${\mathcal
H}(\bbox{r})$ is usually constructed under assumption that it should
be invariant (\ref{SCS})  with respect to the time reversal $T$ and
isorotations $D_\tau$, and covariant (\ref{covariant}) with respect
to the space symmetries $D$ and $P$, see Appendix A in
Ref.~\cite{[Car08]}. All these symmetries are often spontaneously
broken in mean-field theories. As discussed in the Introduction, the
problem of symmetries that are conserved  by $H$ and internally
broken by ${\mathcal H}$ is, in fact,  one of the most important
elements of a mean-field description of many-body systems.

\subsection{Symmetries in the isotopic space}\label{sec8}

As has been discussed in I, the standard case of
no explicit p-n
mixing can be described by the conserved p-n symmetry
given by
\begin{equation}
\hat{c}_3=-i\hat{a}(\hat{\vec{\tau}},\pi 00)
=\hat{\tau}^{(3)}.\label{pnsym}
\end{equation}
That is, $\hat{c}_3$ does not change the third isospin component
but it reverses the sign of $\hat{\tau}^{(1)}$ and $\hat{\tau}^{(2)}$.
Since $\hat{\breve{c}}_{3}$=$-\hat{\tau}^{(3)}$, we obtain from Eq.~(\ref{utc})
that
\bnll{conservt3}
\hat{c}_3 \hat       {\rho}  \hat{c}_3^+ &=&  \phantom{-}\hat       {\rho} , \\
\hat{c}_3 \hat{\breve{\rho}} \hat{\breve{c}}_3^+ &=&            -\hat{\breve{\rho}}.
\enll
Consequently, in the absence of the
explicit  p-n mixing, the {\particle} density matrices
have only the $k$=0 and 3 isospin components, while the {\pairing}
densities have only the $k$=1 and 2
isospin components.

In the presence of an additional p-n exchange symmetry
(charge-reversal transformation $C$ of Eq.~(I-5) multiplied by $i$),
defined as
\begin{equation}
\hat{c}_2=-i\hat{a}(\hat{\vec{\tau}},0\pi 0)
=\hat{\tau}^{(2)},\label{pnex}
\end{equation}
only   the $k$=0 isospin
component  remains for {\particle} density matrix
 whereas  the {\pairing} has only a $k$=1 non-vanishing isospin component.
In other words, in this case proton and neutron densities
 are equal to each other.

\subsection{Symmetries in the position-spin space}\label{sec9}

In this section, we discuss
transformation properties of the generalized density matrices
$\hat{\breve{\mathcal R}}$ under $D(\alpha\beta\gamma)$, $P$, and $T$.
In the case of rotations, the general transformation matrix
(\ref{meu}) is given by $\hat{d}$ of Eq.~(\ref{romat}). Since the
single-particle orbital angular-momentum operator is imaginary,
$\bbox{l}(\bbox{r})$= $-\bbox{l}^{\ast}(\bbox{r})$, by applying
(\ref{utc}) to Eqs.~(\ref{romat}) and (\ref{spinrot}), one obtains:
\begin{equation}
 \hat{\breve{d}}^{\alpha\beta\gamma}(\bbox{r}'s't',\bbox{r}st)
=\hat{d}        ^{\alpha\beta\gamma}(\bbox{r}'s't',\bbox{r}st),
\label{eqbreve}
\end{equation}
i.e.,  the density matrices in both channels, $\hat{\rho}$ and
$\hat{\breve{\rho}}$, transform under rotations in the same
way, and the generalized rotation matrix has a simple form:
\begin{equation}
\hat{\breve{\mathcal D}}(\alpha\beta\gamma)
=\left(\begin{array}{cc}\hat{d}^{\alpha\beta\gamma}& 0\\
 0    & \hat{d}^{\alpha\beta\gamma}\end{array}\right).\label{dgen}
\end{equation}

When applying this symmetry operation to the generalized density matrix
(\ref{trgen}),  we need to use the Hermitian-conjugate matrix,
$\left(\hat{d}^{\alpha\beta\gamma}\right)^+$:
\begin{widetext}
\begin{equation}
\left(\hat{d}^{\alpha\beta\gamma}\right)^+(\bbox{r}'s't',\bbox{r}st)=\delta (\bbox{r}'-\bbox{r})\delta_{t't}
e^{-i\gamma \bbox{l}_z(\bbox{r}')}
e^{-i\beta \bbox{l}_y(\bbox{r}')}
e^{-i\alpha \bbox{l}_z(\bbox{r}')}
\times\hat{a}^+_{s's}(\hat{\bbox{\sigma}},\alpha\beta\gamma)\label{dhc}
\end{equation}
and Pauli matrices that transform as vectors under the spin
matrix $\hat{a}$, i.e.,
\begin{equation}
\hat{a}(\hat{\bbox{\sigma}},\alpha\beta\gamma )\,\hat{{\sigma}}^a\,\hat{a}^+(\hat{\bbox{\sigma}},\alpha\beta\gamma )
=\sum_{b}\mathsf{a}_{ab}(\alpha\beta\gamma)\,\hat{{\sigma}}^b ,
\label{vectr}
\end{equation}
for $a,b=x,y,z$, where the Cartesian rotation matrix reads
\begin{equation}
\mathsf{a}(\alpha\beta\gamma )=\left(\ba{c@{,\quad}c@{,\quad}c}
 \cos{\alpha}\cos{\beta}\cos{\gamma}-\sin{\alpha}\sin{\gamma} & -\cos{\alpha}\cos{\beta}\sin{\gamma}-\sin{\alpha}\cos{\gamma} & \cos{\alpha}\sin{\beta}\\
 \sin{\alpha}\cos{\beta}\cos{\gamma}+\cos{\alpha}\sin{\gamma} & -\sin{\alpha}\cos{\beta}\sin{\gamma}+\cos{\alpha}\cos{\gamma} & \sin{\alpha}\sin{\beta}\\
-\sin{\beta}\cos{\gamma} & \sin{\alpha}\sin{\beta} & \cos{\beta} \ea \right) .
\label{cartrot}
\end{equation}
Similarly, the rotation matrix (\ref{cartrot}) also rotates the
position arguments, $\bbox{r}$ and $\bbox{r}'$, of the density
matrices. Finally, we have
\bnll{rhorot}\!\!\!\!\!\!\!\!
\hat{\rho}^D(\bbox{r}st,\bbox{r}'s't')
&\!\!=\!\!&\tquart\rho_0(\mathsf{a}\bbox{r},\mathsf{a}\bbox{r}')\delta_{ss'}\delta_{tt'}
\!+\! \tquart\delta_{ss'}\vec{\rho}(\mathsf{a}\bbox{r},
\mathsf{a}\bbox{r}')\circ \hat{\vec{\tau}}_{tt'}
\!+\! \tquart\bbox{s}_0(\mathsf{a}\bbox{r},\mathsf{a}\bbox{r}')\cdot (\mathsf{a}\hat{\bbox{\sigma}})_{ss'}\delta_{tt'}
\!+\! \tquart\vec{\bbox{s}}(\mathsf{a}\bbox{r},\mathsf{a}\bbox{r}')\cdot
(\mathsf{a}\hat{\bbox{\sigma}})_{ss'}\circ \hat{\vec{\tau}}_{tt'},
\label{rhorot1}\\\!\!\!\!\!\!\!\!
\hat{\breve{\rho}}^D(\bbox{r}st,\bbox{r}'s't')
&\!\!=\!\!&\tquart\breve{\rho}_0(\mathsf{a}\bbox{r},\mathsf{a}\bbox{r}')\delta_{ss'}\delta_{tt'}
\!+\! \tquart\delta_{ss'}\vec{\breve{\rho}}(\mathsf{a}\bbox{r},
\mathsf{a}\bbox{r}')\circ \hat{\vec{\tau}}_{tt'}
\!+\! \tquart\breve{\bbox{s}}_0(\mathsf{a}\bbox{r},\mathsf{a}\bbox{r}')\cdot (\mathsf{a}\hat{\bbox{\sigma}})_{ss'}\delta_{tt'}
\!+\! \tquart\vec{\breve{\bbox{s}}}(\mathsf{a}\bbox{r},\mathsf{a}\bbox{r}')\cdot
(\mathsf{a}\hat{\bbox{\sigma}})_{ss'}\circ \hat{\vec{\tau}}_{tt'}.
\label{rhorot2}
\enll

The inversion matrix (\ref{invmat}) is evidently real and
symmetric,  and it does not depend on
$\hat{\bbox{\sigma}}$ and $\hat{\vec{\tau}}$. Thus, we also have
$\hat{\breve{p}}(\bbox{r}'s't',\bbox{r}st)=\hat{p}(\bbox{r}'s't',\bbox{r}st)$,
and
\bnll{rhoinv}\!\!\!\!\!\!\!\!
\hat{\rho}^P(\bbox{r}st,\bbox{r}'s't')
&\!\!=\!\!&\tquart\rho_0(-\bbox{r},-\bbox{r}')\delta_{ss'}\delta_{tt'}
\!+\! \tquart\delta_{ss'}\vec{\rho}(-\bbox{r},
-\bbox{r}')\circ \hat{\vec{\tau}}_{tt'}
\!+\! \tquart\bbox{s}_0(-\bbox{r},-\bbox{r}')\cdot \hat{\bbox{\sigma}}_{ss'}\delta_{tt'}
\!+\! \tquart\vec{\bbox{s}}(-\bbox{r},-\bbox{r}')\cdot
\hat{\bbox{\sigma}}_{ss'}\circ \hat{\vec{\tau}}_{tt'},
\label{rhoinv1}\\\!\!\!\!\!\!\!\!
\hat{\breve{\rho}}^P(\bbox{r}st,\bbox{r}'s't')
&\!\!=\!\!&\tquart\breve{\rho}_0(-\bbox{r},-\bbox{r}')\delta_{ss'}\delta_{tt'}
\!+\! \tquart\delta_{ss'}\vec{\breve{\rho}}(-\bbox{r},
-\bbox{r}')\circ \hat{\vec{\tau}}_{tt'}
\!+\! \tquart\breve{\bbox{s}}_0(-\bbox{r},-\bbox{r}')\cdot \hat{\bbox{\sigma}}_{ss'}\delta_{tt'}
\!+\! \tquart\vec{\breve{\bbox{s}}}(-\bbox{r},-\bbox{r}')\cdot
\hat{\bbox{\sigma}}_{ss'}\circ \hat{\vec{\tau}}_{tt'}.
\label{rhoinv2}
\enll

The time reversal $T$ is an antilinear operation and has the form
given in Eq.~(\ref{antlin}) with the corresponding single-particle
time-reversal matrix $\hat{t}$ given in Eq.~(\ref{timat}). It
transforms all the position-dependent densities and the isospin Pauli
matrices $\hat{\vec{\tau}}$ to their complex conjugate partners and
changes signs of  $\bbox{\sigma}$. Therefore,
$\hat{\breve{t}}(\bbox{r}'s't',\bbox{r}st)=
\hat{t}(\bbox{r}'s't',\bbox{r}st)$,
and the time-reversed densities are:
\bnll{rhoti}\!\!\!\!\!\!\!\!
\hat{\rho}^T(\bbox{r}st,\bbox{r}'s't')
&\!\!=\!\!&\tquart\rho^{\ast}_0(\bbox{r},\bbox{r}')\delta_{ss'}\delta_{tt'}
\!+\! \tquart\delta_{ss'}\vec{\rho}^{\ast}(\bbox{r},
\bbox{r}')\circ \hat{\vec{\tau}}^{\ast}_{tt'}
\!-\! \tquart\bbox{s}^{\ast}_0(\bbox{r},\bbox{r}')\cdot \hat{\bbox{\sigma}}_{ss'}\delta_{tt'}
\!-\! \tquart\vec{\bbox{s}}^{\ast}(\bbox{r},\bbox{r}')\cdot
\hat{\bbox{\sigma}}_{ss'}\circ \hat{\vec{\tau}}^{\ast}_{tt'},
\label{rhoti1}\\\!\!\!\!\!\!\!\!
\hat{\breve{\rho}}^T(\bbox{r}st,\bbox{r}'s't')
&\!\!=\!\!&\tquart\breve{\rho}^{\ast}_0(\bbox{r},\bbox{r}')\delta_{ss'}\delta_{tt'}
\!+\! \tquart\delta_{ss'}\vec{\breve{\rho}}^{\ast}(\bbox{r},
\bbox{r}')\circ \hat{\vec{\tau}}^{\ast}_{tt'}
\!-\! \tquart\breve{\bbox{s}}^{\ast}_0(\bbox{r},\bbox{r}')\cdot \hat{\bbox{\sigma}}_{ss'}\delta_{tt'}
\!-\! \tquart\vec{\breve{\bbox{s}}}^{\ast}(\bbox{r},\bbox{r}')\cdot
\hat{\bbox{\sigma}}_{ss'}\circ \hat{\vec{\tau}}^{\ast}_{tt'}.
\label{rhoti2}
\enll
\end{widetext}
From Eqs.~(\ref{rhorot2}), (\ref{rhoinv2}), and (\ref{rhoti2}) we conclude
that the density matrix $\hat{\breve{\rho}}$ in the {\pairing}
channel transforms under all position-spin transformations
considered here in the
same way as the {\particle} density matrix $\hat{\rho}$. In other
words, the generalized transformation matrices of
$\hat{\breve{\mathcal P}}$ and $\hat{\breve{\mathcal
T}}$ have the same general structures as that of rotations,
$\hat{\breve{\mathcal D}}$, given in Eq.~(\ref{dgen}).

\section{Symmetry properties of densities}\label{sec12}

We begin by recalling general symmetry properties of nonlocal
densities, which have been given in I.
Since the {\particle} density matrix (\ref{hc1}) and the Pauli matrices are both
Hermitian, all the {\particle} densities are Hermitian as well:
\bnll{hermicon2}
\rho_k(\bbox{r},\bbox{r}')
&=&\rho^{*}_k(\bbox{r}',\bbox{r}), \\
\bbox{s}_k(\bbox{r},\bbox{r}')
&=&\bbox{s}^{*}_k(\bbox{r}',\bbox{r}),
\enll
for $k=0,1,2,3$; hence, their real parts are symmetric, while the
imaginary parts are antisymmetric, with respect to exchanging
 $\bbox{r}$ and $\bbox{r}'$.

Similarly, transformation properties of Pauli matrices under $TC$
(\ref{hc2}) imply that   {\pairing} densities are either symmetric
(scalar-isovector and vector-isoscalar) or antisymmetric
(scalar-isoscalar and vector-isovector) under the transposition of
their arguments, namely:
\bnll{symepai}
\label{symepai-a}
\breve{\rho}_k        (\bbox{r},\bbox{r}')
&=& (-1)^{t_k+1} \breve{\rho}_k                (\bbox{r}',\bbox{r}), \\
\label{symepai-c}
\breve{\bbox{s}}_k    (\bbox{r},\bbox{r}')
&=& (-1)^{t_k}   \breve{\bbox{s}}_k (\bbox{r}',\bbox{r}),
\enll
for $k=0,1,2,3$, where $t_0=0$ (isoscalars) and $t_{1,2,3}=1$ (isovectors).
Equations (\ref{hermicon2}) and (\ref{symepai}) are fulfilled
independently of any other symmetries conserved by the system;
they result from definitions  of density matrices
$\hat{\rho}$ and $\hat{\breve{\rho}}$.

\subsection{Spherical symmetry}\label{sec8d}

\subsubsection{Spherical and space-inversion symmetries}\label{sec8d1}

Let us suppose that the generalized density matrix is invariant under
the transformations of Eqs.~(\ref{romat})
and (\ref{invmat}) forming the full orthogonal group O(3)=P$\otimes$SO(3)$\supset$SO(3),
which is the direct product of the group of proper rotations SO(3)
and the two-element group P of the space inversion.  It means
that $\hat{\breve{\mathcal R}}^D=\hat{\breve{\mathcal
R}}^P=\hat{\breve{\mathcal R}}^{DP}=\hat{\breve{\mathcal R}}$, and these
symmetries impose the following conditions on nonlocal densities:
\bnll{rinvnon}
\rho_k(\bbox{r},\bbox{r}')&=&\rho_k(\varsigma\mathsf{a}\bbox{r},\varsigma\mathsf{a}\bbox{r}'),\label{rinvnon1}\\
\breve{\rho}_k(\bbox{r},\bbox{r}')&=
&\breve{\rho}_k(\varsigma\mathsf{a}\bbox{r},\varsigma\mathsf{a}\bbox{r}'),\label{rinvnon2}\\
\bbox{s}_k(\bbox{r},\bbox{r}')&=&\mathsf{a}^+\bbox{s}_k(\varsigma\mathsf{a}\bbox{r},\varsigma\mathsf{a}\bbox{r}'),
\label{rinvnon3}\\
\breve{\bbox{s}}_k(\bbox{r},\bbox{r}')&=
&\mathsf{a}^+\breve{\bbox{s}}_k(\varsigma\mathsf{a}\bbox{r},\varsigma\mathsf{a}\bbox{r}') ,
\label{rinvnon4}
\enll
for $k=0,1,2,3$, and for arbitrary Euler angles $\alpha ,\beta
,\gamma$ that are arguments of the rotation matrix $\mathsf{a}$
(\ref{spinrot}). The factor  $\varsigma$ is equal to +1 for rotations and $-1$
for improper rotations.

The full O(3) symmetry imposes quite strong conditions
(\ref{rinvnon}) on the nonlocal densities. Equations~(\ref{rinvnon1})
and (\ref{rinvnon2}) tell us that, due to the Generalized
Cayley-Hamilton (GCH) theorem (see Appendix \ref{sec13}), scalar
densities, $\rho_k$ and $\breve{\rho}_k$, depend on $\bbox{r}$ and
$\bbox{r}'$ through rotational invariants $\bbox{r}\cdot\bbox{r} =
r^2$, $\bbox{r}'\cdot\bbox{r}' = r^{\prime 2}$, and
$\bbox{r}\cdot\bbox{r}'$, i.e.,
\bnll{rinvnons}
\rho_k(\bbox{r},\bbox{r}')&=&\rho_k(r^2,\bbox{r}\cdot\bbox{r}',r^{\prime 2}) , \label{rinvnons1}\\
\breve{\rho}_k(\bbox{r},\bbox{r}')&=&\breve{\rho}_k(r^2,\bbox{r}\cdot\bbox{r}',r^{\prime 2}) , \label{rinvnons2}
\enll
for $k=0,1,2,3$.

Similarly, from Eqs.~(\ref{rinvnon3}) and (\ref{rinvnon4}), we see
that vector densities, ${\bbox{s}}_k$ and $\breve{\bbox{s}}_k$, are
pseudovectors. At the same time, they are functions of two vectors,
$\bbox{r}$ and $\bbox{r}'$. The only pseudovector that can be
constructed from two vectors is their vector product
$\bbox{r}\times\bbox{r}'$ --- therefore, all pseudovector densities,
$\bbox{s}_k$ and $\breve{\bbox{s}}_k$, for $k=0,1,2,3$, have the form
\bnll{rinvnonv}
\label{rinvnon5}
\bbox{s}_k(\bbox{r},\bbox{r}')&=&i(\bbox{r}\times\bbox{r}')s_k(r^2,\bbox{r}\cdot\bbox{r}',r^{\prime 2}), \\
\label{rinvnon6}
\breve{\bbox{s}}_k(\bbox{r},\bbox{r}')&=& (\bbox{r}\times\bbox{r}')\breve{s}_k(r^2,\bbox{r}\cdot\bbox{r}',r^{\prime 2}).
\enll
For the sake of convenience, in the definition (\ref{rinvnon5}) we
have introduced the imaginary unit $i$.

Due to the general symmetry properties (\ref{hermicon2}) and
(\ref{symepai}), scalar functions that define the nonlocal densities
must obey the following conditions:
\bnll{rinvnont-a}
\label{rinvnont1}
        \rho_k(r^2,\bbox{r}\cdot\bbox{r}',r^{\prime 2})&=&                  \rho^*_k(r^{\prime 2},\bbox{r}\cdot\bbox{r}',r^2) , \\
\label{rinvnont2}
\breve{\rho}_k(r^2,\bbox{r}\cdot\bbox{r}',r^{\prime 2})&=&(-1)^{t_k+1}\breve{\rho}_k(r^{\prime 2},\bbox{r}\cdot\bbox{r}',r^2) ,
\enll
and
\bnll{rinvnont-b}
\label{rinvnont5}
           s_k(r^2,\bbox{r}\cdot\bbox{r}',r^{\prime 2})&=&                     s^*_k(r^{\prime 2},\bbox{r}\cdot\bbox{r}',r^2) , \\
\label{rinvnont6}
   \breve{s}_k(r^2,\bbox{r}\cdot\bbox{r}',r^{\prime 2})&=&(-1)^{t_k+1}   \breve{s}_k(r^{\prime 2},\bbox{r}\cdot\bbox{r}',r^2) .
\enll
This means that nonlocal {\particle} densities
(\ref{rinvnont1}) and (\ref{rinvnont5}) are Hermitian,
while nonlocal
isoscalar and isovector {\pairing} densities
(\ref{rinvnont2}) and (\ref{rinvnont6}) are antisymmetric and
symmetric functions of $\bbox{r}$ and $\bbox{r}'$, respectively.

The above symmetry properties of densities imply strong conditions
on  local densities. For instance,
 Eqs.~(\ref{rinvnons})--(\ref{rinvnont-b}) imply that
\bnll{rinvnonl-a}
\label{rinvnonl1}
                         \rho_k(\bbox{r})\equiv                          \rho_k(\bbox{r},\bbox{r})&=&\rho_k(r)=\rho^*_k(r) , \\
\label{rinvnonl2}
             \vec{\breve{\rho}}(\bbox{r})\equiv\vec{\breve{\rho}}_{\phantom{0}}(\bbox{r},\bbox{r})&=&\vec{\breve{\rho}}(r)  ,
\enll
and
\bnll{rinvnonl-b}
\label{rinvnonl5}
                     \bbox{s}_k(\bbox{r})\equiv                      \bbox{s}_k(\bbox{r},\bbox{r})&=&0                    , \\
\label{rinvnonl6}
             \breve{\bbox{s}}_0(\bbox{r})\equiv              \breve{\bbox{s}}_0(\bbox{r},\bbox{r})&=&0                    ,
\enll
i.e., the scalar local {\particle} densities $\rho_k$, and isovector
scalar local {\pairing} densities $\vec{\breve{\rho}}$, depend on
 the radial variable $r$ only, $\rho_k$ are real, and all
vector local {\particle} densities $\bbox{s}_k$ and the isoscalar
vector local {\pairing} density $\breve{\bbox{s}}_0$ vanish. At
this point, we remind the reader that conditions (\ref{symepai})
imply  that local {\pairing} densities
$\breve{\rho}_0$ and
$\vec{\breve{\bbox{s}}}$ always vanish, irrespective of any other
symmetries being imposed or not, see Table IV in I.

All local derivative densities can be derived from
Eqs.~(\ref{rinvnons}) and (\ref{rinvnonv}) by using expressions for
gradients of scalar functions, e.g.,
\bnll{gradscal}
\bbox{\nabla} \rho(r^2,\bbox{r}\cdot\bbox{r}',r^{\prime 2})&=&
2 \frac{\partial{\rho}}{\partial{(r^2)}}\bbox{r}
+\frac{\partial{\rho}}{\partial{(\bbox{r}\cdot\bbox{r}')}}\bbox{r}', \label{gradscal1}\\
\bbox{\nabla}'\rho(r^2,\bbox{r}\cdot\bbox{r}',r^{\prime 2})&=&
 \frac{\partial{\rho}}{\partial{(\bbox{r}\cdot\bbox{r}')}}\bbox{r}
+2\frac{\partial{\rho}}{\partial{(r^{\prime 2})}}\bbox{r}' , \label{gradscal2}
\enll
which are linear combinations of vectors $\bbox{r}$ and $\bbox{r}'$
with scalar coefficients (again illustrating the GCH theorem). In
this way, all local derivative densities can be expressed through
derivatives of the scalar functions $\rho_k$, $\breve{\rho}_k$,
$s_k$, and $\breve{s}_k$.
Alternatively, one can  employ the GCH theorem to build local
scalar, pseudoscalar, vector, pseudovector, and
symmetric-traceless-pseudotensor densities from the single position
vector $\bbox{r}$, and we follow this path below.

Pseudoscalar,
pseudovector, and pseudotensor densities
cannot be built from the position vector $\bbox{r}$.
Therefore, they
must all vanish:
\bnll{pseudol}
\label{pseudolt1}
                        \bbox{T} _k(\bbox{r})&=&0 , \\
\label{pseudolt2}
                 \breve{\bbox{T}}_0(\bbox{r})&=&0 , \\
\label{pseudolf1}
                        \bbox{F} _k(\bbox{r})&=&0 , \\
\label{pseudolf2}
                 \breve{\bbox{F}}_0(\bbox{r})&=&0 , \\
\label{pseudolJ1}
                                J_k(\bbox{r})&=&0 , \\
\label{pseudolJ2}
                    \vec{\breve{J}}(\bbox{r})&=&0 , \\
\label{pseudolJ3}
                \underline{{\mathsf J}}_{kab}(\bbox{r})&=&0 , \\
\label{pseudolJ4}
   \underline{\vec{\breve{{\mathsf J}}}}_{ab}(\bbox{r})&=&0 ,
\enll
for $k=0,1,2,3$.

The local scalar kinetic densities must have
properties analogous to those of scalar densities in
Eqs.~(\ref{rinvnonl-a}), i.e.,
\bnll{rinvkin}
\label{rinvkin1}
                         \tau_k(\bbox{r})\equiv            \tau_k(\bbox{r},\bbox{r})&=&\tau_k(r)=\tau^*_k(r) , \\
\label{rinvkin2}
             \vec{\breve{\tau}}(\bbox{r})\equiv\vec{\breve{\tau}}(\bbox{r},\bbox{r})&=&\vec{\breve{\tau}}(r)  ,
\enll
and all local vector densities must be proportional to $\bbox{r}$, i.e.,
\bnll{rinvj}
\label{rinvj1}
\bbox{j}_k(\bbox{r})&=&\bbox{j}_{kr}(r)\bbox{e}_r=\bbox{j}^*_{kr}(r)\bbox{e}_r , \\
\label{rinvj2}
\breve{\bbox{j}}_0(\bbox{r})&=&\breve{\bbox{j}}_{0r}(r)\bbox{e}_r , \\
\label{rinvj3}
\bbox{J}_k(\bbox{r})&=&\bbox{J}_{kr}(r)\bbox{e}_r=\bbox{J}^*_{kr}(r)\bbox{e}_r , \\
\label{rinvj4}
\vec{\breve{\bbox{J}}}(\bbox{r})&=&\vec{\breve{\bbox{J}}}_r(r)\bbox{e}_r .
\enll
where $\bbox{e}_r=
\frac{\bbox{r}}{r}$ is the unit vector in radial direction.
In addition, the radial components
$\bbox{j}_{kr}$ and
$\bbox{J}_{kr}$ are real.

Conditions on local densities, presented in this section, can be
further restricted by imposing the time-reversal and/or
p-n symmetries, see Table IV of I. The
conditions on local time-even and proton-neutron-symmetric
{\particle} and {\pairing} densities are  exactly those
as in Ref.~\cite{[Dob84]}, whereupon properties of the
{\pairing} densities exactly mirror those of the {\particle}
densities. Such a mirroring does not hold if
time-reversal or p-n symmetries are broken.

For a broken time-reversal symmetry and conserved
p-n symmetry two modifications occur: (i) the isovector
{\pairing} densities $\vec{\breve{\rho}}(r)$,
$\vec{\breve{\tau}}(r)$, and $\vec{\breve{\bbox{J}}}_r(r)$ become
complex, while the {\particle} densities $\rho_k(r)$, $\tau_k(r)$,
and $\bbox{J}_{kr}(r)$  still remain real \cite{[Dob84]}, and (ii) the
current {\particle} density $\bbox{j}_{kr}(r)$ does not vanish. It
is interesting to see that in this case the only non-zero time-odd
density is the current density, i.e., spin polarizations are not
allowed and only the flow of particles in the radial direction (a
breathing mode) is
permitted if the spherical and space-inversion symmetries are present.

It is also interesting to see that the spherical and space-inversion
symmetries impose very strong restrictions on the isoscalar pairing
densities. Indeed, the  isoscalar pairing density
$\breve{\bbox{s}}_0(\bbox{r})$ (\ref{rinvnonl6}) must then vanish
\cite{[Per03b]}. The only allowed isoscalar-pairing channel can be
related to the {\pairing} current density $\breve{\bbox{j}}_{0r}(r)$,
which represents a radial flow of isoscalar pairs within a nucleus.
Such a flow can be, in fact, non-zero either in the time-even or
time-odd case, represented by $\Re(\breve{\bbox{j}}_{0r})(r)$ or
$\Im(\breve{\bbox{j}}_{0r})(r)$, respectively. It corresponds to the
situation in which p-n pairs locally change into the
neutron-proton pairs, or {\it vice versa}, while the sum of densities
thereof remains constant.

\subsubsection{Spherical symmetry alone}\label{sec8d2}

Let us consider the unusual case of the SO(3)$\subset$O(3) symmetry, in which the
spherical symmetry of the generalized density matrix is conserved,
whereas the space inversion symmetry is broken. As compared
to the results presented in the previous section~\ref{sec8d1}, here,
properties of scalar nonlocal densities, Eqs.~(\ref{rinvnons}),
remain the same. However, in Eqs.~(\ref{rinvnon})
$\varsigma$ is always equal to +1, meaning that there is no
difference between vectors and pseudovectors. Thus, vector nonlocal
densities can now have structures that are richer than those of
Eqs.~(\ref{rinvnonv}). Indeed, due to the GCH theorem,
these densities can be linear combinations of the
pseudovector $\bbox{r}\times\bbox{r}'$ {\em and} vectors $\bbox{r}$
and $\bbox{r}'$.

Hence, vector densities in the {\particle} and {\pairing} channels
can now be presented in the form:
\bnll{rinvnon-a}
\label{rinvnon7}
\bbox{s}_k(\bbox{r},\bbox{r}')
&\!\!=\!\!&i(\bbox{r}\times\bbox{r}')s_k(r^2,\bbox{r}\cdot\bbox{r}',r^{\prime 2}) \\
&\!\!+\!\!&\bbox{r} s'  _k(r^2,\bbox{r}\cdot\bbox{r}',r^{\prime 2})
   \!+\!   \bbox{r}'s^{\prime *}_{k}(r^{\prime 2},\bbox{r}\cdot\bbox{r}',r^2), \nonumber \\
\label{rinvnon8}
\breve{\bbox{s}}_k(\bbox{r},\bbox{r}')
&\!\!=\!\!& (\bbox{r}\times\bbox{r}')\breve{s}_k(r^2,\bbox{r}\cdot\bbox{r}',r^{\prime 2}) \\
&\!\!+\!\!&         \bbox{r} \breve{s}'_k(r^2,\bbox{r}\cdot\bbox{r}',r^{\prime 2})
   \!+\!  (-1)^{t_k}\bbox{r}'\breve{s}'_k(r^{\prime 2},\bbox{r}\cdot\bbox{r}',r^2),  \nonumber
\enll
where scalar functions $s_k$ and $\breve{s}_k$ obey previous
conditions (\ref{rinvnont-b}), while scalar functions $s'_k$ and
$\breve{s}'_k$ are arbitrary.

Breaking the parity does not affect the local scalar densities in
both channels and Eqs.~(\ref{rinvnonl-a}) and (\ref{rinvkin}) are still
valid. The same is true for vector current densities of
Eqs.~(\ref{rinvj1}) and (\ref{rinvj2}) being gradients of scalar
densities. On the other hand, the local spin densities no longer
vanish. The isoscalar and isovector components of the {\particle}
spin density are

\begin{equation}
\bbox{s}_k(\bbox{r})=\bbox{s}_{kr}(r)\bbox{e}_r, \label{rinvs1}
\end{equation}
where their radial components $\bbox{s}_{kr}$ are real for $k=0,1,2,3$.
The {\pairing} isoscalar spin density has a complex radial component and reads:
\begin{equation}
\breve{\bbox{s}}_0(\bbox{r})=\breve{\bbox{s}}_{0r}(r)\bbox{e}_r.
\label{rinvs2}
\end{equation}
Spin-kinetic and tensor-kinetic densities have the same structures as
the spin densities of Eqs.~(\ref{rinvs1}) and (\ref{rinvs2}), namely
\bnll{rinvsk}
\bbox{T}_k(\bbox{r})=\bbox{T}^{\ast}_k(\bbox{r})&=&\bbox{T}_{kr}(r)\bbox{e}_r, \label{rinvsk1}\\
\bbox{F}_k(\bbox{r})=\bbox{F}^{\ast}_k(\bbox{r})&=&\bbox{F}_{kr}(r)\bbox{e}_r \label{rinvsk2}
\enll
for all $k$'s and
\bnll{rinvskp}
\breve{\bbox{T}}_0(\bbox{r})&=&\breve{\bbox{T}}_{0r}(r)\bbox{e}_r, \label{rinvskp1}\\
\breve{\bbox{F}}_0(\bbox{r})&=&\breve{\bbox{F}}_{0r}(r)\bbox{e}_r. \label{rinvskp2}
\enll
When the parity is not conserved, the tensor densities have, apart
from the antisymmetric parts represented by vectors of
Eqs.~(\ref{rinvj3}) and (\ref{rinvj4}), also non-vanishing traces,
namely,
\bnll{rinvtens}
J_k(\bbox{r})&=&J_{k}(r)=J^{\ast}_{k}(r) ,\label{rinvtens1}\\
\vec{\breve{J}}(\bbox{r})&=&\vec{\breve{J}}(r) \label{rinvtens2}
\enll
and symmetric traceless parts have the following structure:
\bnll{rinvtent}
\underline{\mathsf J}_{kab}(\bbox{r})&=&
\tfrac{1}{2}\underline{\mathsf J}_{krr}(r)\underline{\mathsf S}_{ab}=
\tfrac{1}{2}\underline{\mathsf J}^{\ast}_{krr}(r)\underline{\mathsf S}_{ab}, \label{rinvtent1}\\
\underline{\vec{\breve{{\mathsf J}}}}_{ab}(\bbox{r})&=&
\tfrac{1}{2}\underline{\vec{\breve{\mathsf J}}}_{krr}(r)\underline{\mathsf S}_{ab}, \label{rinvtent2}
\enll
where the standard
symmetric traceless tensor function of the space vector $\bbox{r}$ is defined as
\beq
\underline{\mathsf S}_{ab}=3\frac{\bbox{r}_a\bbox{r}_b}{r^2}-\delta_{ab}.
\eeq

\subsubsection{Spherical symmetry -- summary}\label{sec8d3}

\newcommand{\fRr}{f_R(r)\bbox{e}_r}
\newcommand{\fIr}{f_I(r)\bbox{e}_r}
\newcommand{\fCr}{f_C(r)\bbox{e}_r}
\newcommand{\fRS}{f_R(r)\underline{\mathsf S}}
\newcommand{\fCS}{f_C(r)\underline{\mathsf S}}
\newcommand{\fIS}{f_I(r)\underline{\mathsf S}}
When the spherical O(3) symmetry is conserved, all the local densities
can be treated as fields depending on the O(3) vector $\bbox{r}$ (see
Appendix \ref{sec13}). Then, vector $\bbox{r}$ itself is the only one
elementary vector and its length squared $r^2$ is the only one
elementary scalar. The scalar densities are functions of $r^2$. The
vector densities are pointing along $\bbox{r}$, and their radial
components depend on $r^2$ only.

The pseudoscalar, pseudovector, and
pseudotensor densities all vanish. They can become non-vanishing when
the parity is broken and only the rotational SO(3) symmetry remains
conserved. Then, they have properties of the scalar, vector, and
tensor densities, respectively. The symmetric tensor densities are
proportional to  outer product $\bbox{r}\otimes\bbox{r}$.

In
general, the {\particle} densities are real whereas the {\pairing}
densities are complex. Additional symmetries (space inversion, time-reversal,
p-n  symmetry), when conserved, can cause
additionally some isoscalars or isovector {\particle} densities to vanish, and
those in the {\pairing} channel  become either purely real or purely
imaginary, or vanish. In Tables \ref{tab1a} and \ref{tab1b}, we list all the
local SO(3)-invariant densities in cases when additional symmetries
are broken or conserved.

\begin{table*}
\renewcommand{\arraystretch}{0.75}
\caption[T3]{Properties of local particle-hole
rotationally symmetric (SO(3)-invariant) densities, depending on the
conserved (C) or broken (B) space-inversion ($P$), proton-neutron
(p-n), or time-reversal ($T$) symmetries. Generic real, imaginary, or
complex functions of the radial variable $r$ are denoted by $f_R(r)$,
$f_I(r)$, or $f_C(r)$, respectively.}
\label{tab1a}
\begin{ruledtabular}
\begin{tabular}{l|llllllll}
symmetry                                &
\multicolumn{8}{c}{conserved (C)
or broken (B)}                                                                   \\
\hline
$P$                                     &   B    &     B  &     B  &     B &  C  &  C  &  C  &  C      \\
p-n                                     &   B    &     B  &     C  &     C     &   B    &     B  &     C  &     C  \\
$T$                                     &   B    &     C  &     B  &     C     &   B    &     C  &     B  &     C    \\
\hline
$\rho_{0,3}$                            &$f_R(r)$&$f_R(r)$&$f_R(r)$&$f_R(r)$&$f_R(r)$&$f_R(r)$&$f_R(r)$&$f_R(r)$       \\
$\rho_1$                            &$f_R(r)$&$f_R(r)$&$0$&$0$&$f_R(r)$&$ f_R(r)  $&$0$&$  0   $    \\
$\rho_2$                          &$f_R(r)$&$0$ &$0$ &$0$ &$f_R(r)$&$  0   $&$0$&$  0   $    \\
$\tau_{0,3}$                            &$f_R(r)$&$f_R(r)$&$f_R(r)$&$f_R(r)$&$f_R(r)$&$f_R(r)$&$f_R(r)$&$f_R(r)$      \\
$\tau_{1}$                            &$f_R(r)$&$f_R(r)$&$0$&$0$&$f_R(r)$&$f_R(r)$&$ 0   $&$  0   $       \\
$\tau_{2}$                           &$f_R(r)$&$0$&$0$&$0$&$f_R(r)$&$0$&$  0   $&$  0   $       \\
${J}_{0,3}$                             &$f_R(r)$&$f_R(r)$&$f_R(r)$&$f_R(r)$&$  0   $&$  0   $&$  0   $ &$0$      \\
${J}_{1}$                             &$f_R(r)$&$f_R(r)$&$0$&$0$&$  0   $&$  0   $&$  0   $    &$0$   \\
$J_{2}$                                &$f_R(r)$&$0$&$0$ &$0$ &$0$&$  0   $&$  0   $   &$0$ \\
$\bbox{s}_{0,3}$                        &$\fRr  $&$0$&$\fRr$&$0$&$  0   $&$  0   $&$  0   $    & $0$ \\
$\bbox{s}_{1}$                        &$\fRr  $&$0$&$0$&$0$&$  0   $&$  0   $&$  0   $     &$0$  \\
$\bbox{s}_{2}$                        &$\fRr  $&$\fRr  $&$0$ & $0$&$  0   $&$  0   $&$  0   $     &$0$  \\
${\bbox{T}}_{0,3}$                      &$\fRr  $&$0$&$\fRr$&$0$&$  0   $&$  0   $&$  0   $    &$0$   \\
${\bbox{T}}_{1}$                      &$\fRr  $&$0$&$0$&$0$&$  0   $&$  0   $&$  0   $     &$0$  \\
${\bbox{T}}_{2}$                      &$\fRr  $&$\fRr  $&$0$ &$0$ &$  0   $&$  0   $&$  0   $     &$0$  \\
${\bbox{F}}_{0,3}$                      &$\fRr  $&$0$&$\fRr$&$0$&$  0   $&$  0   $&$  0   $     &$0$  \\
${\bbox{F}}_{1}$                      &$\fRr  $&$0$&$0$&$0$&$  0   $&$  0   $&$  0   $     &$0$  \\
${\bbox{F}}_{2}$                      &$\fRr  $&$\fRr$&$0$&$0$&$  0   $&$  0   $&$  0   $     &$0$  \\
${\bbox{j}}_{0,3}$                      &$\fRr  $&$0$&$\fRr$&$0$&$\fRr  $&$0$&$\fRr  $&$  0   $       \\
${\bbox{j}}_{1}$                      &$\fRr  $&$0$&$0$&$0$&$\fRr  $&$0$&$  0   $&$  0   $       \\
${\bbox{j}}_{2}$                      &$\fRr  $&$\fRr  $&$0$&$0$&$\fRr  $&$\fRr$&$  0   $&$  0   $       \\
${\bbox{J}}_{0,3}$                      &$\fRr  $&$\fRr$&$\fRr$&$\fRr$&$\fRr  $&$\fRr$&$\fRr  $&$\fRr  $       \\
${\bbox{J}}_{1}$                      &$\fRr  $&$\fRr$&$0$&$0$&$\fRr  $&$\fRr$&$  0   $&$  0   $       \\
${\bbox{J}}_{2}$                      &$\fRr  $&$0  $&$0$&$0$&$\fRr  $&$  0   $&$  0   $    &$0$  \\
$\underline{{\mathsf J}}_{0,3}$         &$\fRS  $&$\fRS  $&$\fRS$&$\fRS$&$  0   $&$  0   $&$  0   $   &$0$   \\
$\underline{{\mathsf J}}_{1}$         &$\fRS  $&$\fRS  $&$0$&$0$&$  0   $&$  0   $&$  0   $     &$0$  \\
$\underline{{\mathsf J}}_{2}$         &$\fRS  $&$0$&$0$&$0$&$  0   $&$  0   $&$  0   $     &$0$  \\
\end{tabular}
\end{ruledtabular}
\renewcommand{\arraystretch}{1}
\end{table*}

\begin{table*}
\caption[T3]{Similar to Table~\protect\ref{tab1a} except for the particle-particle
densities.}
\label{tab1b}
\begin{ruledtabular}
\begin{tabular}{l|llllllll}
symmetry                                &
\multicolumn{8}{c}{conserved (C)
or broken (B)}                                                                   \\
\hline
$P$                                     &   B    &     B  &     B  &     B &  C  &  C  &  C  &  C      \\
p-n                                     &   B    &     B  &     C  &     C     &   B    &     B  &     C  &     C  \\
$T$                                     &   B    &     C  &     B  &     C     &   B    &     C  &     B  &     C    \\
\hline
$\breve{\rho}_1$                        &$f_C(r)$&$f_R(r)$&$f_C(r)$&$f_R(r)$&$f_C(r)$&$f_R(r)$&$f_C(r)$&$f_R(r)$       \\
$\breve{\rho}_2$                        &$f_C(r)$&$f_I(r)$&$f_C(r)$&$f_I(r)$&$f_C(r)$&$f_I(r)$&$f_C(r)$&$f_I(r)$       \\
$\breve{\rho}_3$                        &$f_C(r)$&$f_R(r)$&$0$&$0$&$f_C(r)$&$f_R(r)$&$  0   $&$  0   $      \\
$\breve{\tau}_1$                        &$f_C(r)$&$f_R(r)$&$f_C(r)$&$f_R(r)$&$f_C(r)$&$f_R(r)$&$f_C(r)$&$f_R(r)$      \\
$\breve{\tau}_2$                        &$f_C(r)$&$f_I(r)$&$f_C(r)$&$f_I(r)$&$f_C(r)$&$f_I(r)$&$f_C(r)$&$f_I(r)$       \\
$\breve{\tau}_3$                        &$f_C(r)$&$f_R(r)$&$0$&$0$&$f_C(r)$&$f_R(r)$&$  0   $&$  0   $       \\
${\breve{J}}_1$                       &$f_C(r)$&$f_R(r)$&$f_C(r)$&$f_R(r)$&$  0   $&$  0   $&$  0   $    &$0$   \\
${\breve{J}}_2$                       &$f_C(r)$&$f_I(r)$&$f_C(r)$&$f_I(r)$&$  0   $&$  0   $&$  0   $    &$0$  \\
${\breve{J}}_3$                       &$f_C(r)$&$f_R(r)$&$0$&$0$&$  0   $&$  0   $&$  0   $    &$0$  \\
$\breve{\bbox{s}}_0$                    &$\fCr  $&$\fIr$&$0$&$0$&$  0   $&$  0   $&$  0   $    &$0$  \\
$\breve{\bbox{T}}_0$                    &$\fCr  $&$\fIr$&$0$&$0$&$  0   $&$  0   $&$  0   $   &$0$   \\
$\breve{\bbox{F}}_0$                    &$\fCr  $&$\fIr$&$0$&$0$&$  0   $&$  0   $&$  0   $   &$0$   \\
$\breve{\bbox{j}}_0$                    &$\fCr  $&$\fIr$&$0$&$0$&$\fCr  $&$  0   $&$  0   $    &$0$  \\
$\breve{\bbox{J}}_1$                    &$\fCr  $&$\fRr$&$\fCr$&$\fRr$&$\fCr  $&$\fRr$&$\fCr  $&$\fRr  $      \\
$\breve{\bbox{J}}_2$                    &$\fCr  $&$\fIr$&$\fCr$&$\fIr$&$\fCr  $&$\fIr$&$\fCr  $&$\fIr  $      \\
$\breve{\bbox{J}}_3$                    &$\fCr  $&$\fRr$&$0$&$0$&$\fCr  $&$\fRr$&$  0   $&$  0   $     \\
$\underline{\breve{{\mathsf J}}}_1$ &$\fCS  $&$\fRS$&$\fCS$&$\fRS$&$  0   $&$  0   $&$  0   $   &$0$   \\
$\underline{\breve{{\mathsf J}}}_2$ &$\fCS  $&$\fIS$&$\fCS$&$\fIS$&$  0   $&$  0   $&$  0   $   &$0$   \\
$\underline{\breve{{\mathsf J}}}_3$ &$\fCS  $&$\fRS$&$0$&$0$&$  0   $&$  0   $&$  0   $   & $0$  \\
\end{tabular}
\end{ruledtabular}
\end{table*}

\subsection{Axial symmetry}\label{sec8e}
While the spherical symmetry of the mean field is  often broken,
the axial symmetry is usually  conserved in the presence of time reversal. Here we discuss consequences
of conserved symmetry of
Eq.~(\ref{romat}) with $\beta$=0 and $\gamma$=0, and the
transformation of Eq.~(\ref{signz}), forming together group
O$^{z\perp}$(2)=S$_z\otimes$SO$^{\perp}$(2)$\subset$O(3), which is
the direct product of the rotations SO$^{\perp}$(2) about the
$z$-axis and the reflection S$_z$ in the plane perpendicular to this
axis. For rotations about the $z$-axis, the Cartesian rotation matrix
(\ref{cartrot}) takes the form:
\bn
\mathsf{a}(\alpha 00 )&=&\left(\ba{ccc} \cos{\alpha} & -\sin{\alpha}& 0 \\
                                                            \sin{\alpha}&  \cos{\alpha}& 0 \\
                                                             0 & 0 & 1\ea\right) \nonumber \\
&=&\left(\ba{cc}\mathsf{a}^{\perp}(\alpha ) & 0 \\
                           0 & 1\ea\right).
\label{cartrotz}
\en
It is now convenient to decompose the position vectors as
\begin{equation}
\bbox{r}=\bbox{r}_{\perp}+\bbox{z},
\label{rperp}
\end{equation}
where $\bbox{r}_{\perp}$ is a two-dimensional vector perpendicular to
the $z$-axis and $\bbox{z}$ is the $z$-component vector of
$\bbox{r}$.

Since $-\mathsf{a}^{\perp}(\alpha
)=\mathsf{a}^{\perp}(\alpha +\pi )$, the O$^{z\perp}$(2)=
S$_z\otimes$SO$^{\perp}$(2) symmetry of the generalized density
matrix implies the following conditions for the nonlocal densities:
\bnll{ainvnon}
&&\rho_k(\bbox{r},\bbox{r}')=\rho_k( \mathsf{a}^{\perp}\bbox{r}_{\perp}+\varsigma\bbox{z},
\mathsf{a}^{\perp}\bbox{r}'_{\perp} +\varsigma\bbox{z}'),\label{ainvnon1}\\
&&\breve{\rho}_k(\bbox{r},\bbox{r}')=
\breve{\rho}_k(\mathsf{a}^{\perp}\bbox{r}_{\perp}+\varsigma\bbox{z},
\mathsf{a}^{\perp}\bbox{r}'_{\perp} +\varsigma\bbox{z}'),\label{ainvnon2}\\
&&\bbox{s}_k(\bbox{r},\bbox{r}')=\varsigma(\mathsf{a}^{\perp})^+\bbox{s}_{k\perp}(
\mathsf{a}^{\perp}\bbox{r}_{\perp}+\varsigma\bbox{z},
\mathsf{a}^{\perp}\bbox{r}'_{\perp} +\varsigma\bbox{z}')\nonumber \\
&&+\bbox{s}_{kz}(
\mathsf{a}^{\perp}\bbox{r}_{\perp}+\varsigma\bbox{z},
\mathsf{a}^{\perp}\bbox{r}'_{\perp} +\varsigma\bbox{z}'))\bbox{e}_z,
\label{ainvnon3}\\
&&\breve{\bbox{s}}_k(\bbox{r},\bbox{r}')=\varsigma
(\mathsf{a}^{\perp})^+\breve{\bbox{s}}_{k\perp}(
\mathsf{a}^{\perp}\bbox{r}_{\perp}+\varsigma\bbox{z},
\mathsf{a}^{\perp}\bbox{r}'_{\perp} +\varsigma\bbox{z}')\nonumber \\
&&+\breve{\bbox{s}}_{kz}(
\mathsf{a}^{\perp}\bbox{r}_{\perp}+\varsigma\bbox{z},
\mathsf{a}^{\perp}\bbox{r}'_{\perp} +\varsigma\bbox{z}')\bbox{e}_z,
\label{ainvnon4}
\enll
where $\bbox{e}_z$ is the unit vector ($\bbox{z}=z \bbox{e}_z$) and $\varsigma$
is the sign of the determinant of the orthogonal
O$^{z\perp}$(2) transformation.

\subsubsection{Axial and mirror symmetry}\label{sec8e1}
In this case,  Eqs.~(\ref{ainvnon1})
and (\ref{ainvnon2}) imply that the scalar densities $\rho_k$ and
$\breve{\rho}_k$ depend on $\bbox{r}$ and $\bbox{r}'$ through the O$^{z\perp}(2)$
invariants $\bbox{z}\cdot\bbox{z}=z^2$,
$\bbox{z}'\cdot\bbox{z}'=z^{\prime 2}$, $\bbox{z}\cdot\bbox{z}'=zz'$,
$\bbox{r}_{\perp}\cdot\bbox{r}_{\perp}=r^2_{\perp}$,
$\bbox{r}'_{\perp}\cdot\bbox{r}'_{\perp}=r^{\prime 2}_{\perp}$, and
$\bbox{r}_{\perp}\cdot\bbox{r}'_{\perp}$.
Apart from the invariants, there are two  O$^{z\perp}$(2) pseudoscalars:
$\bbox{z}\cdot (\bbox{r}_{\perp}\times\bbox{r}'_{\perp})$ and
$\bbox{z}'\cdot (\bbox{r}_{\perp}\times\bbox{r}'_{\perp})=(zz'/z^2)\bbox{z}\cdot (\bbox{r}_{\perp}\times\bbox{r}'_{\perp})$.
The spin nonlocal densities have the following
transformation properties under the O$^{z\perp}(2)$.
 Their $z$-components are the
O$^{z\perp}(2)$ scalars. Their perpendicular components
are the SO$^{\perp}(2)$ vectors and
S$_z$ pseudoscalars (i.e., they change sign under $S_z$).
Because the spin densities are the O(3)  pseudovectors, their components that are
parallel  to the $z$-axis should be linear combinations
of $\bbox{r}_{\perp}\times\bbox{r}'_{\perp}$, $[\bbox{z}\cdot (\bbox{r}_{\perp}\times\bbox{r}'_{\perp})]\bbox{z}$,
and $[\bbox{z}'\cdot (\bbox{r}_{\perp}\times\bbox{r}'_{\perp})]\bbox{z}'$,
while the perpendicular components should be linear combinations of
$\bbox{z}\times\bbox{r}_{\perp}$,
$\bbox{z}'\times\bbox{r}'_{\perp}$,  $[\bbox{z}\cdot (\bbox{r}_{\perp}\times\bbox{r}'_{\perp})]\bbox{r}_{\perp}$, and
$[\bbox{z}'\cdot (\bbox{r}_{\perp}\times\bbox{r}'_{\perp})]\bbox{r}'_{\perp}$. Consequently, the Hermitian
nonlocal spin densities in the {\particle} channel should
have the following structure:
\bnll{psvecnon}
\bbox{s}_{kz}(\bbox{r},\bbox{r}')\bbox{e}_z
&=&i(\bbox{r}_{\perp}\times\bbox{r}'_{\perp})\varrho_{kz}(\bbox{r},\bbox{r}') \nonumber \\
&&+[\bbox{z}\cdot (\bbox{r}_{\perp}\times\bbox{r}'_{\perp})]\bbox{z}\varrho_{kz}'(\bbox{r},\bbox{r}') \nonumber \\
&&-[\bbox{z}'\cdot (\bbox{r}_{\perp}\times\bbox{r}'_{\perp})]\bbox{z}'\varrho_{kz}^{\prime +}(\bbox{r},\bbox{r}') , \label{psvecnon1}\\
\bbox{s}_{k\perp}(\bbox{r},\bbox{r}')
&=&(\bbox{z}\times\bbox{r}_{\perp})\varrho_{k\perp}(\bbox{r},\bbox{r}')  \nonumber \\
&&+(\bbox{z}'\times\bbox{r}'_{\perp}){\varrho}^+_{k\perp}(\bbox{r},\bbox{r}') \nonumber \\
&&+[\bbox{z}\cdot (\bbox{r}_{\perp}\times\bbox{r}'_{\perp})]\bbox{r}_{\perp}\varrho_{k\perp}'(\bbox{r},\bbox{r}') \nonumber \\
&&-[\bbox{z}'\cdot (\bbox{r}_{\perp}\times\bbox{r}'_{\perp})]\bbox{r}'_{\perp}\varrho_{k\perp}^{\prime +}(\bbox{r},\bbox{r}'),\label{psvecnon2}
\enll
where $\varrho_{kz}$, $\varrho_{kz}'$, $\varrho_{k\perp}$, and $\varrho_{k\perp}'$ are scalar functions. The pseudovector nonlocal
densities in the {\pairing} channel are either symmetric or
antisymmetric in $\bbox{r}$ and $\bbox{r}'$. Therefore, we have
\bnll{psvecnonp}
\breve{\bbox{s}}_{0z}(\bbox{r},\bbox{r}')\bbox{e}_z
&=&(\bbox{r}_{\perp}\times\bbox{r}'_{\perp})\breve{\varrho}_{0z}(\bbox{r},\bbox{r}') \nonumber \\
&&+[\bbox{z}\cdot (\bbox{r}_{\perp}\times\bbox{r}'_{\perp})]\bbox{z}\breve{\varrho}_{0z}'(\bbox{r},\bbox{r}') \nonumber \\
&&-[\bbox{z}'\cdot (\bbox{r}_{\perp}\times\bbox{r}'_{\perp})]\bbox{z}'\breve{\varrho}_{0z}'(\bbox{r}',\bbox{r})
, \label{psvecnonp1}\\
\vec{\breve{\bbox{s}}}_{z}(\bbox{r},\bbox{r}')\bbox{e}_z
&=&(\bbox{r}_{\perp}\times\bbox{r}'_{\perp})\vec{\breve{\varrho}}_{z}(\bbox{r},\bbox{r}') \nonumber \\
&&+[\bbox{z}\cdot (\bbox{r}_{\perp}\times\bbox{r}'_{\perp})]\bbox{z}\vec{\breve{\varrho}}_{z}'(\bbox{r},\bbox{r}') \nonumber \\
&&+[\bbox{z}'\cdot (\bbox{r}_{\perp}\times\bbox{r}'_{\perp})]\bbox{z}'\vec{\breve{\varrho}}_{z}'(\bbox{r}',\bbox{r}),
\label{psvecnonp2}\\
\breve{\bbox{s}}_{0\perp}(\bbox{r},\bbox{r}')
&=&(\bbox{z}\times\bbox{r}_{\perp})\breve{\varrho}_{0\perp}(\bbox{r},\bbox{r}')  \nonumber \\
&&+(\bbox{z}'\times\bbox{r}'_{\perp}){\breve{\varrho}}_{0\perp}(\bbox{r}',\bbox{r})\nonumber \\
&&+[\bbox{z}\cdot (\bbox{r}_{\perp}\times\bbox{r}'_{\perp})]\bbox{r}_{\perp}\breve{\varrho}_{0\perp}'(\bbox{r},\bbox{r}') \nonumber \\
&&-[\bbox{z}'\cdot (\bbox{r}_{\perp}\times\bbox{r}'_{\perp})]\bbox{r}'_{\perp}\breve{\varrho}_{0\perp}'(\bbox{r}',\bbox{r}),
\label{psvecnonp3}\\
\vec{\breve{\bbox{s}}}_{\perp}(\bbox{r},\bbox{r}')
&=&(\bbox{z}\times\bbox{r}_{\perp})\vec{\breve{\varrho}}_{\perp}(\bbox{r},\bbox{r}') \nonumber \\
&&-(\bbox{z}'\times\bbox{r}'_{\perp})\vec{{\breve{\varrho}}}_{\perp}(\bbox{r}',\bbox{r}) \nonumber \\
&&+[\bbox{z}\cdot (\bbox{r}_{\perp}\times\bbox{r}'_{\perp})]\bbox{r}_{\perp}\vec{\breve{\varrho}}_{\perp}'(\bbox{r},\bbox{r}') \nonumber \\
&&+[\bbox{z}'\cdot (\bbox{r}_{\perp}\times\bbox{r}'_{\perp})]\bbox{r}'_{\perp}\vec{\breve{\varrho}}_{\perp}'(\bbox{r}',\bbox{r}),
\label{psvecnonp4}
\enll
where $\breve{\varrho}_{0\perp}$, $\breve{\varrho}'_{0\perp}$, $\vec{\breve{\varrho}}_{\perp}$, $\vec{\breve{\varrho}}'_{\perp}$,
 $\breve{\varrho}'_{0z}$, and $\vec{\breve{\varrho}}'_{0\perp}$
are arbitrary scalar functions, while  $\breve{\varrho}_{0z}$ and
$\vec{\breve{\varrho}}_{z}$ are antisymmetric and symmetric scalars,
respectively.

The local scalar {\particle} and {\pairing} densities are functions of two
invariants, $z^2$ and $r_{\perp}^2$. The particle density
$\rho_k(z^2,r_{\perp}^2)$ for $k=0,1,2,3$ is real whereas the pairing
density $\vec{\breve{\rho}}(z^2,r_{\perp}^2)$ is complex in general;
hence,
\bnll{rinvnonl-c}
\label{rinvnonl3}
                         \rho_k(\bbox{r})\equiv \rho_k(\bbox{r},\bbox{r})&=&\rho_k(z^2,r_{\perp}^2)=\rho^*_k(z^2,r_{\perp}^2) , \\
\label{rinvnonl4}
             \vec{\breve{\rho}}(\bbox{r})\equiv\vec{\breve{\rho}}_{\phantom{0}}(\bbox{r},\bbox{r})&=&\vec{\breve{\rho}}(z^2,r_{\perp}^2)  .
\enll

Differential operators $\bbox{\nabla}_z$ and
$\bbox{\nabla}_\perp$ have the same transformation properties under
the O$^{z\perp}$(2) transformations  as $\bbox{z}$ and
$\bbox{r}_{\perp}$, respectively. That is, $\bbox{\nabla}_z$ and $\bbox{z}$ are S$_z$
pseudo-invariants and SO$^{\perp}$(2) invariants, whereas
$\bbox{\nabla}_\perp$ and $\bbox{r}_{\perp}$
are S$_z$ invariants and SO$^{\perp}$(2) vectors.
Counterparts of Eqs.~(\ref{gradscal1}) and (\ref{gradscal2}) for
gradients of scalar functions are now linear combinations of vectors $\bbox{z}$,
$\bbox{z}'$, $\bbox{r}_\perp$, and $\bbox{r}'_\perp$ with scalar
coefficients:
\bnll{gradax}
&&(\bbox{\nabla}_z+\bbox{\nabla}_{\perp})\rho(z^2,zz',z'^2,r_{\perp}^2,\bbox{r}_{\perp}\cdot\bbox{r}'_{\perp},r_{\perp}^{\prime 2}) \nonumber \\
&&\hspace{2cm}=2\frac{\partial{\rho}}{\partial{(z^2)}}\bbox{z}+\frac{\partial{\rho}}{\partial{(zz')}}\bbox{z}' \nonumber \\
&&\hspace{2cm}+2\frac{\partial{\rho}}{\partial{(r^{2}_\perp)}}\bbox{r}_\perp+ \frac{\partial{\rho}}{\partial{(\bbox{r}_\perp\cdot\bbox{r}'_\perp)}}\bbox{r}'_\perp ,
\label{gradax1}\\
&&(\bbox{\nabla}'_z+\bbox{\nabla}'_{\perp})\rho(z^2,zz',z'^2,r_{\perp}^2,\bbox{r}_{\perp}\cdot\bbox{r}'_{\perp},r_{\perp}^{\prime 2}) \nonumber \\
&&\hspace{2cm}=\frac{\partial{\rho}}{\partial{(zz')}}\bbox{z}+2\frac{\partial{\rho}}{\partial{(z'^2)}}\bbox{z}' \nonumber \\
&&\hspace{2cm}+ \frac{\partial{\rho}}{\partial{(\bbox{r}_\perp\cdot\bbox{r}'_\perp)}}\bbox{r}_\perp +2\frac{\partial{\rho}}{\partial{(r^{\prime 2}_\perp)}}\bbox{r}'_\perp.
\label{gradax2}
\enll

Both terms of the operator:
\beq\label{lapl}
\bbox{\nabla}\cdot\bbox{\nabla}'=\bbox{\nabla}_z\cdot\bbox{\nabla}'_z+\bbox{\nabla}_\perp\cdot\bbox{\nabla}'_\perp
\eeq
are O$^{z\perp}$(2) scalars.
Therefore, the local scalar kinetic densities:
\bnll{lockin}
\!\!\!\!\!\!\!\!\tau_{k}(z^2,r_{\perp}^2)
&=&[(\bbox{\nabla}_z\cdot\bbox{\nabla}'_z+\bbox{\nabla}_\perp\cdot\bbox{\nabla}'_\perp)\rho_k(\bbox{r},\bbox{r}')]_{\bbox{r}=
\bbox{r}'}, \label{lockin1}\\
\!\!\!\!\!\!\!\!\vec{\breve{\tau}}(z^2,r_{\perp}^2)
&=&[(\bbox{\nabla}_z\cdot\bbox{\nabla}'_z+\bbox{\nabla}_\perp\cdot\bbox{\nabla}'_\perp)
\vec{\breve{\rho}}(\bbox{r},\bbox{r}')]_{\bbox{r}=\bbox{r}'}, \label{lockin2}
\enll
can be expressed as sums of the two O$^{z\perp}$(2) scalars. Since
the operator (\ref{lapl}) is Hermitian, again the {\particle}
densities are real and the {\pairing} ones are complex.

It is seen from Eqs.~(\ref{psvecnon}) and (\ref{psvecnonp} that the spin densities in
both channels are parallel to the vector product $\bbox{z}\times\bbox{r}_{\perp}$ and
take the form:
\bnll{locsp}
\bbox{s}_{k}(\bbox{r}) &=& \varrho_{k\perp}(z^2,r_\perp ^2)(\bbox{z}\times\bbox{r}_{\perp}), \label{locsp1} \\
 \breve{\bbox{s}}_{0} &=& \breve{\varrho}_{0\perp}(z^2,r_\perp ^2)(\bbox{z}\times\bbox{r}_{\perp}), \label{locsp2}
 \enll
with real $\varrho_{k\perp}$ and complex $\breve{\varrho}_{0\perp}$.
Applying Eq.~(\ref{gradax}) to Eqs.~(\ref{psvecnon}) and (\ref{psvecnonp}), we find that the
spin-kinetic and tensor-kinetic densities in both channels can be written as:
\bnll{spkin}
\bbox{T}_k(\bbox{r})&=&[(\bbox{\nabla}_z\cdot\bbox{\nabla}'_z+\bbox{\nabla}_\perp\cdot\bbox{\nabla}'_\perp)\bbox{s}_k(\bbox{r},\bbox{r}')]_{\bbox{r}=
\bbox{r}'} \nonumber \\
&=&\vartheta_k (z^2,r_\perp ^2)(\bbox{z}\times\bbox{r}_{\perp}),\label{spkin1}\\
\breve{\bbox{T}}_0(\bbox{r})&=&[(\bbox{\nabla}_z\cdot\bbox{\nabla}'_z+\bbox{\nabla}_\perp\cdot\bbox{\nabla}'_\perp)
\breve{\bbox{s}}_0(\bbox{r},\bbox{r}')]_{\bbox{r}=
\bbox{r}'} \nonumber \\
&=&\breve{\vartheta}_0(z^2,r_\perp ^2)(\bbox{z}\times\bbox{r}_{\perp}),\label{spkin2}
\enll
and
\bnll{tenkin}
\bbox{F}_k(\bbox{r})&\!=\!&\half [(\bbox{\nabla}_z+\bbox{\nabla}_{\perp})((\bbox{\nabla}'_z+\bbox{\nabla}'_{\perp})\cdot\bbox{s}_k(\bbox{r},\bbox{r}'))
\nonumber \\
&+&(\bbox{\nabla}'_z+\bbox{\nabla}'_{\perp})((\bbox{\nabla}_z+\bbox{\nabla}_{\perp})\cdot\bbox{s}_k(\bbox{r},\bbox{r}'))]_{\bbox{r}=\bbox{r}'} \nonumber \\
&\!=\!&\varphi_k (z^2,r_\perp ^2)(\bbox{z}\times\bbox{r}_{\perp}),\label{tenkin1}\\
\breve{\bbox{F}}_0(\bbox{r})&\!=\!&\half [(\bbox{\nabla}_z+\bbox{\nabla}_{\perp})((\bbox{\nabla}'_z+\bbox{\nabla}'_{\perp})\cdot
\breve{\bbox{s}}_0(\bbox{r},\bbox{r}')) \nonumber \\
&+&(\bbox{\nabla}'_z+\bbox{\nabla}'_{\perp})((\bbox{\nabla}_z+\bbox{\nabla}_{\perp})\cdot\breve{\bbox{s}}_0(\bbox{r},\bbox{r}'))]_{\bbox{r}=\bbox{r}'}
\nonumber \\
&\!=\!&\breve{\varphi}_0(z^2,r_\perp ^2)(\bbox{z}\times\bbox{r}_{\perp}).\label{tenkin2}
\enll
The scalar functions $\vartheta_k$ and $\varphi_k$ are real whereas $\breve{\vartheta}_0$ and $\breve{\varphi}_0$ are, in general, complex.
Equations (\ref{gradax}) imply  that the vector current densities in both channels are spanned
by vectors $\bbox{z}$ and $\bbox{r}_{\perp}$, and read
\bnll{locurr}
\bbox{j}_k(\bbox{r}) &\!=\!&\iota_{kz} (z^2,r_{\perp}^2)\bbox{z}+\iota_{k\perp}(z^2,r_{\perp}^2)\bbox{r}_{\perp}, \label{locurr1}\\
\breve{\bbox{j}}_0(\bbox{r}) &\!=\!& \breve{\iota}_{0z}(z^2,r_{\perp}^2)\bbox{z}
+\breve{\iota}_{0\perp}(z^2,r_{\perp}^2)\bbox{r}_{\perp}, \label{locurr2}
\enll
where $\iota_{kz}$, $\iota_{k\perp}$, $\breve{\iota}_{0z}$, and
$\breve{\iota}_{0\perp}$ are scalar functions. The {\particle}
currents are real and the {\pairing} current  is complex.

The  spin-current pseudotensor densities can be  decomposed according to:
\bnll{tenspc}
\mathsf{J}_k (\bbox{r}) &=& \mathsf{J}_k^z(\bbox{r})+\mathsf{J}_k^{\perp}(\bbox{r})+\mathsf{J}^{z\perp}_k(\bbox{r}), \label{tenspc1}\\
\vec{\breve{\mathsf{J}}}(\bbox{r}) &=& \vec{\breve{\mathsf{J}}}^z(\bbox{r})
+\vec{\breve{\mathsf{J}}}^{\perp}(\bbox{r})+\vec{\breve{\mathsf{J}}}^{z\perp}(\bbox{r}), \label{tenspc2}
\enll
where
\bnll{tenspcz}
\mathsf{J}_k^z(\bbox{r}) &=&\frac{1}{2i}[(\bbox{\nabla}_z-\bbox{\nabla}'_z)\otimes \bbox{s}_{kz}(\bbox{r},\bbox{r}')\bbox{e}_z]_{\bbox{r}=\bbox{r}'},
\label{tenspcz1}\\
\mathsf{J}_k^{\perp}(\bbox{r}) &=&\frac{1}{2i}[(\bbox{\nabla}_{\perp}-\bbox{\nabla}'_{\perp})\otimes
\bbox{s}_{k\perp}(\bbox{r},\bbox{r}')]_{\bbox{r}=\bbox{r}'}, \label{tenspcz2}\\
\mathsf{J}_k^{z\perp}(\bbox{r}) &=&\frac{1}{2i}[(\bbox{\nabla}_z-\bbox{\nabla}'_z)\otimes \bbox{s}_{k\perp}(\bbox{r},\bbox{r}') \nonumber \\
&&+(\bbox{\nabla}_{\perp}-\bbox{\nabla}'_{\perp})\otimes \bbox{s}_{kz}(\bbox{r},\bbox{r}')\bbox{e}_z]_{\bbox{r}=\bbox{r}'},\label{tenspcz3}\\
\vec{\breve{\mathsf{J}}}^z(\bbox{r}) &=&
\frac{1}{2i}[(\bbox{\nabla}_z-\bbox{\nabla}'_z)\otimes \vec{\breve{\bbox{s}}}_{z}(\bbox{r},\bbox{r}')\bbox{e}_z]_{\bbox{r}=\bbox{r}'},
\label{tenspcz4}\\
\vec{\breve{\mathsf{J}}}^{\perp}(\bbox{r}) &=&\frac{1}{2i}[(\bbox{\nabla}_{\perp}-\bbox{\nabla}'_{\perp})\otimes
\vec{\breve{\bbox{s}}}_{\perp}(\bbox{r},\bbox{r}')]_{\bbox{r}=\bbox{r}'}, \label{tenspcz5}\\
\vec{\breve{\mathsf{J}}}^{z\perp}(\bbox{r}) &=&\frac{1}{2i}[(\bbox{\nabla}_z-\bbox{\nabla}'_z)\otimes \vec{\breve{\bbox{s}}}_{\perp}(\bbox{r},\bbox{r}') \nonumber \\
&&+(\bbox{\nabla}_{\perp}-\bbox{\nabla}'_{\perp})\otimes \vec{\breve{\bbox{s}}}_{z}(\bbox{r},\bbox{r}')\bbox{e}_z]_{\bbox{r}=\bbox{r}'}.\label{tenspcz6}
\enll
The
traces of the spin-current pseudotensors:
\bnll{so2ztrace}
J_k(\bbox{r}) &=& J^z_k(\bbox{r})+J^{\perp}_k(\bbox{r}), \label{so2ztrace1} \\
\vec{\breve{J}}(\bbox{r}) &=&\vec{\breve{J}}^z(\bbox{r})+\vec{\breve{J}}^{\perp}(\bbox{r}), \label{so2ztrace2}
\enll
where
\bnll{o2trace}
J^z_k(\bbox{r}) &=& \frac{1}{2i}[(\bbox{\nabla}_z-\bbox{\nabla}'_z)\cdot
\bbox{s}_{kz}(\bbox{r},\bbox{r}')\bbox{e}_z]_{\bbox{r}=\bbox{r}'}, \label{o2trace1} \\
J^{\perp}_k(\bbox{r}) &=& \frac{1}{2i}[(\bbox{\nabla}_{\perp}-\bbox{\nabla}'_{\perp})\cdot
\bbox{s}_{k\perp}(\bbox{r},\bbox{r}')]_{\bbox{r}=\bbox{r}'} ,
 \label{o2trace2}\\
\vec{\breve{J}}^z(\bbox{r}) &=& \frac{1}{2i}[(\bbox{\nabla}_z-\bbox{\nabla}'_z)\cdot
\vec{\breve{\bbox{s}}}_{z}(\bbox{r},\bbox{r}')\bbox{e}_z]_{\bbox{r}=\bbox{r}'}, \label{o2trace3} \\
\vec{\breve{J}}^{\perp}(\bbox{r}) &=&\frac{1}{2i}[(\bbox{\nabla}_{\perp}-\bbox{\nabla}'_{\perp})\cdot
\vec{\breve{\bbox{s}}}_{\perp}(\bbox{r},\bbox{r}')]_{\bbox{r}=\bbox{r}'},  \label{o2trace4}
\enll
are the
scalar products of gradient operators and spin densities.
Since $J_k$ and $\vec{\breve{J}}$ are sums of
O$^{z\perp}$(2) pseudoscalars, they  cannot be constructed from two
vectors $\bbox{z}$ and $\bbox{r}_{\perp}$; hence, all spin currents
(\ref{o2trace}) must vanish.
On the other hand, the O(3) vectors
coming from the antisymmetric parts of spin-current pseudotensors:
\bnll{o2rot}
\bbox{J}_{k}(\bbox{r}) &=& \frac{1}{2i}[(\bbox{\nabla}_z-\bbox{\nabla}'_z)\times
\bbox{s}_{k\perp}(\bbox{r},\bbox{r}') \nonumber \\
&&+(\bbox{\nabla}_{\perp}-\bbox{\nabla}'_{\perp})\times
\bbox{s}_{k}(\bbox{r},\bbox{r}')]_{\bbox{r}=\bbox{r}'}, \label{o2rot1} \\
\vec{\breve{\bbox{J}}}(\bbox{r}) &=& \frac{1}{2i}[(\bbox{\nabla}_z-\bbox{\nabla}'_z)\times
\vec{\breve{\bbox{s}}}_{\perp}(\bbox{r},\bbox{r}') \nonumber \\
&&+(\bbox{\nabla}_{\perp}-\bbox{\nabla}'_{\perp})\times
\vec{\breve{\bbox{s}}}(\bbox{r},\bbox{r}')]_{\bbox{r}=\bbox{r}'}, \label{o2rot2}
\enll
do not vanish.
They can be decomposed in the same way as
the current vectors (\ref{locurr}):
\bnll{o2rotcomp}
\bbox{J}_{k}(\bbox{r})&=&\upsilon_{kz}(z^2,r_{\perp}^2)\bbox{z} +\upsilon_{k\perp}(z^2,r_{\perp}^2)\bbox{r}_{\perp}, \label{o2rotcomp1}\\
\vec{\breve{\bbox{J}}}(\bbox{r})&=&\vec{\breve{\upsilon}}_{z}(z^2,r_{\perp}^2)\bbox{z}
+\vec{\breve{\upsilon}}_{\perp}(z^2,r_{\perp}^2)\bbox{r}_{\perp}. \label{o2rotcomp2}
\enll
The scalar functions $\upsilon_{kz}$ and $\upsilon_{k\perp}$ are real whereas
$\vec{\breve{\upsilon}}_z$ and $\vec{\breve{\upsilon}}_{\perp}$ are complex. Finally, the traceless symmetric parts of the spin-current densities (\ref{tenspc}) are:
\bnll{o2tensor}
\underline{\mathsf J}_{k}^z(\bbox{r}) &=& 0, \label{o2tensor1}\\
\vec{\breve{\underline{\mathsf J}}}^z(\bbox{r}) &=&0, \label{o2tensor2}\\
\underline{\mathsf J}_k^{\perp}(\bbox{r}) &=& \kappa_k^{\perp}(z^2,r_{\perp}^2)(\underline{\bbox{r}_{\perp}\otimes
(\bbox{z}\times\bbox{r}_{\perp})}), \label{o2tensor3} \\
\underline{\mathsf J}_k^{z\perp}(\bbox{r}) &=& \kappa_k^{z\perp}(z^2,r_{\perp}^2)(\underline{\bbox{z}\otimes
(\bbox{z}\times\bbox{r}_{\perp})}), \label{o2tensor4} \\
\vec{\breve{\underline{\mathsf J}}}^{\perp}(\bbox{r}) &=& \vec{\breve{\kappa}}^{\perp}(z^2,r_{\perp}^2)(\underline{\bbox{r}_{\perp}\otimes
(\bbox{z}\times\bbox{r}_{\perp})}), \label{o2tensor5} \\
\vec{\breve{\underline{\mathsf J}}}^{z\perp}(\bbox{r}) &=& \vec{\breve{\kappa}}^{z\perp}(z^2,r_{\perp}^2)(\underline{\bbox{z}\otimes
(\bbox{z}\times\bbox{r}_{\perp})}), \label{o2tensor6}
\enll
where
\bnll{upsten}
\underline{\bbox{r}_{\perp}\otimes(\bbox{z}\times\bbox{r}_{\perp})}&=&
\frac{1}{2}\left[\bbox{r}_{\perp}\otimes(\bbox{z}\times\bbox{r}_{\perp}) \right. \nonumber \\
&&+\left.(\bbox{z}\times\bbox{r}_{\perp})\otimes\bbox{r}_{\perp}\right], \label{upsten1}\\
\underline{\bbox{z}\otimes(\bbox{z}\times\bbox{r}_{\perp})}&=&
\frac{1}{2}\left[\bbox{z}\otimes(\bbox{z}\times\bbox{r}_{\perp}) \right.\nonumber \\
&&+\left.(\bbox{z}\times\bbox{r}_{\perp})\otimes\bbox{z}\right] \label{upsten2}
\enll
are the symmetrized outer products of the vector product $\bbox{z}\times\bbox{r}_{\perp}$, and vectors $\bbox{r}_{\perp}$ and $\bbox{z}$, respectively.
As usual, the scalar functions $\kappa_k$ are real, whereas $\vec{\breve{\kappa}}$ are complex.

\subsubsection{Axial symmetry alone}\label{sec8e2}
From the above discussion we see that the mirror symmetry imposes
quite strong conditions on the nonlocal and local density functions.
When the generalized density matrix  is invariant only under the
SO$^{\perp}$(2)$\subset$SO(3) group of rotations about the $z$-axis,
the transformation rules of Eqs.~(\ref{ainvnon}) with $\varsigma =+1$
are fulfilled. This means that coordinate $\bbox{r}_z=z$  is the SO$^{\perp}$(2)
invariant, and there is no difference between the SO(3) pseudovectors and
vectors. The scalar nonlocal densities are thus functions of $z$,
$r_\perp$, $z'$, $r'_\perp$, and
$\bbox{r}_{\perp}\cdot\bbox{r}'_{\perp}$.

The spin densities are
SO$^{\perp}$(2) vectors and can take  more general forms than those
of Eqs.~(\ref{psvecnon}) and (\ref{psvecnonp}); namely, the Hermitian
spin densities in the {\particle} channel are:
\bnll{vecnon}
\!\!\!\!\bbox{s}_{kz}(\bbox{r},\bbox{r}')\bbox{e}_z
&\!\!=\!\!&i(\bbox{r}_{\perp}\times\bbox{r}'_{\perp})\varrho_{kz}(\bbox{r},\bbox{r}') \nonumber \\
&\!\!\!\!&+\bbox{z}\varrho_{kz}^{\prime}(\bbox{r},\bbox{r}')+
\bbox{z}'\varrho_{kz}^{\prime\ast}(\bbox{r}',\bbox{r}) ,\label{vecnon1}\\
\bbox{s}_{k\perp}(\bbox{r},\bbox{r}')
&\!\!=\!\!&(\bbox{z}\times\bbox{r}_{\perp})\varrho_{k\perp}(\bbox{r},\bbox{r}')+
(\bbox{z}'\times\bbox{r}'_{\perp}){\varrho}^{\ast}_{k\perp}(\bbox{r}',\bbox{r})\nonumber \\
&\!\!\!\!&+ \bbox{r}_{\perp}\varrho_{k\perp}^{\prime}(\bbox{r},\bbox{r}')
+ \bbox{r}'_{\perp}\varrho^{\prime\ast}_{k\perp}(\bbox{r}',\bbox{r}),\label{vecnon2}
\enll
where $\varrho_{kz}$ is  Hermitian and $\varrho^{\prime}_{kz}$, ${\varrho}_{k\perp}$, and
${\varrho}^{\prime}_{k\perp}$ are arbitrary scalars.

The spin nonlocal densities in the {\pairing} channel are either
symmetric or antisymmetric in $\bbox{r}$ and $\bbox{r}'$:
\bnll{vecnonp}
\!\!\!\!\breve{\bbox{s}}_{0z}(\bbox{r},\bbox{r}')\bbox{e}_z
&\!\!=\!\!&(\bbox{r}_{\perp}\times\bbox{r}'_{\perp})\breve{\varrho}_{0z}(\bbox{r},\bbox{r}')\nonumber \\
&\!\!\!\!&+\bbox{z}\breve{\varrho}'_{0z}(\bbox{r},\bbox{r}')+\bbox{z}'\breve{\varrho}'_{0z}(\bbox{r}',\bbox{r}),
\label{vecnonp1}\\
\vec{\breve{\bbox{s}}}_{z}(\bbox{r},\bbox{r}')\bbox{e}_z
&\!\!=\!\!&(\bbox{r}_{\perp}\times\bbox{r}'_{\perp})\vec{\breve{\varrho}}_{z}(\bbox{r},\bbox{r}') \nonumber \\
&\!\!\!\!&+\bbox{z}\vec{\breve{\varrho}}'_{z}(\bbox{r},\bbox{r}')-\bbox{z}'\vec{\breve{\varrho}}'_{z}(\bbox{r}',\bbox{r}),
\label{vecnonp2}\\
\breve{\bbox{s}}_{0\perp}(\bbox{r},\bbox{r}')
&\!\!=\!\!&(\bbox{z}\times\bbox{r}_{\perp})\breve{\varrho}_{0\perp}(\bbox{r},\bbox{r}')
+(\bbox{z}'\times\bbox{r}'_{\perp})\breve{\varrho}_{0\perp}(\bbox{r}',\bbox{r}),
\nonumber \\
&\!\!\!\!&+\bbox{r}_{\perp}{\breve{\varrho}}'_{0\perp}(\bbox{r},\bbox{r}')+\bbox{r}'_{\perp}{\breve{\varrho}}'_{0\perp}(\bbox{r}',\bbox{r}),
\label{vecnonp3}\\
\vec{\breve{\bbox{s}}}_{\perp}(\bbox{r},\bbox{r}')
&\!\!=\!\!&(\bbox{z}\times\bbox{r}_{\perp})\vec{\breve{\varrho}}_{\perp}(\bbox{r},\bbox{r}')
-(\bbox{z}'\times\bbox{r}'_{\perp})\vec{\breve{\varrho}}_{\perp}(\bbox{r}',\bbox{r})
\nonumber \\
&\!\!\!\!&+\bbox{r}_{\perp}\vec{{\breve{\varrho}}}'_{\perp}(\bbox{r},\bbox{r}')
- \bbox{r}'_{\perp}\vec{{\breve{\varrho}}}'_{\perp}(\bbox{r}',\bbox{r}),
\label{vecnonp4}
\enll
where $\breve{\varrho}_{0z}$ is  antisymmetric,
$\vec{\breve{\varrho}}_{z}$ is symmetric, and
$\breve{\varrho}'_{0z}$, $\breve{\varrho}_{0\perp}$,
$\breve{\varrho}'_{0\perp}$, $\vec{\breve{\varrho}}'_{z}$,
$\vec{\breve{\varrho}}_{\perp}$, and $\vec{\breve{\varrho}}'_{\perp}$
are arbitrary complex scalar functions.

In the case of broken mirror
symmetry, the real local {\particle} densities
$\rho_k(z,r_{\perp})=\rho^{\ast}_k(z,r_{\perp})$  and
the complex local isovector {\pairing} density
$\vec{\breve{\rho}}(z,r_{\perp})$ are functions of the two
cylindrical coordinates, $z$ and $r_{\perp}$, i.e., they depend
on the sign of $z$. The same is true for  the scalar kinetic
densities (\ref{lockin}), i.e.,  the real {\particle} densities
$\tau_k(z,r_{\perp})= \tau^{\ast}_k(z,r_{\perp})$ and complex
{\pairing} isovector kinetic density
$\vec{\breve{\tau}}(z,r_{\perp})$.

At this point, it becomes convenient to use the cylindrical coordinates $r_\perp$, $\phi$,
and $z$, and the corresponding unit vectors
$\bbox{e}_{\perp}$,
$\bbox{e}_\phi$, and $\bbox{e}_z$. The {\particle} isoscalar and isovector real
spin densities and the {\pairing} isoscalar complex spin density are:
\bn
\bbox{s}_{k}(\bbox{r})&=&\bbox{s}_{kr_\perp}(z,r_\perp )\bbox{e}_{\perp}
\nonumber \\
&+&\bbox{s}_{k\phi}(z,r_\perp )\bbox{e}_{\phi}
+\bbox{s}_{kz}(z,r_\perp )\bbox{e}_z\label{so2sp1}
\en
for $k=0,1,2,3$, and
\bn
\breve{\bbox{s}}_{0}(\bbox{r})&=&\breve{\bbox{s}}_{0r_\perp}(z,r_\perp )\bbox{e}_{\perp}\nonumber \\
&+&\breve{\bbox{s}}_{0\phi}(z,r_\perp )\bbox{e}_{\phi}
+\breve{\bbox{s}}_{0z}(z,r_\perp )\bbox{e}_z.\label{so2sp2}
\en
All differential local densities can be calculated by using the gradient
formulae:
\bnll{gradso2}
&&(\bbox{\nabla}_z+\bbox{\nabla}_{\perp})\rho(z,z',r_{\perp},\bbox{r}_{\perp}\cdot\bbox{r}'_{\perp},r_{\perp}^{\prime})
 \nonumber \\
&&\hspace{1cm}=\frac{\partial{\rho}}{\partial{z}}\bbox{e}_z+\frac{\partial{\rho}}{\partial{r_\perp}}\bbox{e}_\perp+ \frac{\partial{\rho}}{\partial{(\bbox{r}_\perp\cdot\bbox{r}'_\perp)}}\bbox{r}'_\perp ,
\label{gradso21}\\
&&(\bbox{\nabla}'_z+\bbox{\nabla}'_{\perp})\rho(z,z',r_{\perp},\bbox{r}_{\perp}\cdot\bbox{r}'_{\perp},r_{\perp}^{\prime 2})
 \nonumber \\
&&\hspace{1cm}=\frac{\partial{\rho}}{\partial{z'}}\bbox{e}_z+ \frac{\partial{\rho}}{\partial{(\bbox{r}_\perp\cdot\bbox{r}'_\perp)}}\bbox{r}_\perp +\frac{\partial{\rho}}{\partial{r^{\prime}_\perp}}\bbox{e}'_\perp.
\label{gradso22}
\enll

The spin-kinetic densities (\ref{spkin}) have all non-vanishing  cylindrical components:
\bnll{so2spkin}
\bbox{T}_{k}(\bbox{r})&=&\bbox{T}_{kr_\perp}(z,r_\perp )\bbox{e}_{\perp}
\nonumber \\
&+&\bbox{T}_{k\phi}(z,r_\perp )\bbox{e}_{\phi}
+\bbox{T}_{kz}(z,r_\perp )\bbox{e}_z,\label{so2spkin1}\\
\breve{\bbox{T}}_{0}(\bbox{r})&=&\breve{\bbox{T}}_{0r_\perp}(z,r_\perp )\bbox{e}_{\perp}
\nonumber \\
&+&\breve{\bbox{T}}_{0\phi}(z,r_\perp )\bbox{e}_{\phi}
+\breve{\bbox{T}}_{0z}(z,r_\perp )\bbox{e}_z.\label{so2spkin2}
\enll
As usual, the {\particle} spin-kinetic densities are real and the
{\pairing} ones are complex.

The tensor-kinetic densities
(\ref{tenkin1}) and (\ref{tenkin2}) have the same structure as
the vectors (\ref{so2spkin}):
\bnll{so2tenkin}
\bbox{F}_{k}(\bbox{r})&=&\bbox{F}_{kr_\perp}(z,r_\perp )\bbox{e}_{\perp}
\nonumber \\
&+&\bbox{F}_{k\phi}(z,r_\perp )\bbox{e}_{\phi}
+\bbox{F}_{kz}(z,r_\perp )\bbox{e}_z,\label{so2tenkin1}\\
\breve{\bbox{F}}_{0}(\bbox{r})&=&\breve{\bbox{F}}_{0r_\perp}(z,r_\perp )\bbox{e}_{\perp}
\nonumber \\
&+&\breve{\bbox{F}}_{0\phi}(z,r_\perp )\bbox{e}_{\phi}
+\breve{\bbox{F}}_{0z}(z,r_\perp )\bbox{e}_z.\label{so2tenkin2}
\enll
The current densities, being proportional to the gradients of scalar
functions, have only the
$r_\perp$- and $z$-components:
\bnll{so2locurr}
\bbox{j}_k(\bbox{r}) &= & \bbox{j}_{kz}(z,r_{\perp})\bbox{e}_z+\bbox{j}_{kr_\perp}(z,r_{\perp})
\bbox{e}_\perp , \label{so2locurr1}\\
\breve{\bbox{j}}_0(\bbox{r}) &= & \breve{\bbox{j}}_{0z}(z,r_{\perp})\bbox{e}_z
+\breve{\bbox{j}}_{0r_\perp}(z,r_{\perp})\bbox{e}_\perp . \label{so2locurr2}
\enll

Finally, due to   the breaking of mirror symmetry, the spin-current densities can  have rich structures.
The traces of
spin-current tensors (\ref{o2trace}) do not vanish, and they
decompose into the sums of two  SO$^{\perp}$(2)
scalars:
\bnll{so2trace}
J_k(\bbox{r}) &=& J^z_k(z,r_{\perp})+J^{\perp}_k(z,r_{\perp}), \label{so2trace1} \\
\vec{\breve{J}}(\bbox{r}) &=&\vec{\breve{J}}^z(z,r_{\perp})+\vec{\breve{J}}^{\perp}(z,r_{\perp}). \label{so2trace2}
\enll
The antisymmetric parts (\ref{o2rot}) of
the spin-current  tensors (\ref{tenspcz})  form the
vectors with nonzero transverse
components:
\bnll{so2rot}
&&\bbox{J}_k(\bbox{r}) =\bbox{J}_k^{\ast}(\bbox{r})= \bbox{J}_{kr_{\perp}}(z,r_{\perp})\bbox{e}_{\perp} \nonumber \\
&&+\bbox{J}_{k\phi}(z,r_{\perp})\bbox{e}_{\phi}+\bbox{J}_{kz}(z,r_{\perp})\bbox{e}_z ,\label{so2rot1} \\
&&\vec{\breve{\bbox{J}}}(\bbox{r}) = \vec{\breve{\bbox{J}}}_{r_{\perp}}(z,r_{\perp})\bbox{e}_{\perp} \nonumber \\
&&+\vec{\breve{\bbox{J}}}_{\phi}(z,r_{\perp})\bbox{e}_{\phi}+\vec{\breve{\bbox{J}}}_{z}(z,r_{\perp})\bbox{e}_z .\label{so2rot2}
\enll
The symmetric traceless parts of the tensors (\ref{tenspcz}) vanish:
\bnll{symtrlz}
\underline{\mathsf{J}}^z_{kzz}&=& 0 ,\label{symtrlz1}\\
\underline{\vec{\breve{\mathsf{J}}}}^z_{zz}&=& 0 .\label{symtrlz2}
\enll
The remaining traceless symmetric tensors are:
\bnll{symtrlp}
\underline{\mathsf{J}}^{\perp}_{kab}&=&\underline{\mathsf{J}}^{\perp\ast}_{kab}\nonumber \\
&=&\underline{\mathsf{K}}^{\perp}_{k\perp\perp}(z,r_{\perp})\underline{\mathsf{P}}^{\perp}_{ab}
 +\underline{\mathsf{J}}^{\perp}_{k\perp\perp}(z,r_{\perp})\underline{\mathsf{S}}^{\perp}_{ab},\label{symtrlp1}\\
\underline{\vec{\breve{\mathsf{J}}}}^{\perp}_{ab}&=&
\underline{\vec{\breve{\mathsf{K}}}}^{\perp}_{\perp\perp}(z,r_{\perp})\underline{\mathsf{P}}^{\perp}_{ab}
+\underline{\vec{\breve{\mathsf{J}}}}^{\perp}_{\perp\perp}(z,r_{\perp})\underline{\mathsf{S}}^{\perp}_{ab}\label{symtrlp2}
\enll
for $a,b =x,y$, where
\beq
\underline{\mathsf{S}}^{\perp}_{ab}= 2\frac{\bbox{r}_{\perp a}\bbox{r}_{\perp b}}{r_{\perp}^2}-\delta_{ab} \label{trlp}
\eeq
is the standard symmetric traceless SO$^{\perp}$(2) tensor and
\beq\label{pstrlp}
\underline{\mathsf{P}}^{\perp}_{ab}=\frac{z}{2|z|r_{\perp}^2}(\bbox{r}_{\perp a}\varepsilon_{bzc}+
\bbox{r}_{\perp b}\varepsilon_{azc})\bbox{r}_{\perp c}
\eeq
is a symmetric pseudotensor.
The tensor
$\mathsf{J}^{z\perp}_k$ is traceless by definition. Its symmetric
part has the following structure:
\beq\label{zperp}
\underline{\mathsf{J}}^{z\perp}_{kaz}=\underline{\mathsf{J}}^{z\perp\ast}_{kaz} =
\underline{\mathsf{K}}^{z\perp}_{kz\perp}(z,r_{\perp})\underline{\mathsf{P}}^{z\perp}_{az}+
\underline{\mathsf{J}}^{z\perp}_{kz\perp}(z,r_{\perp})\underline{\mathsf{S}}^{z\perp}_{az}
\eeq
for $a=x,y$, where non-vanishing components of the normalized and
symmetrized outer products $\bbox{z}\otimes\bbox{r}_{\perp}$ and $\bbox{z}\otimes (\bbox{z}\times\bbox{r}_{\perp})$ are:
\beq\label{trlzp}
\underline{\mathsf{S}}^{z\perp}_{az}=
\frac{\bbox{r}_{\perp a}}{r_\perp}\frac{z}{|z|},
\eeq
and
\beq\label{pstrlzp}
\underline{\mathsf{P}}^{z\perp}_{az}=\frac{\varepsilon_{azc}\bbox{r}_{\perp c}}{r_{\perp}}.
\eeq
Similarly,
\beq\label{zperpbr}
\underline{\vec{\breve{\mathsf{J}}}}^{z\perp}_{az}=
\underline{\vec{\breve{\mathsf{K}}}}^{z\perp}_{z\perp}(z,r_{\perp})\underline{\mathsf{P}}^{z\perp}_{az}+
\underline{\vec{\breve{\mathsf{J}}}}^{z\perp}_{z\perp}(z,r_{\perp})\underline{\mathsf{S}}^{z\perp}_{az}.
\eeq
All components of the spin-current tensor in the {\particle} channel
are real whereas those in the {\pairing} channel are complex.

\subsubsection{Axial symmetry -- summary}\label{sec8e3}

In the case of the axial O$^{z\perp}$(2) symmetry, the position
vector $\bbox{r}$ can be  decomposed (\ref{rperp}) into two vectors
$\bbox{z}$ and $\bbox{r}_{\perp}$ having  different
transformation properties under rotations and mirror rotations about
the $z$-axis. Vector $\bbox{r}_{\perp}$ is the SO$^{\perp}$(2) vector, whereas $\bbox{z}$ is the O$^{z\perp}$(2) pseudo-invariant. There are two
O$^{z\perp}$(2) scalars: $z^2$ and $r_{\perp}^2$, and all the local
scalar densities are functions thereof. In this study, we are not
concerned with the question whether the
densities are analytical functions of the invariants. Therefore, it does
not matter whether the argument of densities is $r_{\perp}^2$ or just $r_{\perp}$.

The local vector densities are linear
combinations of vectors $\bbox{z}$ and $\bbox{r}_{\perp}$ with scalar
coefficients. The pseudovector densities are proportional to vector
product $\bbox{z}\times\bbox{r}_{\perp}$ and thus have the
azimuthal direction. While it is not possible to build pseudoscalar
densities from elementary vectors $\bbox{z}$ and
$\bbox{r}_{\perp}$, pseudotensor densities can be constructed. The symmetric pseudotensor
densities are linear combinations of the symmetrized outer products $\underline{\bbox{r}_{\perp}\otimes (\bbox{z}\times\bbox{r}_{\perp})}$
and $\underline{\bbox{z}\otimes (\bbox{z}\times\bbox{r}_{\perp})}$. The symmetry properties of  local axially symmetric
(O$^{z\perp}$(2)-invariant) densities are listed in Tables \ref{tab2a} and \ref{tab2b}.

If the mirror symmetry is broken,
the two SO$^{\perp}$(2) invariants are $z$ and $r_{\perp}$. The
scalar and pseudoscalar densities  are now  functions
thereof.  The vector and pseudovector densities now have  non-vanishing components
along all the three vectors, $\bbox{e}_{\perp}$, $\bbox{e}_{\phi}$,
and $\bbox{e}_z$. The traceless symmetric tensor densities are linear combinations of  the pseudotensors $\underline{\mathsf{P}}^{\perp}$
and $\underline{\mathsf{P}}^{z\perp}$, and  tensors $\underline{\mathsf{S}}^{\perp}$  and $\underline{\mathsf{S}}^{z\perp}$.
Properties of the SO$^{\perp}$(2)-invariant
local densities are listed in Tables \ref{tab3a} and \ref{tab3b}.

\newcommand{\fR} {f_R(z,r_{\perp})}
\newcommand{\fI} {f_I(z,r_{\perp})}
\newcommand{\fC} {f_C(z,r_{\perp})}
\newcommand{\epefez}{\bbox{e}_{\perp},\bbox{e}_{\phi},\bbox{e}_z}
\newcommand{\SpSz}{\underline{\mathsf{S}}^{\perp},\underline{\mathsf{S}}^{z\perp}}
\newcommand{\PpPz}{\underline{\mathsf{P}}^{\perp},\underline{\mathsf{P}}^{z\perp}}
\newcommand{\rp}{\bbox{r}_{\perp}}
\newcommand{\z}{\bbox{z}}
\newcommand{\zrp}{\bbox{z}\times\bbox{r}_{\perp}}
\newcommand{\fRsq}{f_R(z^2,r_{\perp}^2)}
\newcommand{\fIsq}{f_I(z^2,r_{\perp}^2)}
\newcommand{\fCsq}{f_C(z^2,r_{\perp}^2)}

\begin{table*}
\renewcommand{\arraystretch}{0.7}
\caption[T3]{Properties of the local axial and mirror-symmetric
(O$^{z\perp}$(2)-invariant) particle-hole
densities, depending on the conserved (C) or broken (B) proton-neutron
(p-n), and time-reversal ($T$) symmetries. The $z$-simplex ($S_z$)
symmetry is conserved. Vector and pseudovector densities can be expanded in a basis of three vectors:
$\bbox{r}_{\perp}$, $\bbox{z}$ and $\bbox{z}\times\bbox{r}_{\perp}$. The symmetrized
outer products $\underline{\bbox{r}_{\perp}\otimes (\bbox{z}\times\bbox{r}_{\perp})}$
and $\underline{\bbox{z}\otimes (\bbox{z}\times\bbox{r}_{\perp})}$ form a basis for the
pseudotensor densities.
The expansion coefficients are real, imaginary, and complex functions
of the two O$^{z\perp}$(2) scalars: $z^2$ and $r_{\perp}^2$.
Generic real, imaginary, and complex coefficients are  denoted by
$f_R(z^2,r_{\perp}^2)$, $f_I(z^2,r_{\perp}^2)$, and
$f_C(z^2,r_{\perp}^2)$,  respectively.}
\label{tab2a}
\begin{ruledtabular}
\begin{tabular}{l|lllll}
symmetry                                &      &
\multicolumn{4}{c}{conserved (C)
or broken (B)}                                                                   \\
\hline
$S_z$               &                    &   C    &     C  &     C  &     C       \\
p-n                       &              &   B    &     B  &     C  &     C       \\
$T$                         &            &   B    &     C  &     B  &     C       \\
\hline
     & basis &\multicolumn{4}{c} {coefficients}  \\
\hline
$\rho_{0,3}$    &$ 1$                        &$  \fRsq   $&$\fRsq   $&$  \fRsq   $&$  \fRsq   $     \\
$\rho_{1}$        & $ 1 $                    &$  \fRsq   $&$\fRsq   $&$  0   $&$  0   $     \\
$\rho_{2}$         & $ 1 $                   &$  \fRsq   $&$0   $&$  0   $&$  0   $     \\
$\tau_{0,3}$        & $ 1 $                  &$  \fRsq   $&$\fRsq   $&$  \fRsq   $&$  \fRsq   $     \\
$\tau_{1}$             & $ 1 $              &$  \fRsq   $&$\fRsq   $&$  0   $&$  0   $     \\
$\tau_{2}$           & $ 1 $                 &$  \fRsq  $&$0   $&$  0   $&$  0   $     \\
${J}_{0,3}$           & $ 0 $                  &$  0   $&$  0   $&$  0   $&$  0   $     \\
${J}_{1}$               & $ 0 $              &$  0   $&$  0   $&$  0   $&$  0   $     \\
${J}_{2}$               & $ 0 $              &$  0   $&$  0   $&$  0   $&$  0   $     \\
$\bbox{s}_{0,3}$   &$\zrp $               &$\fRsq   $&$0 $&$\fRsq  $   &$  0   $     \\
$\bbox{s}_{1}$   &$\zrp $                       &$\fRsq $&$0 $&$0  $&$  0   $     \\
$\bbox{s}_{2}$    &$\zrp $                      &$\fRsq $&$\fRsq $   &$  0   $&$0$     \\
${\bbox{T}}_{0,3}$   &$\zrp $                     &$  \fRsq   $&$ 0 $&$  \fRsq   $&$  0   $     \\
${\bbox{T}}_{1}$    &$\zrp $                    &$  \fRsq   $&$ 0 $&$   0 $&$  0   $     \\
${\bbox{T}}_{2}$    &$\zrp $                    &$  \fRsq   $&$\fRsq $&$  0   $&$  0   $     \\
${\bbox{F}}_{0,3}$    &$\zrp $                    &$  \fRsq   $&$ 0 $&$  \fRsq   $&$  0   $     \\
${\bbox{F}}_{1}$   &$\zrp $                     &$  \fRsq   $&$ 0 $&$  0   $&$  0   $     \\
${\bbox{F}}_{2}$   &$\zrp $                     &$  \fRsq   $&$\fRsq $&$  0   $&$  0   $     \\
${\bbox{j}}_{0,3}$    &$\rp ,\z $                   &$  \fRsq   $&$ 0 $&$  \fRsq   $&$  0   $     \\
${\bbox{j}}_{1}$  &$\rp ,\z $                      &$  \fRsq   $&$ 0 $&$  0   $&$  0   $     \\
${\bbox{j}}_{2}$      &$\rp ,\z $                  &$  \fRsq   $&$\fRsq $&$  0   $&$  0   $     \\
${\bbox{J}}_{0,3}$   &$\rp ,\z $                     &$  \fRsq   $&$\fRsq $&$  \fRsq   $&$  \fRsq   $     \\
${\bbox{J}}_{1}$   &$\rp ,\z $                     &$  \fRsq   $&$\fRsq $&$ 0   $&$  0   $     \\
${\bbox{J}}_{2}$   &$\rp ,\z $                     &$  \fRsq   $&$0$&$  0   $&$  0   $     \\
$\underline{{\mathsf J}}_{0,3}$  & $\underline{\bbox{r}_{\perp}\otimes (\bbox{z}\times\bbox{r}_{\perp})}$,
 $\underline{\bbox{z}\otimes (\bbox{z}\times\bbox{r}_{\perp})}$        &$  \fRsq   $&$  \fRsq   $&$  \fRsq   $&$  \fRsq   $     \\
$\underline{{\mathsf J}}_{1}$  & $\underline{\bbox{r}_{\perp}\otimes (\bbox{z}\times\bbox{r}_{\perp})}$,
 $\underline{\bbox{z}\otimes (\bbox{z}\times\bbox{r}_{\perp})}$         &$  \fRsq   $&$  \fRsq   $&$  0   $&$  0   $     \\
$\underline{{\mathsf J}}_{2}$  & $\underline{\bbox{r}_{\perp}\otimes (\bbox{z}\times\bbox{r}_{\perp})}$,
 $\underline{\bbox{z}\otimes (\bbox{z}\times\bbox{r}_{\perp})}$         &$  \fRsq   $&$  0   $&$  0   $&$  0   $     \\
\end{tabular}
\end{ruledtabular}
\renewcommand{\arraystretch}{1}
\end{table*}

\begin{table*}
\caption[T3]{Similar to Table \protect\ref{tab2a} except for the particle-particle
densities.}
\label{tab2b}
\begin{ruledtabular}
\begin{tabular}{l|lllll}
symmetry                                &      &
\multicolumn{4}{c}{conserved (C)
or broken (B)}                                                                   \\
\hline
$S_z$               &                    &   C    &     C  &     C  &     C       \\
p-n                       &              &   B    &     B  &     C  &     C       \\
$T$                         &            &   B    &     C  &     B  &     C       \\
\hline
     & basis &\multicolumn{4}{c} {coefficients}  \\
\hline
$\breve{\rho}_1$    & $ 1$                  &$  \fCsq   $&$\fRsq   $&$  \fCsq   $&$  \fRsq   $     \\
$\breve{\rho}_2$  & $ 1$                         &$  \fCsq   $&$\fIsq   $&$  \fCsq   $&$  \fIsq   $     \\
$\breve{\rho}_3$   & $ 1$                        &$  \fCsq   $&$\fRsq   $&$  0   $&$  0   $     \\
$\breve{\tau}_1$  & $ 1$                         &$  \fCsq   $&$\fRsq   $&$  \fCsq   $&$  \fRsq   $     \\
$\breve{\tau}_2$   & $ 1$                        &$  \fCsq   $&$\fIsq   $&$  \fCsq   $&$  \fIsq   $     \\
$\breve{\tau}_3$  & $ 1$                         &$  \fCsq   $&$\fRsq   $&$  0   $&$  0   $     \\
$\breve{J}_1$   & $ 0$                       &$  0   $&$  0  $&$  0   $&$  0   $     \\
$\breve{J}_2$   & $ 0$                       &$  0   $&$  0   $&$  0   $&$ 0    $     \\
$\breve{J}_3$   & $ 0$                       &$  0   $&$  0   $&$  0   $&$  0   $     \\
$\breve{\bbox{s}}_0$   &$\zrp $                    &$  \fCsq   $&$\fIsq $&$  0   $&$  0   $     \\
$\breve{\bbox{T}}_0$  &$\zrp $                     &$  \fCsq   $&$\fIsq $&$  0   $&$  0   $     \\
$\breve{\bbox{F}}_0$   &$\zrp $                    &$  \fCsq   $&$\fIsq $&$  0   $&$  0   $     \\
$\breve{\bbox{j}}_0$     &$\rp ,\z $             &$  \fCsq   $&$\fIsq$&$  0   $&$  0   $     \\
$\breve{\bbox{J}}_1$    &$\rp ,\z $                    &$  \fCsq   $&$\fRsq$&$  \fCsq   $&$  \fRsq   $     \\
$\breve{\bbox{J}}_2$    &$\rp ,\z $                    &$  \fCsq   $&$\fIsq$&$  \fCsq   $&$  \fIsq   $     \\
$\breve{\bbox{J}}_3$    &$\rp ,\z $                   &$  \fCsq   $&$\fRsq$&$  0   $&$  0   $     \\
$\underline{\breve{{\mathsf J}}}_1$   & $\underline{\bbox{r}_{\perp}\otimes (\bbox{z}\times\bbox{r}_{\perp})}$,
 $\underline{\bbox{z}\otimes (\bbox{z}\times\bbox{r}_{\perp})}$  &$  \fCsq   $&$  \fRsq  $&$  \fCsq   $&$  \fRsq   $     \\
$\underline{\breve{{\mathsf J}}}_2$  & $\underline{\bbox{r}_{\perp}\otimes (\bbox{z}\times\bbox{r}_{\perp})}$,
 $\underline{\bbox{z}\otimes (\bbox{z}\times\bbox{r}_{\perp})}$  &$  \fCsq   $&$  \fIsq   $&$  \fCsq   $&$  \fIsq   $     \\
$\underline{\breve{{\mathsf J}}}_3$  & $\underline{\bbox{r}_{\perp}\otimes (\bbox{z}\times\bbox{r}_{\perp})}$,
 $\underline{\bbox{z}\otimes (\bbox{z}\times\bbox{r}_{\perp})}$  &$  \fCsq   $&$  \fRsq   $&$  0   $&$  0   $     \\
\end{tabular}
\end{ruledtabular}
\end{table*}

\begin{table*}
\renewcommand{\arraystretch}{0.7}
\caption[T3]{Properties of local axially symmetric (SO$^{\perp}$(2)-invariant)
particle-hole densities, depending on
the conserved (C) or broken (B) p-n and
time-reversal symmetries. The $z$-simplex ($S_z$) symmetry is
broken. The vector, pseudovector, or pseudotensor densities can be expanded
in the vector
($\bbox{e}_{\perp},\bbox{e}_{\phi},\bbox{e}_z$) or tensor
($\underline{\mathsf{P}}^{\perp},\underline{\mathsf{P}}^{z\perp},\underline{\mathsf{S}}^{\perp},\underline{\mathsf{S}}^{z\perp}$)
bases. The expansion coefficients are real, imaginary, and complex
functions of cylindrical coordinates $r_{\perp}$ and $z$. Generic
real, imaginary, and complex coefficients are denoted by
$f_R(z,r_{\perp})$, $f_I(z,r_{\perp})$, or $f_C(z,r_{\perp})$,
respectively.}
\label{tab3a}
\begin{ruledtabular}
\begin{tabular}{l|lllll}
symmetry                                &      &
\multicolumn{4}{c}{conserved (C)
or broken (B)}                                                                   \\
\hline
$S_z$               &                    &   B    &     B  &     B  &     B       \\
p-n                       &              &   B    &     B  &     C  &     C       \\
$T$                         &            &   B    &     C  &     B  &     C       \\
\hline
     & basis &\multicolumn{4}{c} {coefficients}  \\
\hline
$\rho_{0,3}$    &$ 1$                        &$  \fR   $&$\fR   $&$  \fR   $&$  \fR   $     \\
$\rho_{1}$        & $ 1 $                    &$  \fR   $&$\fR   $&$  0   $&$  0   $     \\
$\rho_{2}$         & $ 1 $                   &$  \fR   $&$0   $&$  0   $&$  0   $     \\
$\tau_{0,3}$        & $ 1 $                  &$  \fR   $&$\fR   $&$  \fR   $&$  \fR   $     \\
$\tau_{1}$             & $ 1 $              &$  \fR   $&$\fR   $&$  0   $&$  0   $     \\
$\tau_{2}$           & $ 1 $                 &$  \fR   $&$0   $&$  0   $&$  0   $     \\
${J}_{0,3}$           & $ 1 $                  &$  \fR   $&$  \fR   $&$  \fR   $&$  \fR   $     \\
${J}_{1}$               & $ 1 $              &$  \fR   $&$  \fR   $&$  0   $&$  0   $     \\
${J}_{2}$               & $ 1 $              &$  \fR   $&$  0   $&$  0   $&$  0   $     \\
$\bbox{s}_{0,3}$   &$\epefez $               &$\fR   $&$0 $&$\fR  $   &$  0   $     \\
$\bbox{s}_{1}$   &$\epefez $                       &$\fR $&$0 $&$0  $&$  0   $     \\
$\bbox{s}_{2}$    &$\epefez $                      &$\fR $&$\fR $   &$  0   $&$0$     \\
${\bbox{T}}_{0,3}$   &$\epefez $                     &$  \fR   $&$ 0 $&$  \fR   $&$  0   $     \\
${\bbox{T}}_{1}$    &$\epefez $                    &$  \fR   $&$ 0 $&$   0 $&$  0   $     \\
${\bbox{T}}_{2}$    &$\epefez $                    &$  \fR   $&$\fR $&$  0   $&$  0   $     \\
${\bbox{F}}_{0,3}$    &$\epefez $                    &$  \fR   $&$ 0 $&$  \fR   $&$  0   $     \\
${\bbox{F}}_{1}$   &$\epefez $                     &$  \fR   $&$ 0 $&$  0   $&$  0   $     \\
${\bbox{F}}_{2}$   &$\epefez $                     &$  \fR   $&$\fR $&$  0   $&$  0   $     \\
${\bbox{j}}_{0,3}$    &$\epefez $                   &$  \fR   $&$ 0 $&$  \fR   $&$  0   $     \\
${\bbox{j}}_{1}$  &$\epefez $                      &$  \fR   $&$ 0 $&$  0   $&$  0   $     \\
${\bbox{j}}_{2}$      &$\epefez $                  &$  \fR   $&$\fR$&$  0   $&$  0   $     \\
${\bbox{J}}_{0,3}$   &$\epefez $                     &$  \fR   $&$\fR$&$  \fR   $&$  \fR   $     \\
${\bbox{J}}_{1}$   &$\epefez $                     &$  \fR   $&$\fR$&$ 0   $&$  0   $     \\
${\bbox{J}}_{2}$   &$\epefez $                     &$  \fR   $&$0$&$  0   $&$  0   $     \\
$\underline{{\mathsf J}}_{0,3}$  & $\PpPz$,$\SpSz $       &$  \fR   $&$  \fR   $&$  \fR   $&$  \fR   $     \\
$\underline{{\mathsf J}}_{1}$  & $\PpPz$,$\SpSz $        &$  \fR   $&$  \fR   $&$  0   $&$  0   $     \\
$\underline{{\mathsf J}}_{2}$  & $\PpPz$,$\SpSz $        &$  \fR   $&$  0   $&$  0   $&$  0   $     \\
\end{tabular}
\end{ruledtabular}
\renewcommand{\arraystretch}{1}
\end{table*}

\begin{table*}
\caption[T3]{Similar to  Table \protect\ref{tab3a} except for the particle-particle
densities.}
\label{tab3b}
\begin{ruledtabular}
\begin{tabular}{l|lllll}
symmetry                                &      &
\multicolumn{4}{c}{conserved (C)
or broken (B)}                                                                   \\
\hline
$S_z$               &                    &   B    &     B  &     B  &     B       \\
p-n                       &              &   B    &     B  &     C  &     C       \\
$T$                         &            &   B    &     C  &     B  &     C       \\
\hline
     & basis &\multicolumn{4}{c} {coefficients}  \\
\hline
$\breve{\rho}_1$    & $ 1$                  &$  \fC   $&$\fR   $&$  \fC   $&$  \fR   $     \\
$\breve{\rho}_2$  & $ 1$                         &$  \fC   $&$\fI   $&$  \fC   $&$  \fI   $     \\
$\breve{\rho}_3$   & $ 1$                        &$  \fC   $&$\fR   $&$  0   $&$  0   $     \\
$\breve{\tau}_1$  & $ 1$                         &$  \fC   $&$\fR   $&$  \fC   $&$  \fR   $     \\
$\breve{\tau}_2$   & $ 1$                        &$  \fC   $&$\fI   $&$  \fC   $&$  \fI   $     \\
$\breve{\tau}_3$  & $ 1$                         &$  \fC   $&$\fR   $&$  0   $&$  0   $     \\
$\breve{J}_1$   & $ 1$                       &$  \fC   $&$  \fR  $&$  \fC   $&$  \fR   $     \\
$\breve{J}_2$   & $ 1$                       &$  \fC   $&$  \fI   $&$  \fC   $&$  \fI   $     \\
$\breve{J}_3$   & $ 1$                       &$  \fC   $&$  \fR   $&$  0   $&$  0   $     \\
$\breve{\bbox{s}}_0$   &$\epefez $                    &$  \fC   $&$\fI $&$  0   $&$  0   $     \\
$\breve{\bbox{T}}_0$  &$\epefez $                     &$  \fC   $&$\fI $&$  0   $&$  0   $     \\
$\breve{\bbox{F}}_0$   &$\epefez $                    &$  \fC   $&$\fI $&$  0   $&$  0   $     \\
$\breve{\bbox{j}}_0$     &$\epefez $                  &$  \fC   $&$\fI$&$  0   $&$  0   $     \\
$\breve{\bbox{J}}_1$    &$\epefez $                   &$  \fC   $&$\fR$&$  \fC   $&$  \fR   $     \\
$\breve{\bbox{J}}_2$    &$\epefez $                   &$  \fC   $&$\fI$&$  \fC   $&$  \fI   $     \\
$\breve{\bbox{J}}_3$    &$\epefez $                   &$  \fC   $&$\fR$&$  0   $&$  0   $     \\
$\underline{\breve{{\mathsf J}}}_1$ &$\PpPz$,$\SpSz $ &$  \fC   $&$  \fR   $&$  \fC   $&$  \fR   $     \\
$\underline{\breve{{\mathsf J}}}_2$&$\PpPz$,$\SpSz $ &$  \fC   $&$  \fI   $&$  \fC   $&$  \fI   $     \\
$\underline{\breve{{\mathsf J}}}_3$&$\PpPz$,$\SpSz $ &$  \fC   $&$  \fR   $&$  0   $&$  0   $     \\
\end{tabular}
\end{ruledtabular}
\end{table*}

\subsection{Symmetry D$_{2\mathrm{h}}$}\label{sec8f}

The identity, inversion, three signatures, three simplexes, and their
negative partners form the symmetry group
D$_{2\mathrm{h}}^\mathrm{D}$. The Cartesian rotation matrices
of Eq.~(\ref{cartrot}) for the identity (labeled by $u$) and
signature operations are all diagonal:
\begin{equation}
\mathsf{a}^a = \left(\ba{ccc}\mathsf{a}^a_x & 0 & 0 \\
                                           0 & \mathsf{a}^a_y & 0 \\
                                           0 & 0 & \mathsf{a}^a_z \ea\right)
\label{asign}
\end{equation}
for $a= u,x,y,z$. They can be written explicitly as:
\bnll{auxyz}
\mathsf{a}^u=\mathsf{a}(000)&=& \left(\ba{ccc}1 & 0 & 0 \\
                                           0 & 1 & 0 \\
                                           0 & 0 & 1  \ea\right), \label{auxyz1}\\
\mathsf{a}^x=\mathsf{a}(0\pi\pi )&=& \left(\ba{ccc}1 & 0 & 0 \\
                                           0 & -1 & 0 \\
                                           0 & 0 & -1  \ea\right), \label{auxyz2}\\
\mathsf{a}^y=\mathsf{a}(0\pi 0)&=& \left(\ba{ccc}-1 & 0 & 0 \\
                                           0 & 1 & 0 \\
                                           0 & 0 & -1  \ea\right) ,\label{auxyz3}\\
\mathsf{a}^z=\mathsf{a}(\pi 00)&=& \left(\ba{ccc}-1 & 0 & 0 \\
                                           0 & -1 & 0 \\
                                           0 & 0 & 1  \ea\right). \label{auxyz4}
\enll
Let us suppose that the generalized density matrix is invariant under
the generalized transformation matrix $\hat{\breve{\mathcal U}}^a$,
($a=u,x,y,z$), belonging to the D$_{2\mathrm{h}}^\mathrm{D}$ of
transformations.
According to Eqs.~(\ref{rinvnon}), the
transformation rules for nonlocal densities are:
\bnll{dinvnon}
\rho_k(\bbox{r},\bbox{r}')&=&\rho_k(\varsigma\mathsf{a}^a\bbox{r},\varsigma\mathsf{a}^a\bbox{r}'),\label{dinvnon1}\\
\breve{\rho}_k(\bbox{r},\bbox{r}')&=
&\breve{\rho}_k(\varsigma\mathsf{a}^a\bbox{r},\varsigma\mathsf{a}^a\bbox{r}'),\label{dnvnon2}\\
\bbox{s}_{kb}(\bbox{r},\bbox{r}')&=&\mathsf{a}^a_b\bbox{s}_{kb}(\varsigma\mathsf{a}^a\bbox{r},\varsigma\mathsf{a}^a\bbox{r}'),
\label{dinvnon3}\\
\breve{\bbox{s}}_{kb}(\bbox{r},\bbox{r}')&=
&\mathsf{a}^a_b\breve{\bbox{s}}_{kb}(\varsigma\mathsf{a}^a\bbox{r},\varsigma\mathsf{a}^a\bbox{r}')
\label{dinvnon4}
\enll
for any $a=u,x,y,z$ and $b=x,y,z$; $\varsigma =+1$ for identity and
signature operations and $\varsigma =-1$ for inversion and simplex operations.

The transformations (\ref{dinvnon})
constitute symmetry conditions for densities under changes of signs
of their arguments; hence, they  relate the values of densities
between  different regions of space. For the
{\particle} densities, this problem has been extensively  discussed in Refs.~\cite{[Dob00a],[Dob00b]}. In the following, we extend the previous discussion to the pairing channel and
provide  general expressions for the {\pairing} densities.

The {\particle} local density matrices fulfill the following symmetry conditions:
\bnll{dinvloc}
\rho_k(\bbox{r})&=&\rho^{\ast}_k(\bbox{r})=\rho_k(\varsigma\mathsf{a}^a\bbox{r}),\label{dinvloc1}\\
\bbox{s}_{kb}(\bbox{r})&=&\bbox{s}^{\ast}_{kb}(\bbox{r})
=\mathsf{a}^a_b\bbox{s}_{kb}(\varsigma\mathsf{a}^a\bbox{r}) \label{dinvloc2}
\enll
for $k=0,1,2,3$, $a=u,x,y,z$ and $b=x,y,z$. The analogous expression
for the {\pairing} densities are:
\bnll{dinvlocp}
\vec{\breve{\rho}}(\bbox{r})&=&\vec{\breve{\rho}}(\varsigma\mathsf{a}^a\bbox{r}),\label{dinvlocp1}\\
\bbox{s}_{0b}(\bbox{r})&=&
\mathsf{a}^a_b\bbox{s}_{0b}(\varsigma\mathsf{a}^a\bbox{r}). \label{dinvlocp2}
\enll

The differential local densities in both channels should now be
classified according to irreducible representations of the point group
D$_{\mathrm{2h}}$  \cite{[Dob00a]}. Therefore,
the vector notation used so far  is no longer useful. Instead, in this section we rely on
definitions (\ref{TF}), where the Cartesian components are explicitly
shown.

The {\particle} isoscalar and isovector kinetic densities (\ref{tau1})
transform according to:
\begin{equation}
\tau_{kbc}(\bbox{r})=\tau^{\ast}_{kbc}(\bbox{r}) =\mathsf{a}^a_b\mathsf{a}^a_c
\tau_{kbc}(\varsigma\mathsf{a}^a\bbox{r}).
\label{tau2}
\end{equation}
The complex isovector component of {\pairing} kinetic densities (\ref{tau3a})  obeys
analogous symmetry conditions:
\bn
\vec{\breve{\tau}}_{bc}(\bbox{r})
&=&\mathsf{a}^a_b\mathsf{a}^a_c
\vec{\breve{\tau}}_{bc}(\varsigma\mathsf{a}^a\bbox{r}).
 \label{tau3}
\en
The symmetries of the kinetic scalar densities (\ref{scalar-k})
can be obtained from Eqs.~(\ref{tau4a}) and (\ref{tau4b}).

The spin-kinetic and tensor-kinetic densities (\ref{T1a}) and (\ref{T1b})
fulfill the symmetry conditions:
\bnll{T2}
\bbox{T}_{kbcd}(\bbox{r})&=&\bbox{T}^{\ast}_{kbcd}(\bbox{r}) =
\mathsf{a}^a_b\mathsf{a}^a_c\mathsf{a}^a_d\bbox{T}_{kbcd}(\varsigma\mathsf{a}^a\bbox{r}), \label{T2a}\\
\breve{\bbox{T}}_{0bcd}(\bbox{r})&=&\mathsf{a}^a_b\mathsf{a}^a_c\mathsf{a}^a_d
\breve{\bbox{T}}_{0bcd}(\varsigma\mathsf{a}^a\bbox{r}).\label{T2b}
\enll
The transformation properties of the spin-kinetic densities (\ref{vector-T}) and  tensor-kinetic densities
(\ref{vector-F}) are obtained by contractions defined by
Eqs.~(\ref{TF1})--(\ref{TF4}).

Finally, the symmetry conditions of the current and the spin-current
densities in both
channels are:
\bnll{jJsym}
\bbox{j}_{kb}(\bbox{r})&=&\bbox{j}^{\ast}_{kb}(\bbox{r})=
\varsigma\mathsf{a}^a_b\bbox{j}_{kb}(\varsigma\mathsf{a}^a\bbox{r}),\label{jJsym1}\\
\breve{\bbox{j}}_{0b}(\bbox{r})&=&
\varsigma\mathsf{a}^a_b\breve{\bbox{j}}_{0b}(\varsigma\mathsf{a}^a\bbox{r},
\label{jJsym2}\\
\mathsf{J}_{kbc}(\bbox{r})&=&\mathsf{J}^{\ast}_{kbc}(\bbox{r})=
\varsigma\mathsf{a}^a_b\mathsf{a}^a_c\mathsf{J}_{kbc}(\varsigma\mathsf{a}^a\bbox{r}),\label{jJsym3}\\
\vec{\breve{\mathsf{J}}}_{bc}(\bbox{r})&=&\varsigma\mathsf{a}^a_b\mathsf{a}^a_c
\vec{\breve{\mathsf{J}}}_{bc}(\varsigma\mathsf{a}^a\bbox{r}).\label{jJsym4}
\enll

\section{Symmetries of multi-reference transition densities}\label{sec14}
In analogy to Eqs.~(\ref{rho}) and (\ref{rhobreve}), the transition {\particle} and {\pairing} density
matrices are defined, respectively, as
\bn\label{rhot}
\hat{\rho}^{(t)}(\bbox{r}st,\bbox{r}'s't')
&=&\langle \Psi_2 |a_{\bbox{r}'s't'}^{+}a_{\bbox{r}st}|\Psi_1 \rangle , \\
\label{rhobrevet}
\hat{\breve{\rho}}\tra(\bbox{r}st,\bbox{r}'s't')
&=& 4s't'\langle \Psi_2 |a_{\bbox{r}'-s'-t'}a_{\bbox{r}st}|\Psi_1 \rangle ,
\en
where $|\Psi_1\rangle$ and $|\Psi_2\rangle$ are two different
independent-quasiparticle states. The corresponding spin-isospin
scalar and vector transition densities $\rho_k\tra
(\bbox{r},\bbox{r}')$ and $\bbox{s}_k\tra (\bbox{r},\bbox{r}')$, and
$\breve{\rho}_k\tra (\bbox{r},\bbox{r}')$ and $\breve{\bbox{s}}_k\tra
(\bbox{r},\bbox{r}')$ with $k=0,\dots\ ,3$ are defined by the
relations analogous to Eqs.~(\ref{izo}). The local transition
densities are defined in an identical way as the local densities
Eqs.~(\ref{scalar-p})-(\ref{tensor-J}) and are  denoted with the same respective symbols
but with the superscript $(t)$.

What are the differences in the symmetry properties of the transition
density matrices from those of the density matrices
discussed above? First of all, the {\particle} transition density matrix is
not Hermitian:
\bn\label{notherm}
\hat{\rho}^{(t)+}(\bbox{r}st,\bbox{r}'s't')&=&\langle \Psi_1 |a_{\bbox{r}'s't'}^{+}a_{\bbox{r}st}|\Psi_2\rangle
\nonumber \\
&\neq &\langle \Psi_2 |a_{\bbox{r}'s't'}^{+}a_{\bbox{r}st}|\Psi_1\rangle.
\en
Consequently, Eqs.~(\ref{hermicon2}) for the  {\particle} nonlocal
transition densities are not fulfilled. On the other hand, the
antisymmetry property of the {\pairing} density matrix is preserved:
\beq\label{antys}
\hat{\breve{\rho}}\tra(\bbox{r}st,\bbox{r}'s't')=-16ss'tt'\hat{\breve{\rho}}\tra(\bbox{r}'-s'-t',\bbox{r}-s-t).
\eeq
Hence, the {\pairing} nonlocal transition densities obey relations (\ref{symepai}).

The transformation rules for the transition matrices under the single-particle unitary (and antiunitary) transformations $U$ ($U_K$)
follow  the transformation rules
(\ref{tra}) for the creation and annihilation operators
and are given by Eqs.~(\ref{matmul}), namely:
\bnll{matmultr}
{}\hat{\rho}^{(t)U} = \hat{u}\bullet\hat{\rho}\tra\bullet\hat{u}^+, \label{matmultr1} \\
\hat{\breve{\rho}}^{(t)U} = \hat{u}\bullet\hat{\breve{\rho}}\tra\bullet\hat{\breve{u}}^+.
\label{matmultr2}
\enll
Therefore, the
discussion of density matrix symmetries  presented in Sec.~\ref{sec12}
applies to  transition
densities with
the only  difference being that
the {\particle} transition densities are, in general, complex, unless
the time-reversal invariance introduces some restrictions.  The
time-reversal-invariant non-local {\particle} transition densities
obey the following relations:
\bnll{trevtra}
&&\rho_k\tra (\bbox{r},\bbox{r}')=\rho_k^{(t)\ast}(\bbox{r},\bbox{r}'), \label{trevtra1}\\
&&\bbox{s}_k\tra (\bbox{r},\bbox{r}')=-\bbox{s}_k^{(t)\ast} (\bbox{r},\bbox{r}') \label{trevtra2}\\
&&\mathrm{for}\quad k=0,\ 1,\ 3,\nonumber \\
\mathrm{and}&& \nonumber \\
&&\rho_2\tra (\bbox{r},\bbox{r}')=-\rho_2^{(t)\ast}(\bbox{r},\bbox{r}'), \label{trevtra3}\\
&&\bbox{s}_2\tra (\bbox{r},\bbox{r}')=\bbox{s}_2^{(t)\ast} (\bbox{r},\bbox{r}'), \label{trevtra4}
\enll
which means that some transition densities are real while others are purely
imaginary. Symmetry properties  of the local {\particle}
transition densities  are
catalogued in Tables \ref{tab1tra}, \ref{tab2tra}, and \ref{tab3tra}. For
the local {\pairing} transition densities, the results of Tables \ref{tab1b},
\ref{tab2b}, and \ref{tab3b} apply.
\begin{table*}
\renewcommand{\arraystretch}{0.75}
\caption[T3]{Properties of local particle-hole
rotationally symmetric (SO(3)-invariant) transition densities, depending on the
conserved (C) or broken (B) space-inversion ($P$), proton-neutron
(p-n), or time-reversal ($T$) symmetries. Generic real, imaginary, or
complex functions of the radial variable $r$ are denoted by $f_R(r)$,
$f_I(r)$, or $f_C(r)$, respectively.}
\label{tab1tra}
\begin{ruledtabular}
\begin{tabular}{l|llllllll}
symmetry                                &
\multicolumn{8}{c}{conserved (C)
or broken (B)}                                                                   \\
\hline
$P$                                     &   B    &     B  &     B  &     B &  C  &  C  &  C  &  C      \\
p-n                                     &   B    &     B  &     C  &     C     &   B    &     B  &     C  &     C  \\
$T$                                     &   B    &     C  &     B  &     C     &   B    &     C  &     B  &     C    \\
\hline
$\rho_{0,3}\tra$                            &$f_C(r)$&$f_R(r)$&$f_C(r)$&$f_R(r)$&$f_C(r)$&$f_R(r)$&$f_C(r)$&$f_R(r)$       \\
$\rho_1\tra$                            &$f_C(r)$&$f_R(r)$&$0$&$0$&$f_C(r)$&$ f_R(r)   $&$ 0 $&$  0   $    \\
$\rho_2\tra$                          &$f_C(r)$&$f_I(r)$ &$0$ &$0$ &$f_C(r)$&$  f_I(r)   $&$0$&$  0   $    \\
$\tau_{0,3}\tra$                            &$f_C(r)$&$f_R(r)$&$f_C(r)$&$f_R(r)$&$f_C(r)$&$f_R(r)$&$f_C(r)$&$f_R(r)$      \\
$\tau_{1}\tra$                            &$f_C(r)$&$f_R(r)$&$0$&$0$&$f_C(r)$&$f_R(r)$&$ 0   $&$  0   $       \\
$\tau_{2}\tra$                           &$f_C(r)$&$f_I(r)$&$0$&$0$&$f_C(r)$&$f_I(r)$&$  0   $&$  0   $       \\
${J}_{0,3}\tra$                             &$f_C(r)$&$f_R(r)$&$f_C(r)$&$f_R(r)$&$  0   $&$  0   $&$  0   $ &$0$      \\
${J}_{1}\tra$                             &$f_C(r)$&$f_R(r)$&$0$&$0$&$  0   $&$  0   $&$  0   $    &$0$   \\
$J_{2}\tra$                                &$f_C(r)$&$f_I(r)$&$0$ &$0$ &$0$&$  0   $&$  0   $   &$0$ \\
$\bbox{s}_{0,3}\tra$                        &$\fCr  $&$\fIr$&$\fCr$&$\fIr$&$  0   $&$  0   $&$  0   $    & $0$ \\
$\bbox{s}_{1}\tra$                        &$\fCr  $&$\fIr$&$0$&$0$&$  0   $&$  0   $&$  0   $     &$0$  \\
$\bbox{s}_{2}\tra$                        &$\fCr  $&$\fRr  $&$0$ & $0$&$  0   $&$  0   $&$  0   $     &$0$  \\
${\bbox{T}}_{0,3}\tra$                      &$\fCr  $&$\fIr$&$\fCr$&$\fIr$&$  0   $&$  0   $&$  0   $    &$0$   \\
${\bbox{T}}_{1}\tra$                      &$\fCr  $&$\fIr$&$0$&$0$&$  0   $&$  0   $&$  0   $     &$0$  \\
${\bbox{T}}_{2}\tra$                      &$\fCr  $&$\fRr  $&$0$ &$0$ &$  0   $&$  0   $&$  0   $     &$0$  \\
${\bbox{F}}_{0,3}\tra$                      &$\fCr  $&$\fIr$&$\fCr$&$\fIr$&$  0   $&$  0   $&$  0   $     &$0$  \\
${\bbox{F}}_{1}\tra$                      &$\fCr  $&$\fIr$&$0$&$0$&$  0   $&$  0   $&$  0   $     &$0$  \\
${\bbox{F}}_{2}\tra$                      &$\fCr  $&$\fRr$&$0$&$0$&$  0   $&$  0   $&$  0   $     &$0$  \\
${\bbox{j}}_{0,3}\tra$                      &$\fCr  $&$\fIr$&$\fCr$&$\fIr$&$\fCr  $&$\fIr$&$\fCr  $&$  \fIr   $       \\
${\bbox{j}}_{1}\tra$                      &$\fCr  $&$\fIr$&$0$&$0$&$\fCr  $&$\fIr$&$  0   $&$  0   $       \\
${\bbox{j}}_{2}\tra$                      &$\fCr  $&$\fRr  $&$0$&$0$&$\fCr  $&$\fRr$&$  0   $&$  0   $       \\
${\bbox{J}}_{0,3}\tra$                      &$\fCr  $&$\fRr$&$\fCr$&$\fRr$&$\fCr  $&$\fRr$&$\fCr  $&$\fRr  $       \\
${\bbox{J}}_{1}\tra$                      &$\fCr  $&$\fRr$&$0$&$0$&$\fCr  $&$\fRr$&$  0   $&$  0   $       \\
${\bbox{J}}_{2}\tra$                      &$\fCr  $&$\fIr  $&$0$&$0$&$\fCr  $&$ \fIr   $&$  0   $    &$0$  \\
$\underline{{\mathsf J}}_{0,3}\tra$         &$\fCS  $&$\fRS  $&$\fCS$&$\fRS$&$  0   $&$  0   $&$  0   $   &$0$   \\
$\underline{{\mathsf J}}_{1}\tra$         &$\fCS  $&$\fRS  $&$0$&$0$&$  0   $&$  0   $&$  0   $     &$0$  \\
$\underline{{\mathsf J}}_{2}\tra$         &$\fCS  $&$\fIS$&$0$&$0$&$  0   $&$  0   $&$  0   $     &$0$  \\
\end{tabular}
\end{ruledtabular}
\renewcommand{\arraystretch}{1}
\end{table*}
\begin{table*}
\renewcommand{\arraystretch}{0.6}
\caption[T3]{Properties of local axially and mirror symmetric
(O$^{z\perp}$(2)-invariant) particle-hole transition
densities depending on the conserved (C) or broken (B) proton-neutron
(p-n), or time-reversal ($T$) symmetries. The $z$-simplex ($S_z$)
symmetry is conserved. The vector and pseudovector densities take the
form of an expansion in a basis of some of the three vectors:
$\bbox{r}_{\perp}$, $\bbox{z}$ and $\bbox{z}\times\bbox{r}_{\perp}$. The pseudotensor densities
are linear combinations of the two symmetric pseudotensors: $\underline{\bbox{r}_{\perp}\otimes (\bbox{z}\times\bbox{r}_{\perp})}$
and $\underline{\bbox{z}\otimes (\bbox{z}\times\bbox{r}_{\perp})}$.
The expansion coefficients are real, imaginary, or complex functions
of the two O$^{z\perp}$(2) scalars, $z^2$ and $r_{\perp}^2$.
Generic real, imaginary, or complex coefficients are  denoted by
$f_R(z^2,r_{\perp}^2)$, $f_I(z^2,r_{\perp}^2)$, or
$f_C(z^2,r_{\perp}^2)$,  respectively.}
\label{tab2tra}
\begin{ruledtabular}
\begin{tabular}{l|lllll}
symmetry                                &      &
\multicolumn{4}{c}{conserved (C)
or broken (B)}                                                                   \\
\hline
$S_z$               &                    &   C    &     C  &     C  &     C       \\
p-n                       &              &   B    &     B  &     C  &     C       \\
$T$                         &            &   B    &     C  &     B  &     C       \\
\hline
     & basis &\multicolumn{4}{c} {coefficients}  \\
\hline
$\rho_{0,3}\tra$    &$ 1$                        &$  \fCsq   $&$\fRsq   $&$  \fCsq   $&$  \fRsq   $     \\
$\rho_{1}\tra$        & $ 1 $                    &$  \fCsq   $&$\fRsq   $&$  0   $&$  0   $     \\
$\rho_{2}\tra$         & $ 1 $                   &$  \fCsq   $&$\fIsq   $&$  0   $&$  0   $     \\
$\tau_{0,3}\tra$        & $ 1 $                  &$  \fCsq   $&$\fRsq   $&$  \fCsq   $&$  \fRsq   $     \\
$\tau_{1}\tra$             & $ 1 $              &$  \fCsq   $&$\fRsq   $&$  0   $&$  0   $     \\
$\tau_{2}\tra$           & $ 1 $                 &$  \fRsq  $&$\fIsq   $&$  0   $&$  0   $     \\
${J}_{0,3}\tra$           & $ 0 $                  &$  0   $&$  0   $&$  0   $&$  0   $     \\
${J}_{1}\tra$               & $ 0 $              &$  0   $&$  0   $&$  0   $&$  0   $     \\
${J}_{2}\tra$               & $ 0 $              &$  0   $&$  0   $&$  0   $&$  0   $     \\
$\bbox{s}_{0,3}\tra$   &$\zrp $               &$\fCsq   $&$\fIsq $&$\fCsq  $   &$  \fIsq   $     \\
$\bbox{s}_{1}\tra$   &$\zrp $                       &$\fCsq $&$\fIsq $&$0  $&$  0   $     \\
$\bbox{s}_{2}\tra$    &$\zrp $                      &$\fCsq $&$\fRsq $   &$  0   $&$0$     \\
${\bbox{T}}_{0,3}\tra$   &$\zrp $                     &$  \fCsq   $&$ \fIsq $&$  \fCsq   $&$  \fIsq   $     \\
${\bbox{T}}_{1}\tra$    &$\zrp $                    &$  \fCsq   $&$ \fIsq $&$   0 $&$  0   $     \\
${\bbox{T}}_{2}\tra$    &$\zrp $                    &$  \fCsq   $&$\fRsq $&$  0   $&$  0   $     \\
${\bbox{F}}_{0,3}\tra$    &$\zrp $                    &$  \fCsq   $&$\fIsq $&$  \fCsq   $&$  \fIsq   $     \\
${\bbox{F}}_{1}\tra$   &$\zrp $                     &$  \fCsq   $&$ \fIsq $&$  0   $&$  0   $     \\
${\bbox{F}}_{2}\tra$   &$\zrp $                     &$  \fCsq   $&$\fRsq $&$  0   $&$  0   $     \\
${\bbox{j}}_{0,3}\tra$    &$\rp ,\z $                   &$  \fCsq   $&$ \fIsq $&$  \fCsq   $&$  \fIsq   $     \\
${\bbox{j}}_{1}\tra$  &$\rp ,\z $                      &$  \fCsq   $&$ \fIsq $&$  0   $&$  0   $     \\
${\bbox{j}}_{2}\tra$      &$\rp ,\z $                  &$  \fCsq   $&$\fRsq $&$  0   $&$  0   $     \\
${\bbox{J}}_{0,3}\tra$   &$\rp ,\z $                     &$  \fCsq   $&$\fRsq $&$  \fCsq   $&$  \fRsq   $     \\
${\bbox{J}}_{1}\tra$   &$\rp ,\z $                     &$  \fCsq   $&$\fRsq $&$ 0   $&$  0   $     \\
${\bbox{J}}_{2}\tra$   &$\rp ,\z $                     &$  \fCsq   $&$\fIsq$&$  0   $&$  0   $     \\
$\underline{{\mathsf J}}_{0,3}\tra$   & $\underline{\bbox{r}_{\perp}\otimes (\bbox{z}\times\bbox{r}_{\perp})}$,
 $\underline{\bbox{z}\otimes (\bbox{z}\times\bbox{r}_{\perp})}$        &$  \fCsq   $&$  \fRsq   $&$  \fCsq   $&$  \fRsq   $     \\
$\underline{{\mathsf J}}_{1}\tra$   & $\underline{\bbox{r}_{\perp}\otimes (\bbox{z}\times\bbox{r}_{\perp})}$,
 $\underline{\bbox{z}\otimes (\bbox{z}\times\bbox{r}_{\perp})}$         &$  \fCsq   $&$  \fRsq   $&$  0   $&$  0   $     \\
$\underline{{\mathsf J}}_{2}\tra$   & $\underline{\bbox{r}_{\perp}\otimes (\bbox{z}\times\bbox{r}_{\perp})}$,
 $\underline{\bbox{z}\otimes (\bbox{z}\times\bbox{r}_{\perp})}$         &$  \fCsq   $&$  \fIsq   $&$  0   $&$  0   $     \\
\end{tabular}
\end{ruledtabular}
\renewcommand{\arraystretch}{1}
\end{table*}
\begin{table*}
\renewcommand{\arraystretch}{0.7}
\caption[T3]{Properties of local axially symmetric (SO$^{\perp}$(2)-invariant)
particle-hole transition densities depending on
the conserved (C) or broken (B) proton-neutron (p-n), or
time-reversal ($T$) symmetries. The $z$-simplex ($S_z$) symmetry is
broken. The vector, pseudovector, or pseudotensor densities take the
form of an expansion in the vector
($\bbox{e}_{\perp},\bbox{e}_{\phi},\bbox{e}_z$) or tensor
($\underline{\mathsf{P}}^{\perp},\underline{\mathsf{P}}^{z\perp},\underline{\mathsf{S}}^{\perp},\underline{\mathsf{S}}^{z\perp}$)
basis. The expansion coefficients are real, imaginary, or complex
functions of cylindrical coordinates $r_{\perp}$ and $z$. Generic
real, imaginary, or complex coefficients are denoted by
$f_R(z,r_{\perp})$, $f_I(z,r_{\perp})$, or $f_C(z,r_{\perp})$,
respectively.}
\label{tab3tra}
\begin{ruledtabular}
\begin{tabular}{l|lllll}
symmetry                                &      &
\multicolumn{4}{c}{conserved (C)
or broken (B)}                                                                   \\
\hline
$S_z$               &                    &   B    &     B  &     B  &     B       \\
p-n                       &              &   B    &     B  &     C  &     C       \\
$T$                         &            &   B    &     C  &     B  &     C       \\
\hline
     & basis &\multicolumn{4}{c} {coefficients}  \\
\hline
$\rho_{0,3}\tra$    &$ 1$                        &$  \fC   $&$\fR   $&$  \fC   $&$  \fR   $     \\
$\rho_{1}\tra$        & $ 1 $                    &$  \fC   $&$\fR   $&$  0   $&$  0   $     \\
$\rho_{2}\tra$         & $ 1 $                   &$  \fC   $&$\fI   $&$  0   $&$  0   $     \\
$\tau_{0,3}\tra$        & $ 1 $                  &$  \fC   $&$\fR   $&$  \fC   $&$  \fR   $     \\
$\tau_{1}\tra$             & $ 1 $              &$  \fC   $&$\fR   $&$  0   $&$  0   $     \\
$\tau_{2}\tra$           & $ 1 $                 &$  \fC   $&$\fI     $&$  0   $&$  0   $     \\
${J}_{0,3}\tra$           & $ 1 $                  &$  \fC   $&$  \fR   $&$  \fC   $&$  \fR   $     \\
${J}_{1}\tra$               & $ 1 $              &$  \fC   $&$  \fR   $&$  0   $&$  0   $     \\
${J}_{2}\tra$               & $ 1 $              &$  \fC   $&$  \fI   $&$  0   $&$  0   $     \\
$\bbox{s}_{0,3}\tra$   &$\epefez $               &$\fC   $&$\fI $&$\fC  $   &$  \fI   $     \\
$\bbox{s}_{1}\tra$   &$\epefez $                       &$\fC $&$\fI $&$0  $&$  0   $     \\
$\bbox{s}_{2}\tra$    &$\epefez $                      &$\fC $&$\fR $   &$  0   $&$0$     \\
${\bbox{T}}_{0,3}\tra$   &$\epefez $                     &$  \fC   $&$ \fI $&$  \fC   $&$  \fI   $     \\
${\bbox{T}}_{1}\tra$    &$\epefez $                    &$  \fC   $&$ \fI $&$   0 $&$  0   $     \\
${\bbox{T}}_{2}\tra$    &$\epefez $                    &$  \fC   $&$\fR $&$  0   $&$  0   $     \\
${\bbox{F}}_{0,3}\tra$    &$\epefez $                    &$  \fC   $&$ \fI $&$  \fC   $&$  \fI   $     \\
${\bbox{F}}_{1}\tra$   &$\epefez $                     &$  \fC   $&$ \fI $&$  0   $&$  0   $     \\
${\bbox{F}}_{2}\tra$   &$\epefez $                     &$  \fC   $&$\fR $&$  0   $&$  0   $     \\
${\bbox{j}}_{0,3}\tra$    &$\epefez $                   &$  \fC   $&$ \fI $&$  \fC   $&$  \fI   $     \\
${\bbox{j}}_{1}\tra$  &$\epefez $                      &$  \fC   $&$ \fI $&$  0   $&$  0   $     \\
${\bbox{j}}_{2}\tra$      &$\epefez $                  &$  \fC   $&$\fR$&$  0   $&$  0   $     \\
${\bbox{J}}_{0,3}\tra$   &$\epefez $                     &$  \fC   $&$\fR$&$  \fC   $&$  \fR   $     \\
${\bbox{J}}_{1}\tra$   &$\epefez $                     &$  \fC   $&$\fR$&$ 0   $&$  0   $     \\
${\bbox{J}}_{2}\tra$   &$\epefez $                     &$  \fC   $&$\fI$&$  0   $&$  0   $     \\
$\underline{{\mathsf J}}_{0,3}\tra$  & $\PpPz$,$\SpSz $       &$  \fC   $&$  \fR   $&$  \fC   $&$  \fR   $     \\
$\underline{{\mathsf J}}_{1}\tra$  & $\PpPz$,$\SpSz $        &$  \fC   $&$  \fR   $&$  0   $&$  0   $     \\
$\underline{{\mathsf J}}_{2}\tra$  & $\PpPz$,$\SpSz $        &$  \fC   $&$  \fI   $&$  0   $&$  0   $     \\
\end{tabular}
\end{ruledtabular}
\renewcommand{\arraystretch}{1}
\end{table*}

\section{Discussion of pairing channels and examples of previous approaches}\label{sec15}

The form of the most general EDF that is quadratic in local isoscalar
and isovector densities has been proposed in Ref.~I, where the
expressions for the {\particle} and {\pairing} mean fields can be
found. In current applications, pairing interaction is often
approximated by the zero-range pairing force
\cite{[Boc67],[Cha76],[Kad78],[Dob01a]},
\begin{equation}\label{ddpi}
V_{\text{pair}}(\bbox{r},\bbox{r}') =
f_{\text{pair}}(\bbox{r})\delta(\bbox{r}-\bbox{r}'),
\end{equation}
where the density-dependent form factor reads
\begin{equation}
f_{\text{pair}}(\bbox{r}) =
V_0\left\{1+x_0\hat{P}^\sigma
-\left[\frac{\rho_0(\bbox{r})}{\rho_c}\right]^\alpha
(1+x_3\hat{P}^\sigma)\right\} ,
\end{equation}
and $\hat{P}^\sigma$ is the usual spin-exchange operator. When only
the isovector pairing is studied, the exchange parameters $x_0$ and
$x_3$ are usually set to zero. However, in the general case of
coexisting isoscalar and isovector pairing correlations, nonzero
values of $x_0$ and $x_3$ have to be used.

In Ref.~\cite{[Sie06]}, the density-independent, zero-range pairing force,
\begin{equation}\label{dipi}
V_{\text{pair}}(\bbox{r},\bbox{r}') =
\sum_{TS}\left[p_0^T\delta(\bbox{r}-\bbox{r}')+p_2^T\bbox{k}'\delta(\bbox{r}-\bbox{r}')\bbox{k}
\right]\hat\Pi_{TS},
\end{equation}
has been employed to study the interplay between isoscalar and
isovector pairing within an axially symmetric HF+BCS scheme. In
Eq.~(\ref{dipi}), $\hat\Pi_{TS}$ stands for the spin-isospin
projection operator, and $p_0^T$ and $p_2^T$ are coupling strengths
adjusted to the data.

As seen in (I-84) and (I-89),
for the commonly used pairing force (\ref{ddpi}),
only two pairing densities come into play:
the isovector density
$\vec{\breve{\rho}}\ofbboxofr$ and
the
isoscalar  {\pairing} spin density $\breve{\bbox{s}}_0\ofbboxofr$.
The corresponding isovector {\pairing} potential
$\vec{\breve{U}}(\bbox{r})$ is simply  proportional to
$\vec{\breve{\rho}}\ofbboxofr$ while the isoscalar {\pairing} field
$\breve{\bbox{\Sigma}}_0(\bbox{r})$  is proportional to the scalar
product of the quasiparticle's  spin $\hat{\bbox{\sigma}}$ and $\breve{\bbox{s}}_0\ofbboxofr$.
Physically,  $\vec{\breve{\rho}}$ represents the  density of $S$=0,
neutron-neutron, proton-proton, and p-n  pairs while  the vector field
$\breve{\bbox{s}}_0\ofbboxofr$  describes  the spin distribution of
$S$=1 isoscalar p-n pairs. Indeed, when expressing  Eq.~(\ref{isoscdens})
directly in terms of the p-n pairs, one can see that
$\breve{\bbox{s}}_{0z}$ contains the $S$=1, $M_S$=0 component while
$\breve{\bbox{s}}_{0x}$ and $\breve{\bbox{s}}_{0y}$ contain
combinations of  $M_S$=1 and $M_S$=$-$1 pairs. The physical
interpretation of the isoscalar {\pairing} mean-field Hamiltonian
\begin{equation}\label{isoscmf}
\breve{h}_0(\bbox{r})
= \breve{\bbox{\Sigma}}_0\ofbboxofr\cdot\hat{\bbox{\sigma}} \propto
\breve{\bbox{s}}_0\ofbboxofr\cdot\hat{\bbox{\sigma}}
\end{equation}
is the projection of the quasiparticle's spin on the spin of the p-n
pairing field, its local magnitude determined by the HFB equations;
hence, the SCSs are present in the problem.

It is important to emphasize that the isoscalar density
$\breve{\bbox{s}}_0$ contains all magnetic components of the $S$=1 p-n
pairing field. When studying  an individual component separately,
e.g., in the context of the so-called $\alpha-\bar\alpha$ or
$\alpha-\alpha$ pairing \cite{[Goo72]}, one may arrive at erroneous
conclusions that the presence of  isoscalar pairing must be
associated with breaking  certain SCSs, such as axial symmetry or
signature. The usual argument, made originally in Ref.~\cite{[Goo72]}
and then repeated in the literature \cite{[Ter98],[Mul82],[Fra00b]},
is that the individual components of the $S$=1 pair field are not
invariant under rotations. For instance,  the pairing tensor
$\kappa_{1M}$ does not commute with  signature ${\cal R}_a$ ($a$=$x,
y, z$) \cite{[Fra00b]}:
\begin{equation}\label{kcomm}
{\cal R}_a^{-1} \kappa_{1M_a} {\cal R}_a = (-1)^{M_a} \kappa_{1M_a},
\end{equation}
and this has led to a conclusion that the isoscalar pairing must break signature.

\begin{figure}[htb]
\centerline{\includegraphics[width=0.9\columnwidth]{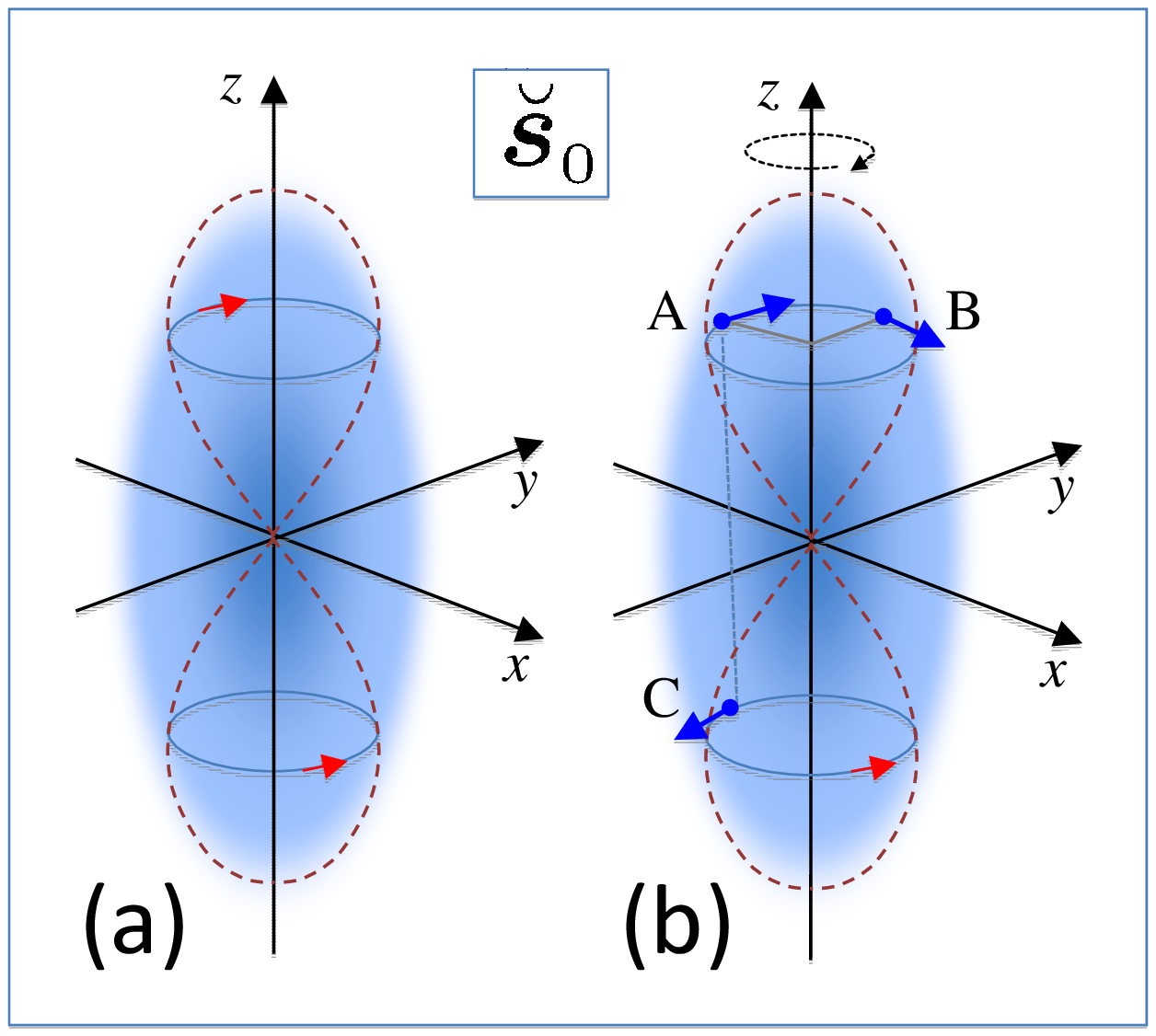}}
\caption{(Color online) (a) Schematic illustration of the isoscalar
vector field $\breve{\bbox{s}}_0$ in the case of conserved axial and
mirror symmetries. The field is solenoidal, with vanishing third
component. (b) Under rotation around the third (symmetry) axis, the
field at  point  $\bbox{r}_A$ is transformed to  position
$\bbox{r}_B$. Likewise, under ${\cal R}_x$, the field is transformed
to   $\bbox{r}_C$. While neither of these operations leave the
individual vector $\breve{\bbox{s}}_0(\bbox{r}_A)$ invariant, the
field as a whole does not change, i.e., it is covariant.
}
\label{fig1}
\end{figure}

Let us consider axial and mirror symmetry as SCS. As seen in Table
\ref{tab2b}, the isoscalar pairing density $\breve{\bbox{s}}_0$
vanishes only if the p-n symmetry is conserved. In the generalized
pairing theory that mixes proton and neutron orbits, the solenoidal
field $\breve{\bbox{s}}_0\ofbboxofr$ is nonzero. The lines of field
$\breve{\bbox{s}}_0$ are schematically shown in the left panel of
Fig.~\ref{fig1}. The right panel shows that while an individual
vector $\breve{\bbox{s}}_0$ at a given point is not invariant with
respect to symmetries ${\cal S}$  such as rotations around the third
axis or signature ${\cal R}_x$, the field
$\breve{\bbox{s}}_0\ofbboxofr$  is perfectly covariant
(\ref{covariant}):
\begin{equation}\label{scovariant}
\breve{\bbox{s}}_0^{\cal S}(\bbox{r})=\breve{\bbox{s}}_0({\cal S}^+ \bbox{r}{\cal S}).
\end{equation}
The scalar product (\ref{isoscmf}) is actually {\it invariant} in
both cases shown in Fig.~\ref{fig1}b. It is interesting to note that
for the geometry of  Fig.~\ref{fig1},  the third component
$\breve{\bbox{s}}_{0z}$ associated with the $M$=0 isoscalar pairing
field vanishes. That is, the solenoidal pairing field is created by
the two components with $M$=$\pm 1$. Therefore, we conclude that the
assumption of axial symmetry, or signature, does not preclude the
existence of isoscalar pairing.

For the finite-range pairing forces and for the general Skyrme {\pairing} functional, other pairing  densities appear in addition to
$\vec{\breve{\rho}}\ofbboxofr$ and
$\breve{\bbox{s}}_0\ofbboxofr$. In the order of importance,  the next
crucial densities are
the  isoscalar vector current density
$\breve{\bbox{j}}_{0}(r)$ and the isovector spin-current tensor density $\vec{\breve{\bbox{J}}}(\bbox{r})$. Both can be generated by the  momentum-dependent term in Eq.~(\ref{dipi}) and can be associated with  the $L$=1
pairing field.

The  isoscalar vector current density
is solely responsible for the isoscalar pairing field
in the spherical, mirror symmetric case when
time-reversal symmetry is broken. Indeed, according to Table~\ref{tab1b}, the $L$=0 density
$\breve{\bbox{s}}_0\ofbboxofr$ vanishes in this limit. It is only
when the mirror symmetry is broken that
$\breve{\bbox{s}}_0\ofbboxofr$ becomes nonzero in spherical nuclei.
Such a scenario could be an interesting possibility in very neutron-rich
nuclei, in which, e.g., the 1$\pi$d$_{5/2}$ and 1$\nu$f$_{5/2}$ or 1$\pi$f$_{7/2}$ and 1$\nu$g$_{7/2}$ orbitals
could appear near the Fermi surface.
The isovector tensor density $\vec{\breve{\bbox{J}}}(\bbox{r})$ is
generally nonzero in spherical nuclei, and it has radial character.

In the axial, parity-conserving  case $\breve{\bbox{j}}_{0}(\bbox{r})$  is perpendicular to $\breve{\bbox{s}}_0\ofbboxofr$.
It is interesting to note that when the  mirror symmetry  is broken, the
densities $\breve{\bbox{s}}_0\ofbboxofr$  and $\breve{\bbox{j}}_0\ofbboxofr$ have all components (radial, azimuthal, and vertical) nonzero but their geometries will differ in general.

In the absence of spin polarization, i.e., for time-reversal
invariant systems, the isoscalar pairing field $\breve{\bbox{s}}_0$ is
purely imaginary. In the presence of rotation, the time-reversal
symmetry is internally broken. In this case, the pairing field is
generally complex. Consequently, assuming the real Bogoliubov
transformation and  real pairing tensor \cite{[Ter98]} may limit the
domain of self-consistent solutions.

Another  factor that may impact
the generality of conclusions of Ref.~\cite{[Ter98]} is the lack of
the p-n symmetry-breaking on the HF level. Such an approximation does
not seem to be justified as the self-consistent polarization between
{\particle} and {\pairing} channels is well known in the isovector
pairing case. Originally, the condition that the {\particle} density matrix preserves the p-n symmetry  has been proposed in Ref.~\cite{[Goo72]} in the context of BCS calculations for $N$=$Z$ nuclei where it was postulated that the expectation value of isospin in the quasiparticle vacuum is zero:
\begin{equation}\label{Tzero}
\langle \Psi| \vec{T} |\Psi\rangle =0.
\end{equation}
We note that while the absence of the p-n mixing in the {\particle}  sector automatically guarantees the condition  (\ref{Tzero}) for the $N$=$Z$ nuclei in the absence of isospin-breaking interactions, the
three  constraints (\ref{Tzero}) are in general  not sufficient for  {\it all} the $t'=-t$ matrix elements of $\hat\rho$ to vanish.

In addition, the independent treatment of time-reversal and isospin symmetries as done in Ref.~\cite{[Goo72]} is not justified. Indeed, as pointed out in Ref.~\cite{[Per04]},  the time-reversal
and the isospin rotations do not commute. This implies that the relative phases between proton and neutron wave functions in a p-n broken quasiparticle state cannot be chosen arbitrarily.

\section{Conclusions}\label{sec11}

In this study, we investigated the symmetries of nucleonic densities
of the generalized nuclear DFT that allows for the arbitrary mixing
of protons and neutrons. We considered the most important self-consistent
symmetries: spherical, axial, space-inversion, and mirror symmetries.
The main conclusions of our work can be summarized as follows:
\begin{enumerate}
\item
The local pairing densities
$\breve{\rho}_0$ (isoscalar pairing density) and
$\vec{\breve{\bbox{s}}}$ (isovector spin density) always vanish.
\item
One can always construct a phase convention for which the local  {\particle}
densities  are purely real \cite{[Dob00a]}.
In the presence of  particular  SCSs,
some  {\particle} densities vanish.
\item
In the absence of SCSs, the local  {\pairing} densities are  complex.
If time reversal is SCS, {\pairing} densities become either purely
real or purely imaginary.
\item
If p-n  symmetry is SCS (no explicit p-n mixing), the
$k$=1 and 2 isospin components of {\particle} densities and $k$=0 and
3 isospin components of   {\pairing} densities vanish.
\item
When O(3) is SCS (spherical,  mirror-symmetric  case), the local
pseudoscalar, pseudovector, and pseudotensor densities vanish.  The
only nonzero isoscalar-pairing density is the  current density
$\breve{\bbox{j}}_{0}(r)$. All those densities can become nonzero in
the SO(3) limit when  the parity is broken. See Tables \ref{tab1a}
and \ref{tab1b} for a summary.
\item
When SO(2) and $S_z$ are  SCS (axial, mirror-symmetric  case),
pseudoscalar and pseudotensor densities vanish. If  time reversal  is
SCS, {\pairing} densities   become either purely real or purely
imaginary. Properties of local axially symmetric
(O$^{z\perp}$(2)-invariant) densities are listed in Tables
\ref{tab2a} and \ref{tab2b}.
\item
When SO(2) is SCS (axial case),  all densities are generally present.
Properties of the SO(2)-invariant local densities are listed in
Tables \ref{tab3a} and \ref{tab3b}.
\item
When space inversion, three signatures, and three simplexes are SCSs
(D$_{2\mathrm{h}}^\mathrm{D}$ group),  all densities are generally
present. Symmetry properties of {\particle} densities have been
previously discussed in Refs.~\cite{[Dob00a],[Dob00b]}. The analogous
expressions for {\pairing} densities are given in Sec.~\ref{sec8f}.
\item
The symmetry properties of the
transition densities are the same as those of the local  densities.
The only notable  difference is that the p-h transition densities
can be, in general, complex unless the time reversal is SCS
and some additional restrictions are present.
\item
The isoscalar pairing density $\breve{\bbox{s}}_0$  is the main
building block of the $T$=0 pairing field. In axial,
reflection-symmetric  nuclei, this field is solenoidal and gives rise to a
paring potential that preserves signature.
\item
The second most important building block of the isoscalar pairing
field is the  current density $\breve{\bbox{j}}_{0}$. It is solely
responsible for  the isoscalar pairing field in the spherical, mirror
symmetric case when time-reversal symmetry is broken. For axial
systems, this field is solenoidal.
\item
Breaking  space inversion or mirror reflection may have profound
consequences for the existence of isoscalar pairing as many isoscalar
densities vanishing in a parity-conserving limit  can become nonzero.
\end{enumerate}

The symmetry properties discussed in this work provide the necessary,
but not sufficient, conditions for the presence of isoscalar pairing
in nuclei. Whether such fields will appear or not depends, of course,
on the actual form of the EDF and the values of coupling constants.
In general, similar to the isovector pairing channel, a strong
dynamical coupling between {\particle} and {\pairing} channels is
expected. Consequently, in order to fully benefit from  the p-n
symmetry-breaking mechanism, p-n symmetry should be broken already on
the level of {\particle} mean field. This is not what has usually
been done in  existing calculations.

The main results of this paper, summarized  in Tables~I-IX and  in
relations in Sec.~\ref{sec8f}, are  symmetry properties of the local
{\particle} and {\pairing} densities that are building blocks of the
generalized nuclear DFT formalism. These results can be useful when
building a microscopic framework, rooted in the LDA, to describe
various phenomena occurring in $N$$\sim$$Z$ nuclei.

This work was supported in part by the Polish Ministry of Science
under Contract No.~N~N202~328234, by the Academy of Finland and
University of Jyv\"askyl\"a within the FIDIPRO programme, and by the
U.S.\ Department of Energy under Contract Nos.\ DE-FC02-07ER41457
(UNEDF SciDAC Collaboration), DE-FG02-96ER40963 (University of
Tennessee), DE-AC05-00OR22725 with UT-Battelle, LLC (Oak Ridge
National Laboratory), and DE-FG05-87ER40361 (Joint Institute for
Heavy Ion Research).

\appendix
\section{Generalization of the Cayley-Hamilton theorem}\label{sec13}

When analyzing symmetries, we often meet the problem of how to
construct a tensor quantity in terms of another tensor. Such a situation
has been first encountered   in investigations of phenomenological
constituent equations for macroscopic systems (see, e.g.,
\cite{[Tru52]}). The starting point in the analysis is the Cayley-Hamilton theorem
\cite{[Tur46],[Bir65]} by which an arbitrary function $f(A)$ of a
$3\times3$ matrix $A$ can be expressed as
\begin{equation}
\label{Cayley-Hamilton}
f(A)_{ab} = C_0(A)\delta_{ab}
          + C_1(A)A_{ab}
          + C_2(A)A^2_{ab},
\end{equation}
where the scalar functions $C_i(A)=C_i(a_1,a_2,a_3)$ ($i$=0, 1,
or 2) depend only on three independent invariants of $A$: $a_1$ (trace
of $A$), $a_2$ (trace of $A^2$), and $a_3$ (determinant of $A$).
In this case, matrix $A$ is  a rank-2 (reducible)
tensor ($\lambda=2$) and the tensor field $f(A)$ is also a rank-2
(reducible) tensor ($L=2$). Coupling of rank-2 tensors to $L=2$
simply corresponds to multiplying matrices, and then
Eq.~(\ref{Cayley-Hamilton}) can be derived from the Taylor expansion of
$f(A)$ combined with the observation that every matrix obeys its own
characteristic equation (any power $A^n$ ($n>2$)
can be written as a linear combination of  $A^0$, $A^1$, and
$A^2$).

In applications presented in this study, we are interested
in spherical (irreducible) tensors and tensor fields corresponding to the rotation
group O(3),
\begin{equation}
\hat{d}(\alpha\beta\gamma) = e^{i\gamma\vec{j}_z}e^{i\beta\vec{j}_y}
                             e^{i\alpha\vec{j}_z} ,
\end{equation}
cf.~Eq.~(\ref{romat}). In this case, $\chi_{\lambda\mu}$ is called irreducible
spherical tensor of rank $\lambda$ \cite{[Var88]} if in the rotated reference frame
it can be expressed as:
\begin{equation}
\chi'_{\lambda\mu'} \equiv \left[\hat{d}^+(\alpha\beta\gamma)
\chi\hat{d}(\alpha\beta\gamma)\right]_{\lambda\mu'}
= \sum_{\mu=-\lambda}^{\lambda} D^\lambda_{\mu'\mu}(\alpha\beta\gamma)
\chi_{\lambda\mu},
\end{equation}
where $D^\lambda_{\mu'\mu}(\alpha\beta\gamma)$ are the Wigner
functions \cite{[Var88]}.

Now let  $\Phi_{LM}$ be a tensor field of $\chi_{\lambda\mu}$:
\begin{equation}
\Phi'_{LM'}
= \sum_{M=-L}^{L} D^L_{M'M}(\alpha\beta\gamma) \Phi_{LM} ,
\end{equation}
which is a function of $2\lambda+1$ components of $\chi_{\lambda\mu}$,
\begin{equation}
\label{tensor-field2}
\Phi_{LM} =  \Phi_{ LM}\left(\chi_{\lambda\mu}\right),
\end{equation}
such that
\begin{equation}
\label{tensor-field}
\Phi'_{LM'} =  \Phi_{LM'}\left(\chi'_{\lambda\mu'}\right)  .
\end{equation}
In other words, we are interested only in tensor fields being the
isotropic functions of $\chi_{\lambda\mu}$ \cite{[Tru52]}.

Condition (\ref{tensor-field}) is essential: it states
that the rotated tensor field $\Phi'_{LM'}$ can be obtained by
calculating the original tensor field $\Phi_{LM}$ at arguments which
are the rotated tensor components $\chi'_{\lambda\mu'}$ of
$\chi_{\lambda\mu}$. It means that functions (\ref{tensor-field2}),
apart from depending on $\chi_{\lambda\mu}$, do not depend on any
other tensor object (fixed material tensor). If they did, values of
rotated tensor field $\Phi'_{LM'}$ could have been obtained by
rotating the arguments $\chi_{\lambda\mu}$ {\em and} all material tensors
simultaneously, but otherwise the rotation of arguments suffices.

From Eq.~(\ref{Cayley-Hamilton}) we derive  the
general form of the quadrupole $(L=2)$ field $\Phi_{2M}$ being an
isotropic  function of the quadrupole tensor $\chi_{2}$
\cite{[Eis87]}:
\begin{eqnarray}
\label{quad-field}
\Phi_{2M}(\chi_2)\!&=&\!C_1([\chi_2\times\chi_2]_0,[\chi_2\times\chi_2\times\chi_2]_0)\chi_{2M} \nonumber \\
&+&
C_2([\chi_2\times\chi_2]_0,[\chi_2\times\chi_2\times\chi_2]_0)[\chi_{2}\times\chi_2]_{2M}.\nonumber \\
\end{eqnarray}
The notation $[\ \times\  ]_L$ means the vector coupling to
multipolarity $L$. Symbols $\times$ and $\,{\cdot}\,$ for the vector
and scalar products of vectors used previously in the paper are, up
to  coefficients, equivalent to $[\ \times\   ]_1$ and $[\ \times\
]_0$, respectively.

To  generalize  Eq.~(\ref{quad-field}) to arbitrary values of $L$ and
$\lambda$, one should first  establish a (finite) complete
system of $i_{\lambda}$ irreducible
elementary tensors $\varepsilon_{lm}$  characteristic for a given
$\lambda$. (Irreducible means that none of them can be
expressed rationally and integrally in terms of the others.) The highest components $(m=l)$ of the elementary tensors
are called elementary factors. The elementary tensors are constructed
by successive vector couplings of  $n$ $\chi_{\lambda}$'s to different
intermediate $l$'s (meaning that $l<n\lambda$):
\begin{equation}
\label{elem-fact}
\varepsilon_{lm}^i\equiv\varepsilon_{lm}^{(n)}([\mathrm{c}])=[\underbrace{\chi_{\lambda}\times\dots\times\chi_{\lambda}}_{n}]_{lm}^{[\mathrm{c}]},
\end{equation}
for $i=1,\dots ,i_{\lambda}$, where symbol [c], redundant in most
cases, stands for a specific coupling scheme. Tensor
$\chi_{\lambda}=\varepsilon_{\lambda}^{(1)}$ is itself an elementary
tensor. In the particular case of  $l=0$, all independent scalars are
elementary factors. Completeness of the system of elementary factors
does not exclude the  relations  (syzygies) between them. The syzygies can be written in the form:
\begin{equation}
\label{syz}
S_j(\varepsilon_{ll}^1,\dots ,\varepsilon_{l'l'}^{i_{\lambda}})=0,
\end{equation}
for $j=1,\dots ,j_{\lambda}$, where $S_j$ are rational integral
functions. Again, the number of independent syzygies is finite and
depends on $\lambda$.

Having determined the elementary tensors,  we align them (i.e., couple to the maximal
multipolarity) to get a tensor of  rank $L$:
\begin{equation}
\label{aligned}
[\varepsilon_{l}^{i}\times\varepsilon_{l'}^{i'}\times\dots]_{L=l +l'+\dots}
\end{equation}
Alignment of the elementary tensors means  multiplication for the
elementary factors. Because of the existence of syzygies (\ref{syz}), some aligned tensors can be expressed in terms of
 others. Using syzygies, we find a {\em finite number} $k_{\lambda
L}$ of independent aligned tensors $\varphi_{LM}^{k}(\varepsilon
)$, ($k= 1,\dots ,k_{\lambda L}$) called fundamental tensors. It
turns out that an arbitrary  tensor field of rank $L$ being the
isotropic function of tensor $\chi_{\lambda}$ can  always be presented
in the form:
\begin{eqnarray}
\label{L-field}
\Phi_{LM}(\chi_{\lambda})&=&\sum_{k=1}^{k_{\lambda L}}C_k(\varepsilon_0)\varphi_{LM}^{k}(\varepsilon),
\end{eqnarray}
where argument $\varepsilon_0$ of $C_k$ stands for all the
independent scalars. Scalar functions $C_k$ can be, in general,
arbitrary. However, the form of some $C_k$ can be restricted. For
instance, high powers of some scalars $\varepsilon_0^i$  do not appear. This is because that a syzygy can make an expression
$(\varepsilon^i_{00})^n(\varepsilon^{i'}_{l'l'})^{n'}\dots
(\varepsilon^{i''}_{l''l''})^{n''}$ for some $i$ and $i',\,\dots ,i''$,
and $n$ and $n',\,\dots ,n''$ dependent on other elementary factors.
We refer to Eq.~(\ref{L-field}) as the Generalized Cayley-Hamilton
(GCH) theorem.

An explanation of the procedure presented above can be traced back to
the theory of covariants of algebraic forms given by Dickson
\cite{[Dic14]}, which, however, uses quite different notations than
those used here. Let us sketch the  main points of the
theory \cite{[Tur46]}. Let the  $2\lambda$-ic (of order $2\lambda$) algebraic form, binary in variables $x_1,\,
x_2$,  be given by:
\bn
\label{form}
&& F_{2\lambda}(x_1,x_2;\chi_{\lambda -\lambda},\dots ,\chi_{\lambda\lambda}) \nonumber \\
&&=\sum_{\mu=-\lambda}^{\lambda}\left(\ba{c}2\lambda \\ \lambda - \mu \ea\right)^{1/2}
\chi_{\lambda\mu}x_1^{\lambda +\mu}x_2^{\lambda -\mu},
\en
with $\chi_{\lambda\mu}$ ($\mu=-\lambda ,\dots ,\lambda$) being the set of coefficients.
Replacing the variables by a linear non-singular transformation
\begin{equation}
\label{lintrans}
x_i=\sum_{k=1}^2A_{ik}y_k
\end{equation}
for $i=1,2$, in the form of Eq.~(\ref{form}), one obtains:
\bn
\label{trform}
F_{2\lambda}(x_1,x_2;\chi_{\lambda} )&=&G_{2\lambda}(y_1,y_2;\chi_{\lambda}) \nonumber \\
&=&\sum_{\mu=-\lambda}^{\lambda}\left(\!\!\ba{c}2\lambda \\ \lambda - \mu \ea\!\!\right)^{1/2}
\psi{_\lambda\mu}(\chi_{\lambda})y_1^{\lambda +\mu}y_2^{\lambda -\mu}, \nonumber \\
\en
where $\psi_{\lambda\mu}$ is a new set of coefficients.
The $2l$-ic  form,
\bn
\label{cov}
H_{2l}^{(n)}(x_1,x_2;\chi_{\lambda})&=&\sum_{m=-l}^{l}\left(\!\!\ba{c}2l \\ l - m \ea\!\!\right)^{1/2}
\!\!\!\!h_{lm}^{(n)}(\chi_{\lambda})x_1^{l +m}x_2^{l -m}, \nonumber \\
\en
such that
\begin{equation}
\label{covdef}
H_{2l}^{(n)}(y_1,y_2;\psi_{\lambda})=(\mathrm{det}(A_{ik}))^wH_{2l}^{(n)}(x_1,x_2;\chi_{\lambda})
\end{equation}
we call  a (homogeneous) covariant of weight $w$ of $F$. Coefficients
$h_{lm}^{(n)}$ are homogeneous polynomials of order $n$ (called the
degree of the covariant) such that $\lambda n=l+w$. When $l=0$,
$H_0^{(n)}=h_{00}^{(n)}$ is an invariant of $F$. The polynomial in
front of the highest power of $x_1$ in Eq.~(\ref{cov}),
$h_{ll}^{(n)}$ is called a semi-invariant of $F$. The theory of covariants of algebraic forms shows  that an arbitrary
covariant can be expressed rationally and integrally in terms of a
finite, irreducibly complete set of covariants, which can be related
rationally and integrally to each other by a finite system of
independent syzygies (Gordan-Hilbert Finiteness Theorem, see
\cite{[Tur46]}). The number of basis covariants for a few lowest
$\lambda$'s is listed by Olver \cite{[Olv99]}.

What do the covariants of
algebraic forms have in common with the tensor fields as functions of
tensors? It turns out that semi-invariant
$h_{ll}^{(n)}(\chi_{\lambda})$ forms the highest projection of a
tensor of rank $l$ dependent on the tensor of rank $\lambda$ (the
heaviest state of an irreducible representation of SO(3) embedded in
an irreducible representation of SU($2\lambda +1$)). The remaining
polynomials in Eq.~(\ref{cov}) $h_{lm}^{(n)}(\chi_{\lambda})$ are other
 components of the same tensor. Proof of this statement is
outlined in Ref.~\cite{[Roh78a]}.

Constructive proof of Eq.~(\ref{L-field}) can be performed by an
explicit construction of a basis in the space of functions of
$\chi_{\lambda}$ and demonstration that it has the structure of
Eq.~(\ref{L-field}). In case of $\lambda =1$, we know  that an arbitrary tensor field of
rank $L$ as a function of the position vector $\bbox{r}$, takes the
form:
\begin{equation}
\label{field-1}
\Phi_{LM}(\bbox{r})=C_L(r)Y_{LM}\left(\frac{\bbox{r}}{r}\right),
\end{equation}
where $r^2=\bbox{r}$$\cdot$$\bbox{r}$ and $Y_{LM}$ is the
spherical harmonic. From Eq.~(\ref{field-1}) we see
immediately that in case of $\lambda =1$, there are two elementary
tensors $(i_{1}=2)$, namely vector $\bbox{r}$ and scalar $r^2$, and
no syzygy $(j_1=0)$. For every given $L$ there is only one
fundamental tensor $(k_{1L}=1)$ of the form \cite{[Var88]}:
\begin{eqnarray}
\label{sphar}
\varphi_{LM}^{(1)}(\bbox{r})&=&[\underbrace{\bbox{r}\times\bbox{r}\times\dots\times\bbox{r}}_{L}]_{LM} \nonumber \\
&\propto & r^LY_{LM}\left(\frac{\bbox{r}}{r}\right).
\end{eqnarray}
In particular, Eq.~(\ref{field-1}) tells that an arbitrary isotropic
vector field takes the form:
\begin{equation}
\bbox{\Phi}(\bbox{r})= C(r)\bbox{r}.
\end{equation}
Other examples are constructions of the oscillator bases in cases of $\lambda =2$ (see
Ref.~\cite{[Eis87]} and references quoted therein) and $\lambda =3$
\cite{[Roh78a]}.

\bibliographystyle{unsrt}

\end{document}